\documentclass[11pt,a4paper]{article}

\usepackage{jcappub}

\usepackage[english]{babel}
\usepackage{amsmath,amsfonts,amssymb}
\usepackage{comment}
\usepackage{bbold}
\usepackage{color}
\usepackage{hyperref}
\usepackage{booktabs}
\usepackage{graphicx}
\usepackage[font=normalsize]{caption}
\usepackage{ulem}
\usepackage{enumitem}

\usepackage{multirow}
\usepackage{arydshln}

\def\SO{\mathrm{SO}}
\def\SU{\mathrm{SU}}

\def\ADD{\addlinespace[4pt]}
\def\BOX{\hbox to 3cm}

\newcommand{\uset}[1]{\underset{\scalebox{0.65}{$#1$}}{}\hspace{-0.21cm}}

\begin{document}

\newcommand{\PRODUCT}[2]{\prod\limits_{i,j,k,l}\!q_{\alpha_i}(#1)\;q_{\beta_j}(#1)\;q_{\alpha'_k}(#2)\;q_{\beta'_l}(#2)}

\newcommand{\POLYFACTOR}[2]{P_{#1}(\kappa)^{n}\,R_{#1}(\kappa_1,\kappa_2)^{m}\,P_{#2}(\kappa)^{n'}\,R_{#2}(\kappa_1,\kappa_2)^{m'}}

\begin{titlepage}
\vspace*{-15mm}
\vspace*{0.7cm}

\begin{center}

	{\Large {\bf Yukawa ratio predictions in non-renormalizable \\[1.5mm] $\mathrm{SO}(10)$ GUT models}}\\[8mm]

	Stefan Antusch$^{\star}$\footnote{Email: \texttt{stefan.antusch@unibas.ch}},  		Christian Hohl$^\star$\footnote{Email: \texttt{ch.hohl@unibas.ch}}, 
	and Vasja Susi\v{c}$^\star$\footnote{Email: \texttt{vasja.susic@unibas.ch}}

\end{center}

\vspace*{0.20cm}

\centerline{$^{\star}$ \it
Department of Physics, University of Basel,}
\centerline{\it
Klingelbergstr.\ 82, CH-4056 Basel, Switzerland}

\vspace*{1.2cm}

\begin{abstract}
\noindent
Since $\mathrm{SO}(10)$ GUTs unify all fermions of the Standard Model plus a right-chiral neutrino in a representation $\mathbf{16}$ per family, they have the potential to be maximally predictive regarding the ratios between the masses (or Yukawa couplings) of different fermion types, i.e.~the up-type quarks, down-type quarks, charged leptons and neutrinos. 
We analyze the predictivity of classes of $\mathrm{SO}(10)$ (SUSY) GUT models for the fermion mass ratios, where the Yukawa couplings for each family are dominated by a single effective GUT operator of the schematic form $\mathbf{16}^2\cdot\mathbf{45}^n\cdot\mathbf{210}^{m}\cdot\mathbf{H}$, for $\mathbf{H}\in\{\mathbf{10},\mathbf{120},\mathbf{\overline{126}}\}$. This extends previous works to general vacuum expectation value directions for GUT-scale VEVs and to larger Higgs representations.  In addition, we show that the location of the MSSM Higgses in the space of all doublets is a crucial aspect to consider.
We discuss highly predictive cases and illustrate the predictive power in toy models consisting of masses for the 3rd and 2nd fermion family. 
\end{abstract}

\tableofcontents

\end{titlepage}


\section{Introduction}

Grand Unified Theories (GUTs) present an attractive framework for physics Beyond the Standard Model (BSM). Besides gauge coupling unification, they also unify fermions in joint GUT representations, presenting an interesting possibility to address the flavor puzzle, i.e.~the origin of the values of masses, mixings and CP violating phases. The most common GUT models are based on the unifying groups $\mathrm{SU}(5)$ and $\mathrm{SO}(10)$; this paper focuses on the latter choice, where an entire SM family of fermions and an additional right-handed neutrino can be embedded into a single representation $\mathbf{16}$.

From the point of view of the Yukawa sector and the flavor puzzle, one can distinguish two approaches to build unified models:
\begin{enumerate}
\item \textit{Minimal renormalizable models}, where a minimal set of irreducible particle representations in the fermionic and Higgs sectors are postulated. All renormalizable terms admitted by gauge symmetry are written down, and the GUT symmetry breaking is achieved with a minimal set of scalar representations. The bigger symmetry can manifest itself in a smaller number of free parameters for the masses and mixings compared to the Standard Model (SM). Usually, the predictions in this type of models appear as correlations between observables and are often rather complicated. They are typically made apparent only through a numeric fit and subsequent analysis. Regarding the predictions for the ratios of fermion masses, these are generically hidden due to each entry of the Yukawa matrix being generated by a linear combination of GUT operators.

\item \textit{Flavor models (as effective theories or renormalizable realizations)}, where the emphasis is put on explaining the observed mass ratios and mixings and on the maximal predictivity for the yet unmeasured observables. Despite models of this type having a larger particle content, they achieve predictivity by postulating a certain (continuous or discrete) ``family symmetry'', broken spontaneously by so-called ``flavon fields'', to ensure control over the textures of the Yukawa entries. This opens up the possibility that each entry of the Yukawa matrices is dominantly generated from one effective GUT operator, a scenario to which we will refer as ``\textit{single operator dominance}''.
When this condition is satisfied, the group-theoretic Clebsch-Gordan coefficients between the different fermion sectors can give rise to fixed ratios between the Yukawa entries. In the main part of this paper, we will focus on this scenario and the predictivity for the fermion mass ratios from the Clebsch factors of single effective GUT operators.
\end{enumerate}

In the context of $\mathrm{SO}(10)$, an example of a model of the first kind is the minimal renormalizable supersymmetric model  \cite{Bajc:2004xe,Bajc:2005qe,Bajc:2005zf,Bajc:2008dc} with a Higgs sector consisting of the irreducible representations $\mathbf{10}$, $\mathbf{\overline{126}}$, $\mathbf{126}$ and $\mathbf{210}$, where the first two are involved in the Yukawa sector of the model, while the second two ensure a suitable potential for GUT symmetry breaking.
A renormalizable Yukawa term  with a $\mathbf{120}$ is also possible. Some fits of renormalizable $\mathrm{SO}(10)$ Yukawa sectors can be found in 
\cite{Babu:2016bmy,Dueck:2013gca}.
Alternative setups using the $\mathbf{54}$ \cite{Babu:2018tfi}, or an additional vector-like fermion family $\mathbf{16}\oplus\overline{\mathbf{16}}$ \cite{Babu:2016cri} also fall under the approach of ``minimal renormalizable models''.

The second approach of flavor models is more prevalent in the context of $\mathrm{SU}(5)$ GUTs. The simplest examples are models which make use of Clebsch factors between the down and charged lepton sector from the renormalizable Yukawa  operators $\mathbf{10}_F\cdot\mathbf{\overline{5}}_{F}\cdot\mathbf{H}$ using the $\mathbf{\overline{5}}$ or $\mathbf{\overline{45}}$ for the Higgs representation $\mathbf{H}$. If each Yukawa entry comes from a single GUT operator (i.e.~in the case of \textit{single operator dominance}), the ratios of entries in the different sectors at the GUT scale are predicted from the type of operator used. 
The simplest aforementioned examples of operators lead to the well known cases of $b$-$\tau$ unification and the Georgi-Jarlskog factor~\cite{Georgi:1979df}. Clebsch factors arising from more general non-renormalizable operators have been studied in \cite{Antusch:2009gu,Antusch:2013rxa}; this was done through an approach where the non-renormalizable operators are assumed to arise as effective theory operators from a renormalizable theory after integrating out heavy mediator fields. For models built on this approach, see for example \cite{Antusch:2013wn,Antusch:2013kna,Antusch:2013tta,Girardi:2013sza,Antusch:2013rla,Zhao:2014qwa,Antusch:2014poa,Gehrlein:2014wda,Dimou:2015yng,Bjorkeroth:2015ora,Antusch:2017ano,Antusch:2018gnu}.

The purpose of this paper is to systematically explore the predictions of classes of (non-renormalizable) $\mathrm{SO}(10)$ GUT operators for the fermion mass ratios, extending the previous results (cf.\  \cite{Anderson:1993fe}), towards the construction of new $\mathrm{SO}(10)$ GUT flavor models with \textit{single operator dominance}. In contrast to $\mathrm{SU}(5)$, apart from the ubiquitous 3rd family Yukawa unification from an operator $\mathbf{16}_F\cdot\mathbf{16}_{F}\cdot\mathbf{10}_{H}$, existing flavor models in $\mathrm{SO}(10)$ typically use linear combinations of operators at least for the masses of the second and first family (e.g.\ \cite{King:2009tj,Anandakrishnan:2012tj,Bjorkeroth:2017ybg}). 
There are a number of new issues and circumstances arising in $\mathrm{SO}(10)$ compared to $\mathrm{SU}(5)$: For instance, all fermions are now in a single representation $\mathbf{16}$. As a consequence, the $\mathrm{SO}(10)$ symmetry relates not just entries in $\mathbf{Y}_d$ (down sector) and $\mathbf{Y}_e$ (charged lepton sector), but $\mathbf{Y}_u$ (up sector) and $\mathbf{Y}_\nu$ (neutrino sector) as well. This means each operator in principle provides $3$ ratios between sectors instead of $1$, making $\mathrm{SO}(10)$ symmetry much more predictive. Also, effective operators are constructed with the use of representations, whose SM singlet components acquire GUT-scale vacuum expectations values (VEVs). In contrast to $\mathrm{SU}(5)$, where the representations $\mathbf{24}$ and $\mathbf{75}$ have only one SM singlet, the representations $\mathbf{45}$ and $\mathbf{210}$ of $\mathrm{SO}(10)$ have $2$ and $3$ SM singlets, respectively. Therefore the direction of the VEV in a multidimensional space of singlets becomes important. Furthermore, the Yukawa ratio predictions in $\mathrm{SO}(10)$ are affected by where the Minimal Supersymmetric SM (MSSM) Higgs doublets are located with respect to the doublet flavor eigenstates. For assessing the predictivity of fermion mass ratios coming from $\mathrm{SO}(10)$ GUT operators all these aspects have to be taken into account. 

The paper is organized as follow: in Section~\ref{sec:operators} we consider the class of non-renormalizable superpotential operators where the $\mathbf{45}$ or $\mathbf{210}$ in $\SO(10)$ acquire GUT-scale VEVs, and compute their predictions.
In Section~\ref{sec:Higgs-location} we show that the location of MSSM Higgses in the doublet states crucially impacts the Yukawa results and provide some predictive scenarios for model building. In Section~\ref{sec:prescription-for-model-building}, we then combine the previous results into a discussion on model building, and provide $3$ example toy models with the $2nd$ and $3rd$ family effective operators. We then conclude. Additionally, a number of more technical considerations have been relegated to the appendices: Appendix~\ref{appendix:conventions} contains the description of $\mathrm{SO}(10)$ conventions used in this paper, while Appendix~\ref{appendix:operators-mediators} analyzes the different constructions of non-renormalizable operators via mediator fields.

\section{A class of effective Yukawa operators in $\mathrm{SO}(10)$\label{sec:operators}}
We consider in this paper a class of non-renormalizable superpotential operators in $\SO(10)$ GUTs, which consist of the representations 
\begin{align}
	\mathbf{16}_I\cdot\mathbf{16}_J\cdot\mathbf{H}
	\cdot\mathbf{45}^n\cdot\mathbf{210}^m
\end{align}
with arbitrary powers $n$ and $m$ and different choices of $\mathbf{H}$ containing the SM Higgs field. As we shall see later, the way we contract the indices in such an invariant also plays a role. This section is dedicated to motivating these operators, properly analyzing them, as well as obtaining their predictions for Yukawa couplings. We consider these operators in a supersymmetric (SUSY) setting, but the result contained herein can also be applied to the non-supersymmetric case.

First, we fix the notation used in this paper. The irreducible representations of groups will be typed in boldface, and we use the labels $G_{51}$, $G_{422}$ and $G_{321}$ for the groups $\SU(5)\times\mathrm{U}(1)_X$, the Pati-Salam group $\SU(4)_C\times\SU(2)_L\times\SU(2)_R$ and the Standard Model group $\SU(3)_C\times\SU(2)_L\times\mathrm{U}(1)_Y$ group, respectively. The groups $G_{51}$ and $G_{422}$ are both maximal subgroups of $\SO(10)$, which contain the SM group $G_{321}$.

A very convenient property of the group $\SO(10)$ is that SM fermions of one family, alongside a right-handed neutrino, all fit into a single spinorial representation $\mathbf{16}$ of $\SO(10)$: its decomposition into irreducible representations of $G_{321}$ is

\begin{align}
\mathbf{16}&=\underbrace{(\mathbf{3},\mathbf{2},+1/6)\;\oplus\;(\mathbf{\bar{3}},\mathbf{1},-2/3)\oplus(\mathbf{1},\mathbf{1},+1)}_{\mathbf{10}\ \text{of}\ \mathrm{SU}(5)}\oplus \underbrace{(\mathbf{\bar{3}},\mathbf{1},+1/3)\oplus (\mathbf{1},\mathbf{2},-1/2)}_{\mathbf{\bar{5}}\ \text{of}\ \mathrm{SU}(5)}\oplus (\mathbf{1},{\mathbf{1}},0)\\
&\equiv Q\oplus u^c\oplus e^c \oplus d^c \oplus L \oplus \nu^c.
\end{align}

We take the above SM embedding and assume that the fermionic sector of the theory contains $3$ copies of this spinorial representation, which we label by $\mathbf{16}_{I}\equiv\mathbf{16}_{FI}$, where $F$ denotes that the representation is ``fermionic'' and $I$ is a family index with $I=1,2,3$. 

At the renormalizable level, two fermionic representations couple to a single Higgs representation $\mathbf{H}$, where the MSSM Higgs doublet/antidoublet $(\mathbf{1},\mathbf{2},\pm 1/2)$ with an EW scale VEV reside (at least partly\footnote{The MSSM Higgs doublet pair can be merely the lightest such pair in the theory. Since the flavor and mass eigenstates do not coincide, $\mathbf{H}$ may contain only part of the MSSM Higgs. This is to be discussed in detail later in Section~\ref{sec:Higgs-location}.}). Suitable representations $\mathbf{H}$ are determined by considering the decomposition of the tensor product of two fermionic representations:
\begin{align}
\mathbf{16}\,\otimes\, \mathbf{16}&=\mathbf{10}_s\oplus\mathbf{126}_s\oplus\mathbf{120}_a,
\end{align}
where $s$ and $a$ denote whether the representation lives in the symmetric or antisymmetric term of the product, respectively. Each of the representations on the right-hand side contain weak doublets/antidoublets, so they can be used to obtain (MS)SM Yukawa terms. At the renormalizable level we thus have exactly $3$ possible Yukawa terms: 
\begin{align}
&\mathbf{16}_{I}\cdot \mathbf{16}_{J}\cdot\mathbf{10},&
&\mathbf{16}_{I}\cdot \mathbf{16}_{J}\cdot\mathbf{\overline{126}},&
&\mathbf{16}_{I}\cdot \mathbf{16}_{J}\cdot\mathbf{120},\label{eq:renormalizable-operators}
\end{align}
associated with $3\times 3$ Yukawa matrices $Y^{10}_{IJ}$, $Y^{\overline{126}}_{IJ}$ and $Y^{120}_{IJ}$. The matrices $Y^{10}$ and $Y^{\overline{126}}$ are symmetric in the indices $I$ and $J$, while $Y^{120}$ is antisymmetric. These statements are all well known, and numeric fits of these operators have been performed, see e.g.~\cite{Deppisch:2018flu,Babu:2016bmy,Dueck:2013gca}.

We now consider possible extensions of such a Yukawa sector, making use of non-renormalizable operators. Assuming no new content in the fermionic sector, i.e.~the fermionic sector still consists of $3$ copies $\mathbf{16}_{I}$, the Yukawa operators consists of two fermionic factors $\mathbf{16}$, a factor containing the SM Higgs, and possible further factors containing SM singlets acquiring GUT-scale VEVs. In order to obtain new prediction possibilities, the extra GUT-scale factors should not form an invariant by themselves, and should thus be contracting in the invariant with the fermionic and Higgs factors in a non-trivial way.
Renormalizable operators from Eq.~\eqref{eq:renormalizable-operators} are  thus most conveniently extended by factors of (self-conjugate) representations $\mathbf{45}$ and $\mathbf{210}$. It is exactly such a class of operators that we systematically consider in this paper. We limit ourselves to invariants where the contractions of the extra factors is performed in a ``spinorial way'', a detail that we discuss later.

The operators under consideration are schematically written in the following way:
\begin{align}
(\prod_{i=1}^{n}\mathbf{45}_{\alpha_i} \cdot \prod_{j=1}^{m}\mathbf{210}_{\beta_j}\cdot \mathbf{16}_{I})
\cdot \mathbf{H} \cdot 
(\prod_{k=1}^{n'}\mathbf{45}_{\alpha'_k} \cdot \prod_{l=1}^{m'}\mathbf{210}_{\beta'_l}\cdot \mathbf{16}_{J}),\label{eq:invariant-general}
\end{align}
\noindent
where $\mathbf{H}$ stands for a $\mathbf{10}$, $\mathbf{\overline{126}}$, or $\mathbf{120}$, and the integers $n,n',m,m'\geq 0$. These integers denote the number of factors in the product. Note that each of the $n+n'$ factors of the representation $\mathbf{45}$ is equipped with an index $\alpha_i$ or $\alpha'_k$, since we are considering the possibility that we have multiple copies of this representation as part of the Higgs sector, i.e.~they may contain different field degrees of freedom. Analogously, we label the possibly different $m+m'$ factors of the representation $\mathbf{210}$ with indices $\beta_j$ and $\beta'_l$. 
We use the more complicated labels, e.g.~$\alpha_i$ instead of just $i$, for later convenience.
For $n=n'=m=m'=0$ the operator reduces to one of the $3$ usual renormalizable operators in Eq.~\eqref{eq:renormalizable-operators}, depending on $\mathbf{H}$. 

The class of invariants under consideration from Eq.~\eqref{eq:invariant-general} can be most conveniently constructed
by writing the representations $\mathbf{45}$, $\mathbf{210}$ and $\mathbf{H}$ in spinorial form as $32\times 32$ matrices. We describe the detailed group theory conventions and procedures for this in Appendix~\ref{appendix:conventions}. In short, the spinorial form of e.g.~the representation $\mathbf{45}$ has the index structure $\mathbf{45}^{A}{}_{B}$, with both the upper index $A$ and lower index $B$ running from $1$ to $32$. The $\mathbf{210}$ and $\mathbf{H}$ have the same index structure of one upper and one lower index.

The representations $\mathbf{45}$ and $\mathbf{210}$ acquire GUT-scale VEVs; for obtaining Yukawa operators the only relevant states in them are the SM singlets. Crucially, these are found (e.g.~by explicit computation) to be on the diagonal of their $32\times 32$ matrix form, and thus $\langle\mathbf{45}\rangle$ and $\langle\mathbf{210}\rangle$ commute and their order in the invariant is not important. In contrast, their commutation with $\mathbf{H}$ is non-trivial, and thus it is important to distinguish on which side of $\mathbf{H}$ the $\mathbf{45}$ or $\mathbf{210}$ representations lie. If we imagine the invariants from Eq.~\eqref{eq:invariant-general} to be generated from a renormalizable theory by integrating out heavy mediators of the type $\mathbf{16}$ and $\mathbf{\overline{16}}$, the situation can be summarized by stating that external legs of the diagram containing $\mathbf{45}$s and $\mathbf{210}$s commute with each other, but not with the external leg of the field $\mathbf{H}$. This is the reason for distinguishing the GUT VEV representations acting on $\mathbf{16}_I$ and $\mathbf{16}_J$ in Eq.~\eqref{eq:invariant-general} (primed and non-primed indices), while their internal order is not relevant. Considerations regarding mediators are investigated in detail in Appendix~\ref{appendix:operators-mediators}.

Given the discussion above, the invariants in \eqref{eq:invariant-general} can be  written explicitly with index contractions as 
\begin{align}
&
\Big(\prod_{i=1}^{n}\prod_{j=1}^{m}\mathbf{45}_{\alpha_i}\mathbf{210}_{\beta_j}\Big)^{A}{}_{D}\,(\mathbf{16}_{I})^{D}\,C_{AB}\,\mathbf{H}^{B}{}_{E}\,
\Big(\prod_{k=1}^{n'}\prod_{l=1}^{m'}\mathbf{45}_{\alpha'_k}\mathbf{210}_{\beta'_l}\Big)^{E}{}_{F}\,(\mathbf{16}_{J})^{F},
\label{eq:invariant-explicit}
\end{align}
where all capital indices $A,B,D,E,F$ run from $1$ to $32$ and $C_{AB}$ are the components of the ``charge conjugation'' operator, see Appendix~\ref{appendix:conventions}. The products over $i,j,k,l$ in parentheses are understood as ordinary matrix multiplication, e.g.~
\begin{align}
\Big(\prod_{i=1}^{2}\mathbf{45}_{\alpha_i}\Big)^{A}{}_{B}\equiv (\mathbf{45}_{\alpha_1})^{A}{}_{D}\;(\mathbf{45}_{\alpha_2})^{D}{}_{B}.
\end{align}
As already discussed, the GUT VEVs of such objects simply form a diagonal $32\times 32$ matrix. The $\mathbf{10}$, $\mathbf{120}$ and $\mathbf{\overline{126}}$ (the $\mathbf{H}$), on the other hand, acquire an EW scale VEV in their doublet/antidoublet components.

Concerning the GUT-scale VEVs, it is important to note that each $\mathbf{45}$ contains $2$ SM singlet states, while the $\mathbf{210}$ has $3$ SM singlets. Their VEVs can thus have an arbitrary direction in the spaces of singlet fields, i.e.~in the spaces $\mathbb{F}^2$ and $\mathbb{F}^3$, respectively. In a non-SUSY GUT $\mathbb{F}=\mathbb{R}$, while in SUSY GUTs $\mathbb{F}=\mathbb{C}$, since chiral supermultiplets contain complex fields and the real representations thus need to be complexified.

For specifying the singlet states, we make use of a basis adapted to the maximal subgroup $G_{51}$, under which basis states all lie in a single irreducible representation of this subgroup. These decompositions can be found in Table~\ref{tab:decompositions-su5} of Appendix~\ref{appendix:conventions}. We label the VEVs of the basis singlet fields by
\begin{align}
	X_{1}&:=\langle \mathbf{1}\rangle_{\mathbf{45}},&
	X_{2}&:=\langle \mathbf{24}\rangle_{\mathbf{45}},\nonumber\\
	Z_{1}&:=\langle \mathbf{1}\rangle_{\mathbf{210}},&
	Z_{2}&:=\langle \mathbf{24}\rangle_{\mathbf{210}},&
	Z_{3}&:=\langle \mathbf{75}\rangle_{\mathbf{210}}.\label{eq:VEVs-SU5}
\end{align}
The underlying singlet fields of the $X$ and $Z$ VEVs have well-defined transformation properties under $G_{51}$: they belong to the $G_{51}$ irreducible representation in the angled brackets, while their $\SO(10)$ origin is denoted in the index. Note that in a SUSY scenario the VEVs $X$ and $Z$ have in general complex values.

Alternatively, we can use a basis of singlet states adapted to the irreducible representations of the Pati-Salam group $G_{422}$, which is the other regular maximal subgroup of $\SO(10)$ containing $G_{321}$. All associated decompositions of representations are given in Table~\ref{tab:decompositions-patisalam} of Appendix~\ref{appendix:conventions}. We denote the VEVs in the Pati-Salam adapted basis by
\begin{align}
	\tilde{X}_{1}&:=\langle (\mathbf{1},\mathbf{1},\mathbf{3})\rangle_{\mathbf{45}},&
	\tilde{X}_{2}&:=\langle (\mathbf{15},\mathbf{1},\mathbf{1})\rangle_{\mathbf{45}},\nonumber\\
	\tilde{Z}_{1}&:=\langle (\mathbf{1},\mathbf{1},\mathbf{1})\rangle_{\mathbf{210}},&
	\tilde{Z}_{2}&:=\langle (\mathbf{15},\mathbf{1},\mathbf{1})\rangle_{\mathbf{210}},&
	\tilde{Z}_{3}&:=\langle (\mathbf{15},\mathbf{1},\mathbf{3})\rangle_{\mathbf{210}}.\label{eq:VEVs-ps}
\end{align}

\noindent
The relation between the $G_{51}$ and $G_{422}$ adapted bases is explicitly computed to be

\begin{align}
\begin{split}
 \tilde{X}_{1}&= \sqrt{\tfrac{2}{5}}\,X_{1}-\sqrt{\tfrac{3}{5}}\,X_{2}, \\
 \tilde{X}_{2}&= \sqrt{\tfrac{3}{5}}\,X_{1}+\sqrt{\tfrac{2}{5}}\,X_{2}, \\
 \tilde{Z}_{1}&= \tfrac{1}{\sqrt{10}}\,Z_{1}-\sqrt{\tfrac{2}{5}}\,Z_{2}+\tfrac{1}{\sqrt{2}}\,Z_{3}, \\
 \tilde{Z}_{2}&= \sqrt{\tfrac{3}{10}}\,Z_{1}+2 \sqrt{\tfrac{2}{15}}\,Z_{2}+\tfrac{1}{\sqrt{6}}\,Z_{3}, \\
 \tilde{Z}_{3}&= \sqrt{\tfrac{3}{5}}\,Z_{1}-\tfrac{1}{\sqrt{15}}\,Z_{2}-\tfrac{1}{\sqrt{3}}\,Z_{3}.\\
\end{split}
\end{align}
These VEV relations implicitly contain some relative phase conventions for the VEVs. We normalize the $\SU(5)$ and Pati-Salam aligned states so that their VEVs are orthonormal, i.e.
\begin{align}
\langle \mathbf{45}^\ast_{ij}\cdot\mathbf{45}^{ij}\rangle&=|X_1|^2+|X_2|^2\qquad\qquad=|\tilde{X}_1|^2+|\tilde{X}_2|^2,\nonumber\\
\langle \mathbf{210}^\ast_{ijkl}\cdot\mathbf{210}^{ijkl}\rangle&=|Z_1|^2+|Z_2|^2+|Z_3|^2\quad=|\tilde{Z}_1|^2+|\tilde{Z}_2|^2+|\tilde{Z}_3|^2,\label{eq:VEVs-SU5-and-PS}
\end{align}
where the contracted indices are the complex (anti)fundamental indices of the representation $\mathbf{10}$, see Appendix~\ref{appendix:conventions} for details. In the $\mathbf{H}$-representations $\mathbf{10}$, $\mathbf{\overline{126}}$ and $\mathbf{120}$, we have doublets $(\mathbf{1},\mathbf{2}, 1/2)$ and $(\mathbf{1},\mathbf{2}, -1/2)$ of the SM group $G_{321}$; we denote their neutral components by $H^{u}_{x}$ and $H^{d}_{x}$, respectively, where the index $x$ specifies the exact state. There is one doublet-antidoublet pair in each of the representations $\mathbf{10}$ and $\mathbf{\overline{126}}$, but two pairs in $\mathbf{120}$. Furthermore, the representation $\mathbf{\overline{126}}$ also contains a SM singlet. We label these states in the following way under the embedding chain $G_{321}\subseteq G_{51}\subseteq \SO(10)$: the doublets are
\begin{align}
\begin{split}
H^{u}_{1}&\quad\equiv\quad (\mathbf{1},\mathbf{2},+1/2)  \quad\subseteq\quad \mathbf{\phantom{4}5}(2) \quad\subseteq\quad \mathbf{10},\\
H^{u}_{2}&\quad\equiv\quad (\mathbf{1},\mathbf{2},+1/2)  \quad\subseteq\quad \mathbf{\phantom{4}5}(2) \quad\subseteq\quad \mathbf{\overline{126}},\\
H^{u}_{3}&\quad\equiv\quad (\mathbf{1},\mathbf{2},+1/2)  \quad\subseteq\quad \mathbf{\phantom{4}5}(2) \quad\subseteq\quad \mathbf{120},\\
H^{u}_{4}&\quad\equiv\quad (\mathbf{1},\mathbf{2},+1/2)  \quad\subseteq\quad \mathbf{45}(2) \quad\subseteq\quad \mathbf{120},\\
\end{split}
\label{eq:hu-list}
\end{align}
the antidoublets are
\begin{align}
\begin{split}
H^{d}_{1}&\quad\equiv\quad (\mathbf{1},\mathbf{2},-1/2)  \quad\subseteq\quad \mathbf{\phantom{4}\overline{5}}(-2) \quad\subseteq\quad \mathbf{10},\\
H^{d}_{2}&\quad\equiv\quad (\mathbf{1},\mathbf{2},-1/2)  \quad\subseteq\quad \mathbf{\overline{45}}(-2) \quad\subseteq\quad \mathbf{\overline{126}},\\
H^{d}_{3}&\quad\equiv\quad (\mathbf{1},\mathbf{2},-1/2)  \quad\subseteq\quad \mathbf{\phantom{4}\overline{5}}(-2) \quad\subseteq\quad \mathbf{120},\\
H^{d}_{4}&\quad\equiv\quad (\mathbf{1},\mathbf{2},-1/2)  \quad\subseteq\quad \mathbf{\overline{45}}(-2) \quad\subseteq\quad \mathbf{120},\\
\end{split}
\label{eq:hd-list}
\end{align}
and the singlet VEV in the $\mathbf{\overline{126}}$ is denoted by 
\begin{align}
\overline{\Delta}&\quad\equiv\quad \langle(\mathbf{1},\mathbf{1},0)\rangle \quad\subseteq\quad \langle\mathbf{1}(10)\rangle \quad\subseteq\quad \langle\mathbf{\overline{126}}\rangle. 
\end{align}
All the above states $H^{u}_x$, $H^{d}_{x}$ and the VEV $\overline{\Delta}$ are (canonically) normalized so as to be orthonormal, i.e.~analogous to Eq.~\eqref{eq:VEVs-SU5-and-PS} when using the following contracted expressions:
\begin{align}
&\mathbf{10}^{\ast}_{i}\,\mathbf{10}^i,&&\mathbf{120}^{\ast}_{ijk}\,\mathbf{120}^{ijk}, && \mathbf{\overline{126}}^{\ast}_{ijklm}\,\mathbf{\overline{126}}^{ijklm}.
\end{align}

With all definitions at hand, we proceed to the explicit results for the Yukawa terms generated by a non-renormalizable operator of the type specified in Eq.~\eqref{eq:invariant-explicit}. There are, broadly speaking, two lines of inquiry one can follow: 
\begin{enumerate}
\item The acquired GUT-scale VEVs in the
$\mathbf{45}$ and $\mathbf{210}$ are in \textit{discrete} directions corresponding to singlets contained in a single irreducible representation of the maximal subgroups $G_{51}$ and $G_{422}$. These special discrete directions can be obtained for example due to the choice of invariants used in the (super)potential for the Higgs sector.
\item The acquired GUT-scale VEVs have an \textit{arbitrary} (continuous) direction in the space of SM singlets for the representations $\mathbf{45}$ and $\mathbf{210}$.
\end{enumerate}
We state the results for each possibility in a dedicated subsection. A mixed case with some factors having VEVs in discrete directions and some factors in arbitrary directions would in principle also be possible; we do not consider such a case here, but the explicit results can be inferred by combining the results of the discrete and arbitrary direction cases.

\subsection{Discrete directions\label{sec:discrete-directions}}

We assume that each representation $\mathbf{45}_{\alpha_i}$ and $\mathbf{45}_{\alpha'_k}$ in Eq.~\eqref{eq:invariant-general} has a discrete alignment of its VEV along one of the directions $X_1$, $X_2$, $\tilde{X}_1$ or $\tilde{X}_2$ defined in Eq.~\eqref{eq:VEVs-SU5} and \eqref{eq:VEVs-ps}. These directions have well-defined transformation properties in a maximal subgroup, i.e.~their corresponding particle states lie in a single irreducible representation of a maximal subgroup of $\SO(10)$. Each $\mathbf{45}_{\alpha_i}$ and $\mathbf{45}_{\alpha'_k}$ factor therefore has one of $4$ possible alignments, which we can specify in our notation e.g.~by $\alpha_1=X_1$ or $\alpha_1=\tilde{X}_2$.

We analogously assume that each $\mathbf{210}_{\beta_j}$ and $\mathbf{210}_{\beta'_l}$ has one of $6$ possible discrete alignments ($Z$s and $\tilde{Z}$s) with well-defined transformation properties under a maximal subgroup.

This setup in the context of $\SO(10)$ is an extension of the analysis performed in \cite{Anderson:1993fe}, where only the factors $\mathbf{45}$ and $\mathbf{H}=\mathbf{10}$ were considered. We shall not consider in this paper how to obtain such alignments with GUT Higgs potentials after GUT breaking. Constructing the Yukawa sector of a model with operators using discrete VEV alignments predicts (apart from the location of MSSM Higgs doublets, to be discussed later in this section and in Section~\ref{sec:Higgs-location}) ratios of Yukawa matrix entries in different sectors, and is thus
the natural implementation of the predictive Clebsch approach from $\mathrm{SU}(5)$ flavor models, see e.g.~\cite{Antusch:2009gu, Antusch:2013rxa,Antusch:2018gnu}.

As already stated, the $32\times 32$ matrices $\mathbf{45}^{A}{}_{B}$ and $\mathbf{210}^{A}{}_{B}$ acquire the VEVs $X,Z$ and $\tilde{X},\tilde{Z}$ on the diagonal. We label the coefficient alongside the VEV in the diagonal entries (the ``charge'') corresponding to the particle $p$ by $q(p)$, and normalize such that $q(Q)=1$ for the quark doublet $Q$, or in the case when $q(Q)=0$ we normalize by $q(u^c)=1$. Charges for different particles $p$ in the same SM irreducible representations need to be the same due to $G_{321}$ symmetry. Furthermore, the underlying space of states of the spinorial representation is reducible: $\mathbf{32}=\mathbf{16}\oplus\mathbf{\overline{16}}$.
The particles in the $\mathbf{\overline{16}}$ have opposite charges of those in the $\mathbf{16}$, including the $q$ charge discussed above, i.e.~$q(\bar{p})=-q(p)$.

\def\BOXX{\hbox to 1.8cm}

\begin{table}[htb]
	\begin{centering}
	\caption{The charges $q$ of different VEV alignments $X$ of the $\mathbf{45}$ on fermions.
     \label{table:qnumbers-45}}
            \begin{tabular}{crrrr}
            \toprule
            particle&\BOXX{\hss$X_1$}&\BOXX{\hss$X_2$}&\BOXX{\hss$\tilde{X}_1$}&\BOXX{\hss$\tilde{X}_2$}\\
            \midrule
            $Q$     &$1$ &$1$ &$0$&$1$\\
            $u^c$   &$1$ &$-4$&$1$&$-1$\\
            $d^c$   &$-3$&$2$ &$-1$&$-1$\\
            $L$     &$-3$&$-3$&$0$&$-3$\\
            $e^c$   &$1$ &$6$ &$-1$&$3$\\
            $\nu^c$ &$5$ &$0$ &$1$&$3$\\
            \midrule
            $\mathcal{N}$&$-\sqrt{2/5}$&$-2/\sqrt{15}$&$2$&$-\sqrt{2/3}$\\
            \bottomrule
            \end{tabular}
        \par
        \end{centering}
        \end{table}

\begin{table}[htb]
        \begin{centering}
        \caption{The charges $q$ of different VEV alignments $Z$ of the $\mathbf{210}$ on fermions. \label{table:qnumbers-210}}
            \begin{tabular}{crrrrrr}
            \toprule
            particle&
            \BOXX{\hss$Z_1$}&
            \BOXX{\hss$Z_2$}&
            \BOXX{\hss$Z_3$}&
            \BOXX{\hss$\tilde{Z}_1$}&
            \BOXX{\hss$\tilde{Z}_2$}&
            \BOXX{\hss$\tilde{Z}_3$}  \\
            \midrule
            $Q$&$1$&$1$&$1$         &$1$ &$1$ &$0$\\
            $u^c$&$1$&$-4$&$-1$     &$-1$&$1$ &$1$\\
            $d^c$&$-1$&$-6$&$0$     &$-1$&$1$&$-1$\\
            $L$&$-1$&$9$&$0$        &$1$ &$-3$ &$0$\\
            $e^c$&$1$&$6$&$-3$      &$-1$&$-3$ &$3$\\
            $\nu^c$&$-5$&$0$&$0$    &$-1$&$-3$&$-3$\\
            \midrule
            $\mathcal{N}$&$4\sqrt{3/5}$&$-4/\sqrt{15}$&
            $8\sqrt{3}$&$2\sqrt{6}$&$2\sqrt{2}$&$4$\\
            \bottomrule
            \end{tabular}
        \par
        \end{centering}
\end{table}

The above discussion implies that knowing the charges $q$ of the various SM representations is sufficient to reconstruct the VEV matrices $\mathbf{45}^{A}{}_{B}$ and $\mathbf{210}^{A}{}_{B}$. We provide the charges and their normalizing factors $\mathcal{N}$ in Tables~\ref{table:qnumbers-45} and \ref{table:qnumbers-210}. With discrete alignments of all $\alpha_i,\beta_j,\alpha'_k,\beta'_l$, we obtain the following Yukawa terms (written as superpotential operators) from an operator in Eq.~\eqref{eq:invariant-explicit} given fixed family indices $I$ and $J$:
\begin{align}
W&\supset \frac{C}{\Lambda^{n+n'+m+m'}}\bigg(\prod_{i,j,k,l}\mathcal{N}_{\alpha_i}\mathcal{N}_{\beta_j}\mathcal{N}_{\alpha'_k}\mathcal{N}_{\beta'_l}\,X_{\alpha_i}\,X_{\alpha'_k}\,Z_{\beta_j}\, Z_{\beta'_l}\bigg)\;\bigg( \nonumber\\
&\quad\,\, \phantom{+} Q_{I}\,u^c_{J}\,H_u^{\mathbf{H}}\;
\bigg[ \phantom{s^\mathbf{H}}\,C_{ud}^{\mathbf{H}}\,\PRODUCT{Q}{u^c} \bigg] \nonumber\\
&\quad + Q_{J}\,u^c_{I}\,H_u^{\mathbf{H}}\;
\bigg[ s^\mathbf{H}\,C_{ud}^{\mathbf{H}}\,\PRODUCT{u^c}{Q} \bigg] \nonumber\\
&\quad + Q_{I}\,d^c_{J}\,H_d^{\mathbf{H}}\;
\bigg[ \phantom{s^\mathbf{H}}\,C_{ud}^{\mathbf{H}}\,\PRODUCT{Q}{d^c} \bigg] \nonumber\\
&\quad + Q_{J}\,d^c_{I}\,H_d^{\mathbf{H}}\;
\bigg[ s^\mathbf{H}\,C_{ud}^{\mathbf{H}}\,\PRODUCT{d^c}{Q} \bigg] \nonumber\\
&\quad + L_{I}\,e^c_{J}\,H_e^{\mathbf{H}}\;
\bigg[ \phantom{s^\mathbf{H}}\,C_{e\nu}^{\mathbf{H}}\,\PRODUCT{L}{e^c} \bigg] \nonumber\\
&\quad + L_{J}\,e^c_{I}\,H_e^{\mathbf{H}}\;
\bigg[ s^\mathbf{H}\,C_{e\nu}^{\mathbf{H}}\,\PRODUCT{e^c}{L} \bigg] \nonumber\\
&\quad + L_{I}\,\nu^c_{J}\,H_\nu^{\mathbf{H}}\;
\bigg[ \phantom{s^\mathbf{H}}\,C_{e\nu}^{\mathbf{H}}\,\PRODUCT{L}{\nu^c} \bigg] \nonumber\\
&\quad + L_{J}\,\nu^c_{I}\,H_\nu^{\mathbf{H}}\;
\bigg[ s^\mathbf{H}\,C_{e\nu}^{\mathbf{H}}\,\PRODUCT{\nu^c}{L} \bigg] \nonumber\\
&\quad + \nu^c_{I}\,\nu^c_{J}\,\overline{\Delta}\;
\bigg[\; C_{\Delta}^{\mathbf{H}} \,\PRODUCT{\nu^c}{\nu^c}\bigg]\bigg).\label{eq:result-discrete}
\end{align}
Above, each $\mathcal{N}_{\alpha_i}$ labels a normalization factor from Table~\ref{table:qnumbers-45} given one of four possible alignments $\alpha_i$ the representation $\mathbf{45}_{\alpha_i}$ takes, $X_{\alpha_i}$ is the VEV of that representation (it can be any one of the VEVs $\{X_1,X_2,\tilde{X}_1,\tilde{X}_2\}$, depending on the direction, and has the normalization from Eq.~\eqref{eq:VEVs-SU5-and-PS}) and $q_{\alpha_i}(p)$ is the charge of the particle $p$ under the $\alpha_i$ alignment. Analogous definitions hold for quantities with $\alpha'_k$, as well as those with $\beta_j$ and $\beta'_k$ referring to representations $\mathbf{210}$. The $\mathbf{H}$-dependent quantities in Eq.~\eqref{eq:result-discrete} are given in Table~\ref{table:H-quantities}. Lastly, the prefactor $C/\Lambda^{n+n'+m+m'}$ is simply the coefficient in front of the non-renormalizable operator, where $C$ is a dimensionless number and $\Lambda$ is the cutoff scale of the GUT theory.

\begin{table}[htb]
\begin{center}
	\caption{The $\mathbf{H}$-dependent quantities of Eq.~\eqref{eq:result-discrete} for different choices of $\mathbf{H}$.\label{table:H-quantities}}
	\begin{tabular}{lrrr}
		\toprule
		$\mathbf{H}$&\BOX{\hss$\mathbf{10}$}&\BOX{\hss$\mathbf{\overline{126}}$}&\BOX{\hss$\mathbf{120}$}\\
		\midrule
		$s^{\mathbf{H}}$&$1$&$1$&$-1$\\
		$C_{ud}^{\mathbf{H}}$&$\sqrt{2}$&$4\sqrt{10}$&$4$\\\ADD
		$C_{e\nu}^{\mathbf{H}}$&$\sqrt{2}$&$-12\sqrt{10}$&$-4\sqrt{3}$\\\ADD
		$C_{\Delta}^{\mathbf{H}}$&$0$&$-16\sqrt{15}$&$0$\\\ADD\ADD
		$H^{\mathbf{H}}_{u}$&$H^{u}_{1}$&$H^{u}_{2}$&$H^{u}_{4}$\\\ADD
		$H^{\mathbf{H}}_{d}$&$H^{d}_{1}$&$H^{d}_{2}$&$\tfrac{\sqrt{3}}{2}H^{d}_{3}-	\tfrac{1}{2}H^{d}_{4}$\\\ADD
		$H^{\mathbf{H}}_{e}$&$H^{d}_{1}$&$H^{d}_{2}$&$\tfrac{1}{2}H^{d}_{3}+				\tfrac{\sqrt{3}}{2}H^{d}_{4}$\\\ADD
		$H^{\mathbf{H}}_{\nu}$&$H^{u}_{1}$&$H^{u}_{2}$&$H^{u}_{3}$\\
		\bottomrule
	\end{tabular}
\end{center}
\end{table}

We make the following observations regarding this result:
\begin{itemize}
\item Each of the 4 fermion sectors (the up, down, charged lepton and neutrino sectors) have two terms associated with it, since for example the left-handed fields ($Q$,$L$) can come from $\mathbf{16}_{I}$ or $\mathbf{16}_{J}$. We thus have $8$ terms for Dirac-type masses. The last $\overline{\Delta}$ term is the Majorana type mass for right-handed neutrinos, present only when $\mathbf{H}=\mathbf{\overline{126}}$, since only such $\mathbf{H}$ contains also a SM singlet VEV. 
\item The SM quarks and leptons acquire masses only in the EW broken phase, when $H_{u,d,e,\nu}^{\mathbf{H}}$ acquire EW scale VEVs. The value of these VEVs depend on $\mathbf{H}$ and on the doublet-antidoublet mass matrix, and may introduce new parameters from that matrix into the Yukawa sector of the MSSM.
In particular, one first needs to identify which (anti)doublet flavor states $H^{u,d}_i$ the various indices $u$, $d$, $e$ and $\nu$ represent for a given $\mathbf{H}$, cf.~Table~\ref{table:H-quantities}. One needs to then investigate
how the MSSM low mass pair $H_{u,d}$ aligns with the flavor states $H^{u,d}_i$ 
given the doublet mass matrix; the issue of the location of MSSM Higgs states will be discussed in detail in Section~\ref{sec:Higgs-location}, with a few comments already in the next bullet point below.
\item If $I=J$, one has to take the sum of the two terms in each sector. The predictions for the ratios of Yukawa coefficients between the different sectors thus read
{\small
	\begin{align}
	\begin{split}
	\frac{(\mathbf{Y}_e)_{II}}{(\mathbf{Y}_d)_{II}}&=\chi^{\mathbf{H}}_{ed}\,\frac{C^{\mathbf{H}}_{e\nu}}{C^{\mathbf{H}}_{ud}}
	\frac{\PRODUCT{L}{e^c}+s^{\mathbf{H}}\!\!\PRODUCT{e^c}{L}}
	{\PRODUCT{Q}{d^c}+s^{\mathbf{H}}\!\!\PRODUCT{d^c}{Q}},\\
	\frac{(\mathbf{Y}_u)_{II}}{(\mathbf{Y}_d)_{II}}&=\chi^{\mathbf{H}}_{ud}\,
	\frac{\PRODUCT{Q}{u^c}+s^{\mathbf{H}}\!\!\PRODUCT{u^c}{Q}}
	{\PRODUCT{Q}{d^c}+s^{\mathbf{H}}\!\!\PRODUCT{d^c}{Q}},\\
	\frac{(\mathbf{Y}_\nu)_{II}}{(\mathbf{Y}_d)_{II}}&=\chi^{\mathbf{H}}_{\nu d}\,\frac{C^{\mathbf{H}}_{e\nu}}{C^{\mathbf{H}}_{ud}}
	\frac{\PRODUCT{L}{\nu^c}+s^{\mathbf{H}}\!\!\PRODUCT{\nu^c}{L}}
	{\PRODUCT{Q}{d^c}+s^{\mathbf{H}}\!\!\PRODUCT{d^c}{Q}}.\\
	\end{split}
	\label{eq:Yukawa-ratios-discrete}
	\end{align}
}
The factors $\chi^{\mathbf{H}}$ appear due to the possible non-alignment of the states $H^{u,d}_{i}$ with the MSSM fields $H_{u,d}$, as was remarked above.
At this point, we only briefly comment on them:

\begin{itemize}
\item There exist predictive scenarios, which fix the coefficients $\chi^\mathbf{H}$ to numeric values. A convenient list of some such scenarios is provided for the $\mathrm{SO}(10)$ model builder in Table~\ref{table:DT-models-factors}. The various scenarios specify which $\mathbf{H}$ are available for use in the operator, and what the $\chi^{\mathbf{H}}$ coefficients are. Scenarios $\# 5$ and $\# 6$ have two possible Higgs representations $\mathbf{H}$ available for use in operators. The scenario numbers are based on scenarios of specific Higgs sectors in Table~\ref{table:DT-models} of Section~\ref{sec:Higgs-location}, and should be viewed as determining the available $\mathbf{H}$ for all the operators in the Yukawa sector.
\item The coefficients $\chi^{\mathbf{H}}$ depend on the doublet-antidoublet mass matrix, which in turn involves Higgs sector parameters. Specifying any Higgs sector of the model, one can compute $\chi^{\mathbf{H}}$ using the tools provided in Section~\ref{sec:Higgs-location}. In general, the $\chi^{\mathbf{H}}$ will be functions of the Higgs sector parameters; in the previous point, we considered special setups yielding numeric values independent of any parameters. 
\item Alternatively, if one wishes to remain agnostic regarding the specifics of the Higgs sector, one can also take the factors $\chi^{\mathbf{H}}$ to be free complex parameters. The only assumption necessary in this case is that the Higgs sector is sufficiently rich to provide the chosen numerical values for these factors. In such a scenario, the Yukawa ratios from a single operator are not predicted, but since the 3 family Yukawa sector will involve many such operators, a predictive model may still be obtained. Note that these factors would be fixed to the same value in all operators using the same $\mathbf{H}$.  
\item An important observation is that $\chi^{\mathbf{H}}_{ed}=1$ if $\mathbf{H}=\mathbf{10}$ or $\mathbf{H}=\mathbf{\overline{126}}$, regardless of the details of the Higgs sector. The reason is that in this case $H^{\mathbf{H}}_d=H^{\mathbf{H}}_e$, i.e.~the down sector and the charged lepton sector couple to the same flavor state $H^{d}_i$, since there is only one such state in the representation $\mathbf{H}$. In contrast, $\mathbf{H}=\mathbf{120}$ contains two states $H^{d}_{i}$, and they couple with different linear combinations to the down and charged lepton sectors, as seen from Table~\ref{table:H-quantities}, making the coefficient $\chi^{\mathbf{120}}_{ed}$ model dependent.
\end{itemize}

\item Compared to $\mathrm{SU}(5)$ GUTs, where only the charged lepton and down sector are related, $\mathrm{SO}(10)$ relates Yukawa coefficients of all $4$ sectors. Furthermore, a naive expectation in $\mathrm{SO}(10)$ is that with discrete $\mathrm{SU}(5)$-compatible directions of the VEVs, at least the Clebsch ratios between the charged lepton and down sector from the $\mathrm{SU}(5)$ symmetric operators would be reproduced, e.g.~from the classification in \cite{Antusch:2009gu}. Curiously, this turns out not to be the case, with the ratios in principle different due to the presence of two Yukawa terms in each sector in Eq.~\eqref{eq:result-discrete} or \eqref{eq:Yukawa-ratios-discrete}.
This implies that each $\mathrm{SO}(10)$ operator actually combines two operators at the $\mathrm{SU}(5)$ level in a given Yukawa sector, thus modifying the single operator $\mathrm{SU}(5)$ predictions.
\item At the renormalizable level, the product over $i,j,k,l$ in Eq.~\eqref{eq:result-discrete} has no factors, leading to a value of $1$.
The relative Clebsch factors between the sectors are then determined by the ratio $C^{\mathbf{H}}_{e\nu}/C^{\mathbf{H}}_{ud}$, which is $1$ and $-3$ if $\mathbf{H}$ is $\mathbf{10}$ or $\mathbf{\overline{126}}$, corresponding in the $\mathrm{SU}(5)$ context to the $b$-$\tau$ unification and the Georgi-Jarlskog factor \cite{Georgi:1979df}, respectively. As is well known, Yukawa matrices arising from renormalizable operators for $\mathbf{H}$ equal to $\mathbf{10}$ or $\mathbf{\overline{126}}$ are symmetric under the exchange of family indices $I$ and $J$, and anti-symmetric for the $\mathbf{120}$. This result is reproduced in Eq.~\eqref{eq:result-discrete} due to the coefficient $s^{\mathbf{H}}$. In the general non-renormalizable case, where the product over $\alpha_i,\beta_j$ from the charges acting on $\mathbf{16}_I$ is not the same as the product over $\alpha'_k,\beta'_l$ of charges acting on $\mathbf{16}_{J}$, the overall symmetry or anti-symmetry in family indices is lost. In particular, non-renormalizable operators with $\mathbf{H}=\mathbf{120}$ can thus also be used for generating diagonal Yukawa entries when using asymmetric products. 
\item The product of normalizations $\mathcal{N}$ and the product of VEV sizes $X$ or $Z$ can be absorbed into the overall operator coefficient $C/\Lambda^{n+n'+m+m'}$, thus having no impact on the $\mathrm{SO}(10)$ relations between different fermion sectors. The normalizing factors $\mathcal{N}$ become important though when arbitrary VEV directions are considered in the next subsection.
\end{itemize}

\begin{table}[htb]
\begin{center}
	\caption{Examples of scenarios where $\chi^{\mathbf{H}}$ factors have fixed numerical values. The $\#$ case numbers refer to cases of Higgs superpotentials
in Table~\ref{table:DT-models}, to be presented later in the paper. For each case, the available representations $\mathbf{H}$ and the associated $\chi^{\mathbf{H}}$ factors are given.  
	\label{table:DT-models-factors}}
	\begin{tabular}{lrrrr}
	\toprule
	\multicolumn{1}{p{0.5cm}}{\#}&
	\multicolumn{1}{p{0.8cm}}{\hfill$\mathbf{H}$}&
	\multicolumn{1}{p{2cm}}{\hfill$\chi^{\mathbf{H}}_{ed}$}&
	\multicolumn{1}{p{2cm}}{\hfill$\chi^{\mathbf{H}}_{ud}$}&
	\multicolumn{1}{p{2cm}}{\hfill$\chi^{\mathbf{H}}_{\nu d}$}\\
	\midrule
	$1$&$\mathbf{10}$&$1$&$1$&$1$\\\ADD
	$2$&$\mathbf{120}$&$-1/\sqrt{3}$&$-1$&$1/\sqrt{3}$\\\ADD
	$3$&$\mathbf{120}$&$\sqrt{3}$&$1$&$\sqrt{3}$\\\ADD
	$4$&$\mathbf{\overline{126}}$&$1$&$1$&$1$\\\ADD
	$5$&$\mathbf{10}$&$1$&$-1$&$-1$\\
	&$\mathbf{120}$&$-1/\sqrt{3}$&$-1$&$1/\sqrt{3}$\\\ADD
	$6$&$\mathbf{10}$&$1$&$-1$&$-1$\\
	&$\mathbf{120}$&$\sqrt{3}$&$1$&$\sqrt{3}$\\
	\bottomrule
	\end{tabular}
\end{center}
\end{table}

A subclass of operators with $m=m'=0$ and $\mathbf{H}=\mathbf{10}$ has been previously considered in the literature~\cite{Anderson:1993fe}.
To facilitate the comparison of our notation for the $\mathbf{45}$ with that in \cite{Anderson:1993fe}, we provide below a dictionary for translating our notation to theirs:
\begin{align}
X_{1}&\to X,&\tilde{X}_{1}&\to T_{3R},\nonumber\\
X_{2}&\to Y,&\tilde{X}_{2}&\to B-L.
\end{align}
The idea behind the alternative notation is that the $\mathbf{45}$ is the adjoint representation of $\SO(10)$, so one can make use of the gauge boson labeling also for the scalar (or chiral supermultiplet in SUSY) fields analyzed here. In this sense the scalar states correspond to the following gauge boson generators: $X$ for the $\mathrm{U}(1)_X$ charge in $G_{51}$, $Y$ for the hypercharge (a SM generator), $B-L$ to the difference of baryon and lepton number, which is gauged and is represented by the generator $T_{B-L}$ in $\SO(10)$, and $T_{3R}$ to the diagonal generator of the $\SU(2)_R$ factor in the Pati-Salam group $G_{422}$. Since in this paper we also consider the VEVs in the representation $\mathbf{210}$, to which the logic of the notation from \cite{Anderson:1993fe} cannot be extended, we prefer our notation due to greater visual clarity. The $X$ and $Z$ letters refer to $\mathbf{45}$ and $\mathbf{210}$, respectively, and the non-tilde and tilde refer to the $G_{51}$ and $G_{422}$ alignment, respectively.

We conclude this section on discrete VEV alignments by applying the general formulae we derived to a concrete example, which is of particular importance for flavor GUT model building. We are interested in the numerical predictions for the ratios of diagonal Yukawa entries between different Yukawa sectors from Eq.~\eqref{eq:Yukawa-ratios-discrete}, the analog of the $\mathrm{SU}(5)$ predictions from \cite{Antusch:2009gu,Antusch:2013rxa}. 

\begin{figure}[h!]
\includegraphics[width=0.90\textwidth]{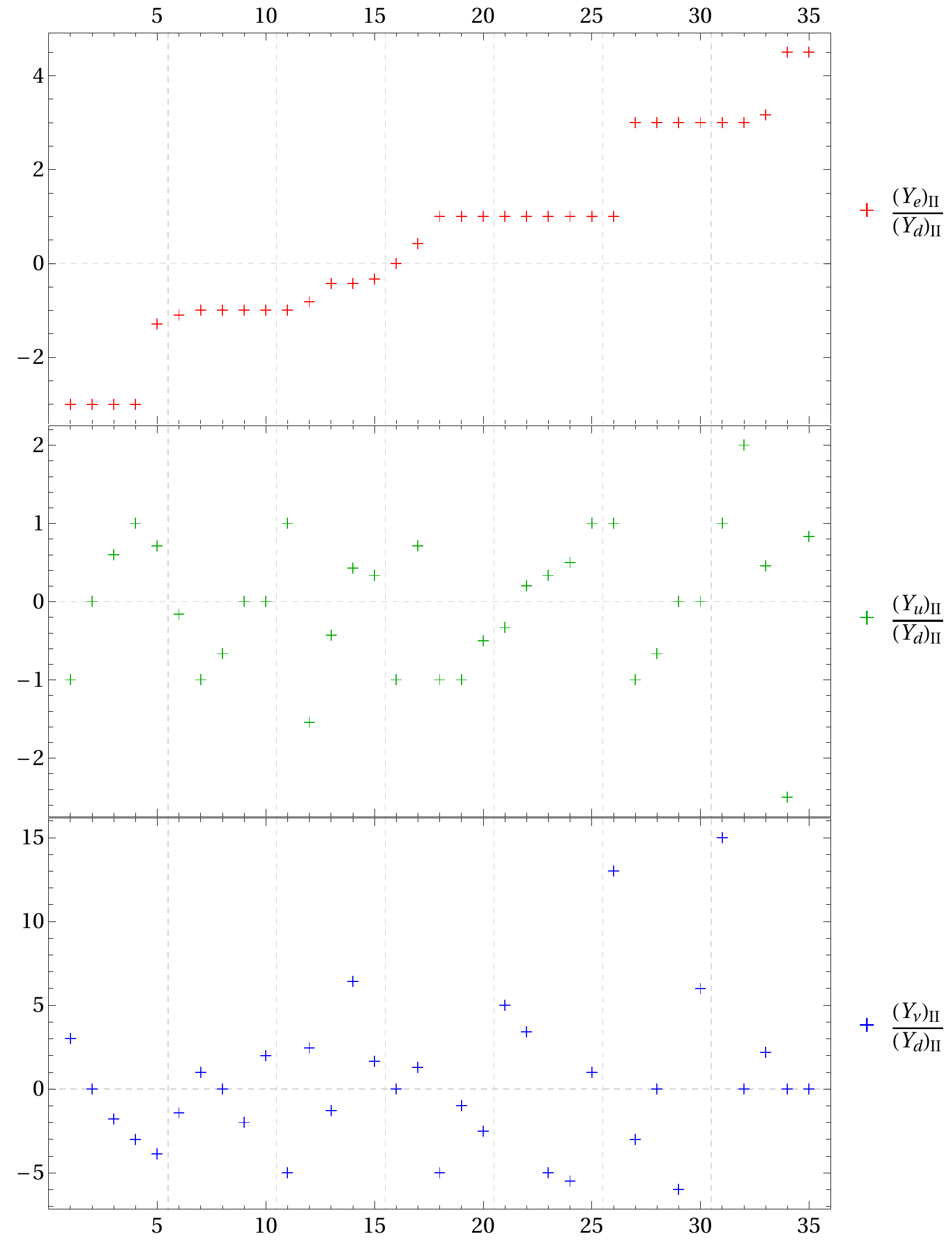}
\caption{A graphical illustration of the Yukawa ratios that are listed in Table~\ref{table:yukawa_ratios}. They arise from superpotential operators as in Eq.~\eqref{eq:invariant-general} with $\mathbf{H} = \mathbf{10}$ and case $\# 1$ in Table~\ref{table:DT-models-factors} for the Higgs sector, using the discrete $\mathrm{SU}(5)$ and Pati-Salam alignments for the singlet VEVs of the $\mathbf{45}$ and $\mathbf{210}$. The cases are limited to operators which are at most of dimension five, i.e. $n+n'+m+m' \leq 2$, and which provide $|(\mathbf{Y}_e)_{II}/(\mathbf{Y}_d)_{II}|\leq8$. The numbering of the cases on the $x$-axis corresponds to the one in Table~\ref{table:yukawa_ratios}. \label{fig:yukawa_ratios}}
\end{figure}

As discussed earlier, the predictions in $\mathrm{SO}(10)$ now relate all Yukawa sectors. As an illustration, we choose the simplest predictive case $\# 1$ from Table~\ref{table:DT-models-factors}, i.e.~$\mathbf{H}=\mathbf{10}$ (with $H_{1}^u=v_u$ and $H_{1}^{d}=v_d$) and all $\chi^\mathbf{H}=1$. We computed for this scenario the ratios of Eq.~\eqref{eq:Yukawa-ratios-discrete} for all operators containing up to $5$ fields, i.e.~$n+n'+m+m'\leq 2$, and all possible combinations of discrete-VEV alignments. The list of Clebsch factor predictions is compiled in Table~\ref{table:yukawa_ratios}, with the ratio values then shown graphically in Figure~\ref{fig:yukawa_ratios} with the same case numbering convention. The table is meant to be used to identify the particular operator giving the prediction, while the figure is very convenient for searching through the possible numerical values for the ratios. The labeling order of the various cases is that of increasing values of ratios $(\mathbf{Y}_e)_{II}/(\mathbf{Y}_d)_{II}$, $(\mathbf{Y}_u)_{II}/(\mathbf{Y}_d)_{II}$ and $(\mathbf{Y}_\nu)_{II}/(\mathbf{Y}_d)_{II}$ (in that order of importance). As can be seen from the table, different cases of operators can predict the same triple of Yukawa ratios, so they are listed under the same number Nr. 

Finally, we remind the reader that the ratios $(\mathbf{Y}_e)_{II}/(\mathbf{Y}_d)_{II}$ in Table~\ref{table:yukawa_ratios} are not merely a reproduction of the $\mathrm{SU}(5)$ Clebsch factors, since one $\mathrm{SO}(10)$ operator actually gives a sum of two $\mathrm{SU}(5)$ operators in each Yukawa sector, as was discussed. As an example, the most common $\mathrm{SU}(5)$ Clebsch ratio $-3/2$ (obtained when the $\mathbf{24}$ of $\mathrm{SU}(5)$ acquires a VEV) is not among the $\mathrm{SO}(10)$ ratios in the table. The case that is naively expected to yield the $-3/2$ ratio involves one $\mathbf{45}$ factor aligned in the $X_2$ direction; it instead yields the $(\mathbf{Y}_e)_{II}/(\mathbf{Y}_d)_{II}$ ratio $1$, corresponding to  the subcase $(X_2|.)$ (cf.~table caption for label definition) of Nr $19$ in Table~\ref{table:yukawa_ratios}.  

\begin{table}[htb]
\caption{{\small
A list of Yukawa ratios generated from Eq.~\eqref{eq:Yukawa-ratios-discrete} with $\mathbf{H}=\mathbf{10}$ and the predictive Higgs sector case $\# 1$ from Table~\ref{table:DT-models-factors}. This is a complete list for operators up to dimension five in $W$ (i.e. $n+n'+m+m' \leq 2$), using the well defined $\mathrm{SU}(5)$ and Pati-Salam alignments from Eq.~\eqref{eq:VEVs-SU5} and \eqref{eq:VEVs-ps} for the singlet VEVs of the $\mathbf{45}$ and $\mathbf{210}$, and which provide $|(\mathbf{Y}_e)_{II}/(\mathbf{Y}_d)_{II}|\leq8$. The last column lists the alignements of the SM singlet VEVs which provide the corresponding tuple of Yukawa ratios, where the vertical bar separates the fields $\{\mathbf{45}_{\alpha_i},\mathbf{210}_{\beta_j}\}$ on the left-hand side and $\{\mathbf{45}_{\alpha'_k},\mathbf{210}_{\beta'_l}\}$ on the right-hand side of the $\mathbf{10}$. A dot means no field on that side of the Higgs.  \label{table:yukawa_ratios}}}
\begin{small}
\renewcommand{\arraystretch}{1.03}
\begin{tabular}{rrlllll}
\toprule
Nr & $\Big(\frac{(\mathbf{Y}_e)_{II}}{(\mathbf{Y}_d)_{II}},\frac{(\mathbf{Y}_u)_{II}}{(\mathbf{Y}_d)_{II}},\frac{(\mathbf{Y}_
\nu)_{II}}{(\mathbf{Y}_d)_{II}}\Big)$ & \multicolumn{5}{l}{$\big(\{\mathbf{45}_{\alpha_i},\mathbf{210}_{\beta_j}\}|\{\mathbf{45}_{\alpha'_k},\mathbf{210}_{\beta'_l}\}\big)$} \\
\midrule
$1$ & $(-3,-1,3)$ & $(\tilde{Z}_3|\cdot)$\,,\hspace{0.3cm}$(\tilde{X}_1,\tilde{X}_2|\cdot)$\,,\hspace{0.3cm}$(X_1,\tilde{Z}_2|\cdot)$\,,\hspace{0.3cm}$(X_2,\tilde{Z}_2|\cdot)$\,,\hspace{0.3cm}$(\tilde{X}_1,\tilde{Z}_2|\cdot)$\,,\vspace{-0.1cm} \\
& & $(\tilde{Z}_1,\tilde{Z}_3|\cdot)$\,,\hspace{0.3cm}$(X_1|\tilde{X}_1)$\,,\hspace{0.3cm}$(X_2|\tilde{X}_1)$\,,\hspace{0.3cm}$(\tilde{X}_1|\tilde{X}_2)$\,,\hspace{0.3cm}$(X_1|\tilde{Z}_2)$\,,\vspace{-0.1cm} \\
& & $(X_2|\tilde{Z}_2)$\,,\hspace{0.3cm}$(\tilde{X}_1|\tilde{Z}_2)$\,,\hspace{0.3cm}$(\tilde{Z}_1|\tilde{Z}_3)$ \\
$2$ & $(-3,0,0)$ & $(Z_3|\cdot)$\,,\hspace{0.3cm}$(X_1,Z_3|\cdot)$\,,\hspace{0.3cm}$(Z_1,Z_3|\cdot)$\,,\hspace{0.3cm}$(X_1|Z_3)$\,,\hspace{0.3cm}$(Z_1|Z_3)$ \\
$3$ & $(-3,\frac{3}{5},-\frac{9}{5})$ & $(Z_2|\cdot)$\,,\hspace{0.3cm}$(X_1,X_2|\cdot)$ \\
$4$ & $(-3,1,-3)$ & $(\tilde{Z}_2|\cdot)$\,,\hspace{0.3cm}$(\tilde{X}_1,\tilde{Z}_3|\cdot)$\,,\hspace{0.3cm}$(\tilde{X}_2,\tilde{Z}_1|\cdot)$\,,\hspace{0.3cm}$(\tilde{X}_2|\tilde{Z}_1)$ \\
$5$ & $(-\frac{9}{7},\frac{5}{7},-\frac{27}{7})$ & $(\tilde{X}_2,Z_2|\cdot)$\,,\hspace{0.3cm}$(\tilde{X}_2|Z_2)$ \\
\hdashline
$6$ & $(-\frac{21}{19},-\frac{3}{19},-\frac{27}{19})$ & $(X_1,Z_2|\cdot)$ \\
$7$ & $(-1,-1,1)$ & $(\tilde{X}_1|Z_1)$ \\
$8$ & $(-1,-\frac{2}{3},0)$ & $(\tilde{X}_1,Z_2|\cdot)$ \\
$9$ & $(-1,0,-2)$ & $(X_1,\tilde{Z}_1|\cdot)$\,,\hspace{0.3cm}$(X_1|\tilde{Z}_1)$ \\
$10$ & $(-1,0,2)$ & $(Z_1,\tilde{Z}_1|\cdot)$\,,\hspace{0.3cm}$(Z_1|\tilde{Z}_1)$ \\
\hdashline
$11$ & $(-1,1,-5)$ & $(\tilde{X}_1,Z_1|\cdot)$ \\
$12$ & $(-\frac{9}{11},-\frac{17}{11},\frac{27}{11})$ & $(X_2,Z_2|\cdot)$ \\
$13$ & $(-\frac{3}{7},-\frac{3}{7},-\frac{9}{7})$ & $(Z_1,Z_2|\cdot)$ \\
$14$ & $(-\frac{3}{7},\frac{3}{7},\frac{45}{7})$ & $(Z_1|Z_2)$ \\
$15$ & $(-\frac{1}{3},\frac{1}{3},\frac{5}{3})$ & $(X_1,\tilde{X}_1|\cdot)$ \\
\hdashline
$16$ & $(0,-1,0)$ & $(\tilde{X}_1|Z_3)$\,,\hspace{0.3cm}$(Z_3|\tilde{Z}_3)$ \\
$17$ & $(\frac{3}{7},\frac{5}{7},\frac{9}{7})$ & $(Z_2,\tilde{Z}_1|\cdot)$\,,\hspace{0.3cm}$(Z_2|\tilde{Z}_1)$ \\
$18$ & $(1,-1,-5)$ & $(Z_1|Z_1)$ \\
$19$ & $(1,-1,-1)$ & $(X_1|\cdot)$\,,\hspace{0.3cm}$(X_2|\cdot)$\,,\hspace{0.3cm}$(\tilde{X}_1|\cdot)$\,,\hspace{0.3cm}$(\tilde{X}_1,\tilde{Z}_1|\cdot)$\,,\hspace{0.3cm}$(\tilde{X}_1|\tilde{Z}_1)$ \\
$20$ & $(1,-\frac{1}{2},-\frac{5}{2})$ & $(X_1|Z_1)$ \\
\hdashline
$21$ & $(1,-\frac{1}{3},5)$ & $(X_1|X_1)$ \\
$22$ & $(1,\frac{1}{5},\frac{17}{5})$ & $(X_1,X_1|\cdot)$ \\
$23$ & $(1,\frac{1}{3},-5)$ & $(X_1,\tilde{Z}_3|\cdot)$\,,\hspace{0.3cm}$(X_1|Z_2)$ \\
$24$ & $(1,\frac{1}{2},-\frac{11}{2})$ & $(X_1,Z_1|\cdot)$ \\
$25$ & $(1,1,1)$ & $(\cdot|\cdot)$\,,\hspace{0.3cm}$(\tilde{X}_1,\tilde{X}_1|\cdot)$\,,\hspace{0.3cm}$(\tilde{Z}_1,\tilde{Z}_1|\cdot)$\,,\hspace{0.3cm}$(\tilde{Z}_1|\tilde{Z}_1)$ \\
\hdashline
$26$ & $(1,1,13)$ & $(Z_1,Z_1|\cdot)$ \\
$27$ & $(3,-1,-3)$ & $(Z_1|\tilde{Z}_3)$ \\
$28$ & $(3,-\frac{2}{3},0)$ & $(Z_2,\tilde{Z}_3|\cdot)$ \\
$29$ & $(3,0,-6)$ & $(\tilde{X}_2,Z_1|\cdot)$\,,\hspace{0.3cm}$(\tilde{X}_2|Z_1)$ \\
$30$ & $(3,0,6)$ & $(X_1,\tilde{X}_2|\cdot)$\,,\hspace{0.3cm}$(X_1|\tilde{X}_2)$ \\
\hdashline
$31$ & $(3,1,15)$ & $(Z_1,\tilde{Z}_3|\cdot)$ \\
$32$ & $(3,2,0)$ & $(X_2,\tilde{X}_1|\cdot)$\,,\hspace{0.3cm}$(Z_3,\tilde{Z}_1|\cdot)$\,,\hspace{0.3cm}$(Z_3|\tilde{Z}_1)$ \\
$33$ & $(\frac{117}{37},\frac{17}{37},\frac{81}{37})$ & $(Z_2,Z_2|\cdot)$ \\
$34$ & $(\frac{9}{2},-\frac{5}{2},0)$ & $(X_2|Z_3)$ \\
$35$ & $(\frac{9}{2},\frac{5}{6},0)$ & $(Z_2|Z_3)$ \\
\bottomrule
\end{tabular}
\renewcommand{\arraystretch}{1.0}
\end{small}
\end{table}

\subsection{Arbitrary directions\label{sec:arbitrary-directions}}
We now perform an analysis of Yukawa couplings generated by operators in Eq.~\eqref{eq:invariant-explicit}, but where we do not assume the VEVs of $\mathbf{45}$ and $\mathbf{210}$ to have discrete directions, but instead can take an arbitrary direction.
Since the two representations have $2$ and $3$ SM singlets, respectively, the VEVs point in an arbitrary direction in $\mathbb{C}^2$ and $\mathbb{C}^3$ (in the SUSY context).

In this type of model building approach, we assume for simplicity to only have a single copy of a $\mathbf{45}$ and $\mathbf{210}$: the VEV direction in the $\mathbf{45}$ is the same for all $\alpha_i$ and $\alpha'_k$, and analogously the VEV directions in the $\mathbf{210}$s are the same for all $\beta_j$ and $\beta'_l$. This setup of (at most) one copy of each representation provides maximal predictivity of such models, since it introduces a minimal set of new continuous parameters related to the arbitrary directions.

Alongside the continuous parameters describing the VEV direction and size, the model is specified by the powers $n,n'$ for the $\mathbf{45}$ and $m,m'$ for $\mathbf{210}$ for each operator added to the Yukawa sector. The arbitrary direction VEV approach does not have a good analog in $\mathrm{SU}(5)$ flavor GUTs, since no low-dimensional irreducible representations of $\mathrm{SU}(5)$ contain more than one SM singlet. This approach is therefore relevant only for bigger GUT groups such as $\mathrm{SO}(10)$ and $\mathrm{E}_{6}$.

Since the overall size of the VEVs can be absorbed into the coefficient in front of an operator, it is convenient to define the following ratios of GUT-scale VEVs: 
\begin{align}
	\kappa&:=X_{2}/X_{1},&
	\kappa_1&:=Z_{2}/Z_{1},&
	\kappa_2&:=Z_{3}/Z_{1}.\label{eq:ratios}
\end{align}
We do not lose any generality by considering the ratios, since the result for $X_{1}=0$ can be recovered by taking the limit $\kappa\to \infty$ given the relation $X_{1}=X_2/\kappa$. An analogous treatment also applies to the case $Z_{1}=0$ in $\kappa_1$ and $\kappa_2$. Note that the $\kappa$-ratios are complex numbers in the SUSY context, and real numbers in the non-SUSY context.

The discrete alignments from Section~\ref{sec:discrete-directions} can be recovered by taking special values for the $\kappa$ ratios. One can easily reproduce them by using the definitions in Eq.~\eqref{eq:ratios} and the connection to Pati-Salam alignments in Eq.~\eqref{eq:VEVs-SU5-and-PS}. For the sake of completeness and to identify special values of $\kappa$ at a glance, we provide them in Table~\ref{table:ratios-alignments}.

\begin{table}[htb]
	\begin{centering}
    \caption{The special values of ratios $\kappa$, $\kappa_1$ and $\kappa_2$ corresponding to discrete VEV alignments in the representations $\mathbf{45}$ and $\mathbf{210}$.\label{table:ratios-alignments}}
		\begin{tabular}{cr@{\hbox to 3cm{}}crr}
        \toprule
        alignment&\hbox to 1.3cm{\hss$\kappa$}&alignment&\hbox to 1.3cm{\hss $\kappa_1$}&\hbox to 1.3cm{\hss $\kappa_2$}\\
        \midrule
        $X_{1}$&$0$&$Z_{1}$&$0$&$0$\\
        $X_{2}$&$\infty$&$Z_{2}$&$\infty$&$0$\\
        &&$Z_{3}$&$0$&$\infty$\\\addlinespace
        $\tilde{X}_{1}$&$-\sqrt{3/2}$&$\tilde{Z}_{1}$&$-2$&$\sqrt{5}$\\
        $\tilde{X}_{2}$&$\sqrt{2/3}$&$\tilde{Z}_{2}$&$4/3$&$\sqrt{5/9}$\\
        &&$\tilde{Z}_{3}$&$-1/3$&$-\sqrt{5/9}$\\
        \bottomrule
    	\end{tabular}
    \par
	\end{centering}
\end{table}

Using the freedom of the overall phase of $\mathbf{45}$ and $\mathbf{210}$, the values $X_{1}$ and $Z_{1}$ can be chosen real without loss of generality; the ratios $\kappa$, $\kappa_1$ and $\kappa_2$, however, always remain complex in the SUSY context.

Using the ratios $\kappa$, $\kappa_1$ and $\kappa_2$, we define for a fermion of type $p$ the order 1 polynomials $P_p(\kappa)$ and $R_p(\kappa_1,\kappa_2)$ by
\begin{align}
P_p(\kappa)&:=\frac{1}{X_{1}}\sum_{i=1}^{2}\mathcal{N}_{X_i}\;q_{X_i}(p)\;X_i\nonumber\\
&=\mathcal{N}_{X_1}\;q_{X_1}(p)+\mathcal{N}_{X_2}\;q_{X_2}(p)\;\kappa,\label{eq:polynomial-p}\\
R_p(\kappa_1,\kappa_2)&:= \frac{1}{Z_1}\sum_{i=1}^{3}\mathcal{N}_{Z_i}\;q_{Z_{i}}(p)\;Z_i\nonumber\\
&=\mathcal{N}_{Z_{1}}\;q_{Z_{1}}(p)+\mathcal{N}_{Z_{2}}\;q_{Z_{2}}(p)\;\kappa_1+\mathcal{N}_{Z_{3}}\;q_{Z_{3}}(p)\;\kappa_2,\label{eq:polynomial-r}
\end{align}
which represent the combined charge for the particle $p$ given the arbitrary direction of the VEV. In the above definition of $P_p$, the expression $\mathcal{N}_{X_i}$ represents the normalization factor $\mathcal{N}$ under the $X_{i}$ alignment, while $q_{X_{i}}(p)$ is the charge under the $X_{i}$ alignment of the particle $p\in\{Q,u^c,d^c,L,e^c,\nu^c\}$ from Tables~\ref{table:qnumbers-45} and \ref{table:qnumbers-210}. The index $i$ goes over all possible VEVs in the representation, i.e.~over all $\mathrm{SU}(5)$ compatible alignments. Analogous definitions hold for the quantities in the definition of $R_p$. The normalizations $\mathcal{N}$ and charges $q$ are those from Tables~\ref{table:qnumbers-45} and \ref{table:qnumbers-210}. 

The operator from Eq.~\eqref{eq:invariant-explicit} with arbitrary alignments in a single copy of $\mathbf{45}$ and $\mathbf{210}$ generates the following Yukawa terms in the superpotential:

\begin{align}
W&\supset \frac{C}{\Lambda^{n+n'+m+m'}}\;X_1^{n+n'}\,Z_1^{m+m'}\;\bigg( \nonumber\\
&\quad\, \phantom{+}\; Q_{I}\,u^c_{J}\,H_u^{\mathbf{H}}\;
\Big[ \phantom{s^\mathbf{H}}\,C_{ud}^{\mathbf{H}}\;\POLYFACTOR{Q}{u^c} \Big] \nonumber\\
&\quad + Q_{J}\,u^c_{I}\,H_u^{\mathbf{H}}\;
\Big[ s^\mathbf{H}\,C_{ud}^{\mathbf{H}}\;\POLYFACTOR{u^c}{Q} \Big] \nonumber\\
&\quad + Q_{I}\,d^c_{J}\,H_d^{\mathbf{H}}\;
\Big[ \phantom{s^\mathbf{H}}\,C_{ud}^{\mathbf{H}}\;\POLYFACTOR{Q}{d^c} \Big] \nonumber\\
&\quad + Q_{J}\,d^c_{I}\,H_d^{\mathbf{H}}\;
\Big[ s^\mathbf{H}\,C_{ud}^{\mathbf{H}}\;\POLYFACTOR{d^c}{Q} \Big]\nonumber\\
&\quad + L_{I}\,e^c_{J}\,H_e^{\mathbf{H}}\;
\Big[ \phantom{s^\mathbf{H}}\,C_{e\nu}^{\mathbf{H}}\;\POLYFACTOR{L}{e^c} \Big] \nonumber\\
&\quad + L_{J}\,e^c_{I}\,H_e^{\mathbf{H}}\;
\Big[ s^\mathbf{H}\,C_{e\nu}^{\mathbf{H}}\;\POLYFACTOR{e^c}{L} \Big] \nonumber\\
&\quad + L_{I}\,\nu^c_{J}\,H_\nu^{\mathbf{H}}\;
\Big[ \phantom{s^\mathbf{H}}\,C_{e\nu}^{\mathbf{H}}\;\POLYFACTOR{L}{\nu^c} \Big] \nonumber\\
&\quad + L_{J}\,\nu^c_{I}\,H_\nu^{\mathbf{H}}\;
\Big[ s^\mathbf{H}\,C_{e\nu}^{\mathbf{H}}\;\POLYFACTOR{\nu^c}{L} \Big] \nonumber\\
&\quad + \nu^c_{I}\,\nu^c_{J}\,\overline{\Delta}\;
\Big[\;C^{\mathbf{H}}_{\Delta}\;
P_{\nu^c}(\kappa)^{n+n'}\,R_{\nu^c}(\kappa_1,\kappa_2)^{m+m'}\Big]\bigg).\label{eq:result-continuous}
\end{align}

The $\mathbf{H}$-dependent quantities in the above expression can again be found in Table~\ref{table:H-quantities}, while the polynomials $P(\kappa)$ and $R(\kappa_1,\kappa_2)$ are those defined in Eq.~\eqref{eq:polynomial-p} and \eqref{eq:polynomial-r}. The coefficient $C/\Lambda^{n+n'+m+m'}$ is again the overall coupling in front of the operator, and two terms contribute to each Yukawa sector.

A final remark on the result: note that the $X_2$, $Z_2$ and $Z_3$ charges for right-handed neutrinos are zero, as can be seen in Tables~\ref{table:qnumbers-45} and \ref{table:qnumbers-210}, making the last term in Eq.~\eqref{eq:result-continuous} especially simple and $\kappa,\kappa_1,\kappa_2$-independent:
\begin{align}
\begin{split}
P_{\nu^c}(\kappa)&=-\sqrt{10},\\ 
R_{\nu^c}(\kappa_1,\kappa_2)&=-4\sqrt{15},\\
\end{split}
\end{align}
yielding 
\begin{align}
\nu^c_{I}\,\nu^c_{J}\,\overline{\Delta}\;
\Big[\;C^{\mathbf{H}}_{\Delta} \,
P_{\nu^c}(\kappa)^{n+n'}\,R_{\nu^c}(\kappa_1,\kappa_2)^{m+m'}\Big]&=\nu^c_{I}\,\nu^c_{J}\,\overline{\Delta}\;\Big[
4\,(-\sqrt{10})^{n+n'}\;(-4\,\sqrt{15})^{m+m'+1}
\Big].
\end{align}

\section{MSSM Higgs location\label{sec:Higgs-location}}

\subsection{General considerations}

We now turn to the issue of the location of the MSSM Higgs doublets,
which crucially influences the Yukawa sector predictions, as we saw in Section~\ref{sec:operators}. 

The terms in the results of Eq.~\eqref{eq:result-discrete} and \eqref{eq:result-continuous} involve the quantities $H^{\mathbf{H}}_{u,d,e,\nu}$, which according to Table~\ref{table:H-quantities} correspond to the $(\mathbf{1},\mathbf{2},+1/2)$ doublet fields $H^{u}_i$ or $(\mathbf{1},\mathbf{2},-1/2)$ anti-doublet fields
$H^{d}_{i}$. The $H^{u,d}_i$ are flavor eigenstates, which do not necessarily correspond to the mass eigenstates of the (anti)doublets. For the Yukawa predictions at the MSSM level, we are ultimately interested how the flavor states $H^{u,d}_{i}$ relate to the MSSM doublets $H_u$ and $H_d$, which are the only light mass eigenstates in the full doublet mass matrix above the GUT scale. All the other doublet mass eigenstates should be heavy, i.e.~at or near the GUT scale, since they should be integrated out at the GUT scale to obtain the MSSM 
as the effective description.

As is well known, the doublet mass matrix in GUT models is linked to the 
mass matrix of the triplets $(\mathbf{3},\mathbf{1},+1/3)$ and antitriplets
$(\mathbf{\bar{3}},\mathbf{1},-1/3)$, since the (weak) doublets come together with (color) triplets in GUT representations. Crucially, these (anti)triplets mediate $D=5$ proton decay in SUSY GUTs and must therefore be heavy not to violate proton decay bounds. We thus have a situation of two mass matrices linked by GUT symmetry: the (anti)doublet mass matrix with one pair of light states and all others heavy, and the triplet mass matrix with all states heavy. The issue of having only one light pair of doublet states with all other doubles and triplets heavy is known as doublet-triplet (DT) splitting, see e.g.~\cite{Nath:2006ut,Babu:2011tw} for a brief overview. The location of the MSSM Higgses is thus related to the question of DT splitting.

At the $\mathrm{SU}(5)$ level, the representations containing both a doublet and a triplet are $\mathbf{5}$ and $\mathbf{45}$, while the representation $\mathbf{50}$ contains a triplet only, but no doublet. Analogous statements hold for the conjugates of these representations. At the $\mathrm{SO}(10)$ level, the location and number of doublets and triplets can be looked up in Table~\ref{tab:decompositions-su5}.

Suppose we have a concrete model of the Higgs sector, whose superpotential $W$ is specified. We fix our notation by defining the doublet-antidoublet $M_D$ and triplet-antitriplet $M_T$ mass matrices via
	\begin{align}
        (M_D)_{kl}&=\frac{\partial^2 W}{\partial D_k\partial\overline{D}_l}\bigg|_{\text{vacuum}},\label{eq:MD-def}\\
        (M_T)_{kl}&=\frac{\partial^2 W}{\partial T_k\partial\overline{T}_l}\bigg|_{\text{vacuum}},\label{eq:MT-def}
	\end{align}
where $D_k$, $\overline{D}_l$, $T_k$ and $\overline{T}_l$ denote the doublet, antidoublet, triplet and antitriplet states, respectively, where the indices $k$ and $l$ go over available states of each type in the Higgs sector. The scalar mass matrices for doublets and antidoublets are then $M_D^\ast M_D^{T}$ and $M_D^{\dagger} M_D$, respectively. 

A necessary condition for DT splitting is that $\det M_D\approx 0$ (the MSSM pair of doublets is almost massless compared to the GUT scale) and $\det M_{T}\neq 0$.
Since the mass eigenmodes of $D$ and $\overline{D}$ are the left and right singular modes of $M_D$, respectively, the null mass pair of the left and right singular mode correspond to $H_u$ and $H_d$ of the MSSM, respectively. They can be written as a linear combination of (anti)doublets:
	\begin{align}
    H_u&=\sum_k a_k\, D_k,\label{eq:linear-combo-Hu}\\
    H_d&=\sum_l b_l\, \overline{D}_l,\label{eq:linear-combo-Hd}
    \end{align}
where $a_k$ and $b_l$ are complex coefficients with $\sum_k|a_k|^2=\sum_{l}|b_l|^2=1$ due to the unitarity of the left and right transition matrices in the singular value decomposition of $M_D$. There is a remaining overall phase ambiguity for the $a_k$, and similarly for $b_k$, associated to the phases of the states $H_u$ and $H_d$. The phase ambiguity is not relevant, since the phases are fixed in the MSSM so that the EW-breaking VEVs of $H_u$ and $H_d$ are real.

In the mass basis, only $H_u$ and $H_d$ can obtain an EW-breaking VEV, since they are the only light states. In the flavor basis $D_k$ and $\overline{D}_l$, adapted to GUT representations, the unitarity of transition matrices from the flavor to the mass eigenbasis then enforces the expansion 
\begin{align}
D_k&=a_k^\ast\,H_u+\ldots,&
\overline{D}_l&=b_l^\ast\,H_d+\ldots,
\label{eq:Hu-Hd-content}
\end{align}
where we omitted the heavy mass eigenstates on the right-hand side. Note that (some of) the $D_k$ flavor states correspond to the states $H^{u}_i$ in Eq.~\eqref{eq:hu-list} (the correspondence depends on the labeling convention for the $D$ states), and similarly (some) $\overline{D}_l$ states correspond to the $H^{d}_i$ states in Eq.~\eqref{eq:hd-list}. It is the $H^{u,d}_i$ states which enter via Table~\ref{table:H-quantities} into the Yukawa operator results of Eq.~\eqref{eq:result-discrete} and Eq.~\eqref{eq:result-continuous}. Once EW symmetry is broken, the doublet flavor eigenstates would thus acquire VEVs
\begin{align}
\langle D_k\rangle &= a_k^\ast\,v_u,&
\langle \overline{D}_l \rangle &= b_l^\ast\,v_d,\label{eq:vui-vdi}
\end{align}
where the usual MSSM definitions apply:
\begin{align}
v_u&:=v\,\sin\beta,&v_d&:=v\,\cos\beta,
\end{align} 
with $v=174\,\mathrm{GeV}$ and $\tan\beta=v_u/v_d$. 

The crucial point for model building is that in the presence of multiple doublets and antidoublets, the flavor eigenstates $D_k$ and $\overline{D}_l$, which include the $H^{u,d}_i$ coupled to SM fermions, depend on the coefficients $a_k$ and $b_l$ through Eq.~\eqref{eq:Hu-Hd-content}. These coefficients in turn 
depend on the superpotential parameters coming into $M_D$. Consequently, there is in principle extra freedom to the $\mathrm{SO}(10)$ constraints in the Yukawa entries due to the $a_i$ and $b_i$ parameters, depending on the particularities of the Higgs sector.  

\subsection{Tools for computing DT splitting\label{sec:DT-tools}}

We have seen that the details of DT splitting are crucial for the Yukawa predictions due to the coefficients $a_k$ and $b_l$, but are also model specific. 

One way to remain agnostic about the Higgs sector, as was discussed in Section~\ref{sec:discrete-directions}, is to simply take $a_k$ and $b_l$ as free complex parameters. The only assumption then required is that the freedom in the Higgs sector parameters is sufficient for the freedom in the $a_k$ and $b_l$ parameters, i.e.~that the unknown Higgs sector is sufficiently rich.

On the other hand, it is useful to the model builder to have the necessary tools for computing the $a_k$ and $b_l$ coefficients in a particular model of the Higgs sector.
We provide in this subsection all the necessary information for the reader to reconstruct the matrices $M_D$ and $M_T$ for the model of their choosing, provided that the Higgs sector of the superpotential consists of renormalizable operators with $\mathrm{SO}(10)$ representations of dimensions $210$ or below (and excluding $\mathbf{144}$, $\mathbf{\overline{144}}$ and $\mathbf{210}'$).
The coefficients $a_k$ and $b_l$ can then be extracted as the coefficients of the normalized left and right null vectors of the matrix $M_D$.

We write below all renormalizable superpotential terms of the $\mathrm{SO(10)}$ Higgs sector, where only one copy of each type of representation up to dimension $210$ (excluding $\mathbf{144}$, $\mathbf{\overline{144}}$ and $\mathbf{210}'$) is considered. We include only the terms relevant for the doublet/triplet mass matrices, while the terms relevant for the breaking of GUT symmetry, which determine the SM singlet VEVs, are not considered. Such a scenario involves the following terms:

\arraycolsep=4pt\def\arraystretch{1.5}
\begin{align}
W\supset&\nonumber\\\addlinespace[-24pt]
&
\begin{array}{rrllrrll}
&\tfrac{1}{2}&m_{10}&\mathbf{10}_{i}\,\mathbf{10}^{i}&
+&\tfrac{1}{2}&m_{120}&\mathbf{120}_{ijk}\,\mathbf{120}^{ijk}\\
+&&m_{126}&\mathbf{\overline{126}}_{ijklm}\,\mathbf{126}^{ijklm}&
+&\tfrac{1}{2}&m_{210}&\mathbf{210}_{ijkl}\,\mathbf{210}^{ijkl}\\
-&\sqrt{\tfrac{5}{3}}&\lambda'_{112}&\mathbf{10}_{i}\,\mathbf{10}_j\,\mathbf{54}^{ij}&
+&\tfrac{\sqrt{15}}{2}&\lambda_{123}&\mathbf{120}_{ijk}\,\mathbf{10}^{i}\,\mathbf{45}^{jk}\\
+&\sqrt{10}&\lambda_{134}&\mathbf{120}_{ijk}\,\mathbf{10}_{l}\,\mathbf{210}^{ijkl}&
-&2\sqrt{15}&\lambda'_{233}&\mathbf{120}^{ijk}\,\mathbf{120}_{ijl}\,\mathbf{54}_{k}{}^{l}\\
+&\sqrt{15}&\lambda_{334}&\mathbf{120}^{ijk} \mathbf{120}_{klm}\,\mathbf{210}_{ij}{}^{lm}&
+&\sqrt{5}&\lambda_{145}&\mathbf{10}_{i}\,\mathbf{210}_{jklm}\,\mathbf{126}^{ijklm}\\
+&\sqrt{5}&\lambda_{14\bar{5}}&\mathbf{10}_{i}\,\mathbf{210}_{jklm}\,\mathbf{\overline{126}}^{ijklm}&
+&\sqrt{15}&\lambda'_{255}&\mathbf{126}_{ijklm}\,\mathbf{126}^{ijkln}\,\mathbf{54}^{m}{}_{n}\\
+&\sqrt{15}&\lambda'_{2\bar{5}\bar{5}}&\mathbf{\overline{126}}_{ijklm}\,\mathbf{\overline{126}}^{ijkln}\,\mathbf{54}^{m}{}_{n}&
+&5\sqrt{10}&\lambda_{25\bar{5}}&\mathbf{\overline{126}}_{ijklm} \mathbf{126}^{ijkln}\,\mathbf{45}^{m}{}_{n}\\
+&10&\lambda_{45\bar{5}}&\mathbf{\overline{126}}_{ijklm}\,\mathbf{126}^{ijkno}\,\mathbf{210}^{lm}{}_{no}&
+&\tfrac{\sqrt{5}}{48\sqrt{2}}&\lambda_{244}&\mathbf{210}^{abcd}\,\mathbf{210}^{ijkl}\,\mathbf{45}^{mn}\,\varepsilon_{abcdijklmn}\\
+&\tfrac{4\sqrt{5}}{\sqrt{3}}&\lambda'_{244}&\mathbf{210}_{ijkl}\,\mathbf{210}^{ijkm}\,\mathbf{54}^{l}{}_{m}&
+&\sqrt{\tfrac{5}{3}}&\lambda_{444}&\mathbf{210}^{ijkl}\,\mathbf{210}_{klmn}\,\mathbf{210}_{ij}{}^{mn}\\
+&5&\lambda_{235}&\mathbf{120}_{ijk}\,\mathbf{45}_{lm}\,\mathbf{126}^{ijklm}&
+&\sqrt{30}&\lambda_{345}&\mathbf{120}_{ijn}\,\mathbf{210}_{klm}{}^{n}\,\mathbf{126}^{ijklm}\\
+&5&\lambda_{23\bar{5}}&\mathbf{120}_{ijk}\,\mathbf{45}_{lm}\,\mathbf{\overline{126}}^{ijklm}&
+&\sqrt{30}&\lambda_{34\bar{5}}&\mathbf{120}_{ijn}\,\mathbf{210}_{klm}{}^{n}\,\mathbf{\overline{126}}^{ijklm}\\
+&&m_{16}&\mathbf{\overline{16}}_{A}\,\mathbf{16}^{A}\\
+&\tfrac{\sqrt{5}}{3\sqrt{2}}&\lambda_{26\bar{6}}&\mathbf{\overline{16}}_A\,\mathbf{45}^{A}{}_{B}\,\mathbf{16}^B&
-&\tfrac{1}{8\sqrt{6}}&\lambda_{46\bar{6}}&\mathbf{\overline{16}}_A\,\mathbf{210}^{A}{}_{B}\,\mathbf{16}^B\\
+&\tfrac{1}{\sqrt{8}}&\lambda_{166}&\mathbf{16}_A\,\mathbf{10}^{A}{}_{B}\,\mathbf{16}^{B}&
-&\tfrac{1}{24\sqrt{10}}&\lambda_{\bar{5}66}&\mathbf{16}_A\,\mathbf{\overline{126}}^{A}{}_{B}\,\mathbf{16}^{B}\\
-&\tfrac{1}{\sqrt{8}}&\lambda_{1\bar{6}\bar{6}}&\mathbf{\overline{16}}_A\,\mathbf{10}^{A}{}_{B}\,\mathbf{\overline{16}}^{B}&
+&\tfrac{1}{24\sqrt{10}}&\lambda_{5\bar{6}\bar{6}}&\mathbf{\overline{16}}_A\,\mathbf{126}^{A}{}_{B}\,\mathbf{\overline{16}}^{B}.\\
\end{array}\label{eq:DT-superpotential-onecopy}
\end{align}

The conventions for the representations and their indices in tensor notation are specified in Appendix~\ref{appendix:conventions}. The fundamental indices have been lowered by $P_{ij}$ from Eq.\eqref{eq:raise-lower-fundamental}, while spinor indices have been lowered by $C_{AB}$ from Eq.~\eqref{eq:lower-raise-spinor}. Also, the spinor forms of the representations, such as $\mathbf{10}^{A}{}_{B}$, have been defined in Eq.~\eqref{eq:spinor-10}-\eqref{eq:spinor-126bar}. One can confirm that in each term all indices are indeed contracted. 

The normalizations of the states and VEVs in these representations is canonical, in the sense that the quadratic invariant formed with the complex conjugate gives orthonormal normalization. For example, if the states in the representation 
$\mathbf{126}$ are labeled by $X_K$, then
\begin{align}
\mathbf{126}^\ast_{ijklm}\,\mathbf{126}^{ijklm}&=\sum_{K}|X_K|^2.
\end{align}

The labeling of the coefficients $\lambda$ in front of the operators follows the following scheme: each $\lambda$ has $3$ numbers of increasing size in the index, which correspond to the three representations forming the invariant. Each number corresponds to the representation with that many fundamental indices, including a possible bar on top of the label if the representation has a bar; the exceptions are the $\mathbf{54}$, which also adds a prime to the symbol $\lambda$, and the representations $\mathbf{16}$ and $\mathbf{\overline{16}}$ corresponding to labels $6$ and $\bar{6}$, respectively. Altogether this provides an efficient labeling scheme for the coefficients; the numeric factors in front are there as part of our convention for later convenience. Eq.~\eqref{eq:DT-superpotential-onecopy} should thus also be viewed as defining our notation for the $m$ and $\lambda$ parameters in front of the operators.

We now specify a basis for all the doublets/antidoublets present in the representations of Eq.~\eqref{eq:DT-superpotential-onecopy}:
\begin{align}
\begin{split}
D_{k}&=\begin{pmatrix}
D_{\mathbf{10}\supset\mathbf{5}}&
D_{\mathbf{120}\supset\mathbf{5}}&
D_{\mathbf{120}\supset\mathbf{45}}&
D_{\mathbf{126}\supset\mathbf{45}}&
D_{\mathbf{\overline{126}}\supset\mathbf{5}}&
D_{\mathbf{210}\supset\mathbf{5}}&
D_{\mathbf{\overline{16}}\supset\mathbf{5}}\\
\end{pmatrix},\\
\overline{D}_{l}&=\begin{pmatrix}
\overline{D}_{\mathbf{10}\supset\mathbf{\overline{5}}}&
\overline{D}_{\mathbf{120}\supset\mathbf{\overline{5}}}&
\overline{D}_{\mathbf{120}\supset\mathbf{\overline{45}}}&
\overline{D}_{\mathbf{\overline{126}}\supset\mathbf{\overline{45}}}&
\overline{D}_{\mathbf{126}\supset\mathbf{\overline{5}}}&
\overline{D}_{\mathbf{210}\supset\mathbf{\overline{5}}}&
\overline{D}_{\mathbf{16}\supset\mathbf{\overline{5}}}\\
\end{pmatrix},\\
\end{split}\label{eq:labels-doublets}
\end{align}
in an obvious notation based on which $\mathrm{SO}(10)$ representation 
and $\mathrm{SU}(5)$ subrepresentation they are located in. The explicit relation of such a basis to the more specific $H^{u,d}_i$ notation from Eq.~\eqref{eq:hu-list} and \eqref{eq:hd-list} relevant for Table~\ref{table:H-quantities} is the following: 
\begin{align}
H^{u}_{1}&=D_1=D_{\mathbf{10}\supset\mathbf{5}},&
H^{d}_{1}&=\overline{D}_1=\overline{D}_{\mathbf{10}\supset\mathbf{\overline{5}}},\nonumber\\
H^{u}_{2}&=D_5=D_{\mathbf{\overline{126}}\supset\mathbf{5}},&
H^{d}_{2}&=\overline{D}_4=\overline{D}_{\mathbf{\overline{126}}\supset\mathbf{\overline{45}}},\nonumber\\
H^{u}_{3}&=D_2=D_{\mathbf{120}\supset\mathbf{5}},&
H^{d}_{3}&=\overline{D}_2=\overline{D}_{\mathbf{120}\supset\mathbf{\overline{5}}},\nonumber\\
H^{u}_{4}&=D_3=D_{\mathbf{120}\supset\mathbf{45}},&
H^{d}_{4}&=\overline{D}_3=\overline{D}_{\mathbf{120}\supset\mathbf{\overline{45}}}.
\end{align}

\noindent
Similarly, we use the triplet/antitriplet basis
\begin{align}
\begin{split}
T_{k}&=\begin{pmatrix}
T_{\mathbf{10}\supset\mathbf{5}}&
T_{\mathbf{120}\supset\mathbf{5}}&
T_{\mathbf{120}\supset\mathbf{45}}&
T_{\mathbf{126}\supset\mathbf{45}}&
T_{\mathbf{\overline{126}}\supset\mathbf{5}}&
T_{\mathbf{210}\supset\mathbf{5}}&
T_{\mathbf{\overline{16}}\supset\mathbf{5}}&
T_{\mathbf{\overline{126}}\supset\mathbf{\mathbf{50}}}\\
\end{pmatrix},\\
\overline{T}_{l}&=\begin{pmatrix}
\overline{T}_{\mathbf{10}\supset\mathbf{\overline{5}}}&
\overline{T}_{\mathbf{120}\supset\mathbf{\overline{5}}}&
\overline{T}_{\mathbf{120}\supset\mathbf{\overline{45}}}&
\overline{T}_{\mathbf{\overline{126}}\supset\mathbf{\overline{45}}}&
\overline{T}_{\mathbf{126}\supset\mathbf{\overline{5}}}&
\overline{T}_{\mathbf{210}\supset\mathbf{\overline{5}}}&
\overline{T}_{\mathbf{16}\supset\mathbf{\overline{5}}}&
\overline{T}_{\mathbf{126}\supset\mathbf{\overline{50}}}\\
\end{pmatrix},\\
\end{split}\label{eq:labels-triplets}
\end{align}
where an extra (anti)triplet state from the $\mathbf{\overline{126}}$ ($\mathbf{126}$) contained in the $\mathbf{50}$ ($\mathbf{\overline{50}}$) of $\mathrm{SU}(5)$ is now present. 

Given these bases, $D_k$ and $\overline{D}_l$ consist of $7$ states each, while $T_k$ and $\overline{T}_l$ consist of $8$ states, so that $M_{D}$ and $M_{T}$ from Eq.~\eqref{eq:MD-def} and \eqref{eq:MT-def}, respectively, are $7\times 7$ and $8\times 8$ matrices. They can be compactly written as a single $8\times 8$ matrix $M_{D|T}$ in the following way:

\begingroup
\allowdisplaybreaks
{\scriptsize
\begin{align}
M_{D|T}=&\qquad
\begin{pmatrix}
 m_{10} & 0 & 0 & 0 & 0 & \lambda_{145} V_{126} & \lambda_{166} V_{16} & 0 \\
 0 & m_{120} & 0 & 0 & 0 & \lambda_{345} V_{126} & 0 & 0 \\
 0 & 0 & m_{120} & 0 & 0 & 0 & 0 & 0 \\
 0 & 0 & 0 & m_{126} & 0 & 0 & 0 & 0 \\
 0 & 0 & 0 & 0 & m_{126} & \lambda_{45\bar{5}}\,V_{126} & \lambda_{\bar{5}66}\,V_{16} & 0 \\
 \lambda_{14\bar{5}}\, V_{\overline{126}} & \lambda_{34\bar{5}}\, V_{\overline{126}} & 0 & 0 & \lambda_{45\bar{5}}\, V_{\overline{126}} & m_{210} & \lambda_{46\bar{6}}\, V_{\overline{16}} & 0 \\
\lambda_{1\bar{6}\bar{6}}\, V_{\overline{16}} & 0 & 0 & 0 & \lambda_{5\bar{6}\bar{6}}\, V_{\overline{16}} & \lambda_{46\bar{6}}\,V_{16} & m_{16} & 0 \\
 0 & 0 & 0 & 0 & 0 & 0 & 0 & m_{126} \\
\end{pmatrix} \nonumber\\
&+V_{45}\,
\begin{pmatrix}
 0 & -\lambda_{123} & 0 & 0 & 0 & 0 & 0 & 0 \\
 \lambda_{123} & 0 & 0 & 0 & \sqrt{3} \lambda_{235} & 0 & 0 & 0 \\
 0 & 0 & 0 & \lambda_{23\bar{5}} & 0 & 0 & 0 & 0 \\
 0 & 0 & -\lambda_{235} & \lambda_{25\bar{5}} & 0 & 0 & 0 & 0 \\
 0 & -\sqrt{3} \lambda_{23\bar{5}} & 0 & 0 & -\lambda_{25\bar{5}} & 0 & 0 & 0 \\
 0 & 0 & 0 & 0 & 0 & \lambda_{244} & 0 & 0 \\
 0 & 0 & 0 & 0 & 0 & 0 & \lambda_{26\bar{6}} & 0 \\
 0 & 0 & 0 & 0 & 0 & 0 & 0 & -\lambda_{25\bar{5}} \\
\end{pmatrix} \nonumber \\
&+ W_{45}\,
\left(\begin{smallmatrix}
 0 & -\tfrac{1}{4} \sqrt{\tfrac{3}{2}} \eta_{1} \lambda_{123} & -\tfrac{5}{4 \sqrt{2}} \eta_{2} \lambda_{123}& 0 & 0 & 0 & 0 & 0 \\
 \tfrac{1}{4} \sqrt{\tfrac{3}{2}} \eta_{1} \lambda_{123} & 0 & 0 & \tfrac{5}{4 \sqrt{2}} \eta_{2} \lambda_{23\bar{5}}& \tfrac{3}{4 \sqrt{2}} \eta_{1} \lambda_{235}& 0 &
   0 & 0 \\
 \tfrac{5}{4 \sqrt{2}} \eta_{2} \lambda_{123}& 0 & 0 & \tfrac{13}{4 \sqrt{6}} \eta_{3} \lambda_{23\bar{5}}& -\tfrac{5}{4 \sqrt{6}} \eta_{2} \lambda_{235}& 0 & 0 &
   -\tfrac{5}{3}\lambda_{235} \\
 0 & -\tfrac{5}{4 \sqrt{2}} \eta_{2} \lambda_{235}& -\tfrac{13}{4 \sqrt{6}} \eta_{3} \lambda_{235}& -\sqrt{\tfrac{3}{2}} \eta_{1} \lambda_{25\bar{5}} & 0 & 0 & 0 & 0
   \\
 0 & -\tfrac{3}{4 \sqrt{2}} \eta_{1} \lambda_{23\bar{5}}& \tfrac{5}{4 \sqrt{6}} \eta_{2} \lambda_{23\bar{5}}& 0 & \sqrt{\tfrac{3}{2}} \eta_{1} \lambda_{25\bar{5}} & 0 & 0 & 0 \\
 0 & 0 & 0 & 0 & 0 & -\sqrt{\tfrac{3}{2}} \eta_{1} \lambda_{244} & 0 & 0 \\
 0 & 0 & 0 & 0 & 0 & 0 & \sqrt{\tfrac{2}{3}} \eta_{1} \lambda_{26\bar{6}} & 0 \\
 0 & 0 & \tfrac{5}{3} \lambda_{23\bar{5}}& 0 & 0 & 0 & 0 & -\sqrt{\tfrac{2}{3}} \lambda_{25\bar{5}} \\
\end{smallmatrix}\right) \nonumber\\
&+ W_{54}
\begin{pmatrix}
\eta_{1} \lambda'_{112} & 0 & 0 & 0 & 0 & 0 & 0 & 0 \\
 0 & \eta_{1} \lambda'_{233} & -\tfrac{5 \eta_{2} \lambda'_{233}}{\sqrt{3}} & 0 & 0 & 0 & 0 & 0 \\
 0 & -\tfrac{5 \eta_{2} \lambda'_{233}}{\sqrt{3}} & \tfrac{13 \eta_{3} \lambda'_{233}}{3} & 0 & 0 & 0 & 0 & 0 \\
 0 & 0 & 0 & 0 & \eta_{2} \lambda'_{255} & 0 & 0 & -2 \sqrt{\tfrac{2}{3}} \lambda'_{255} \\
 0 & 0 & 0 & \eta_{2} \lambda'_{2\bar{5}\bar{5}} & 0 & 0 & 0 & 0 \\
 0 & 0 & 0 & 0 & 0 & \eta_{1} \lambda'_{244} & 0 & 0 \\
 0 & 0 & 0 & 0 & 0 & 0 & 0 & 0 \\
 0 & 0 & 0 & -2 \sqrt{\tfrac{2}{3}} \lambda'_{2\bar{5}\bar{5}} & 0 & 0 & 0 & 0 \\
\end{pmatrix} \nonumber\\
&+V_{210}\,
\begin{pmatrix}
 0 & \lambda_{134} & 0 & 0 & \sqrt{\tfrac{3}{5}} \lambda_{145} & 0 & 0 & 0 \\
 \lambda_{134} & \lambda_{334} & 0 & 0 & -\tfrac{1}{2} \sqrt{\tfrac{3}{5}} \lambda_{345} & 0 & 0 & 0 \\
 0 & 0 & -\tfrac{1}{3}\lambda_{334} & -\tfrac{1}{2 \sqrt{5}} \lambda_{34\bar{5}}& 0 & 0 & 0 & 0 \\
 0 & 0 & -\tfrac{1}{2 \sqrt{5}} \lambda_{345}& 0 & 0 & 0 & 0 & 0 \\
 \sqrt{\tfrac{3}{5}} \lambda_{14\bar{5}} & -\tfrac{1}{2} \sqrt{\tfrac{3}{5}} \lambda_{34\bar{5}} & 0 & 0 & \tfrac{2}{\sqrt{15}}\lambda_{45\bar{5}} & 0 & 0 & 0 \\
 0 & 0 & 0 & 0 & 0 & \lambda_{444} & 0 & 0 \\
 0 & 0 & 0 & 0 & 0 & 0 & \tfrac{1}{2 \sqrt{10}}\lambda_{46\bar{6}} & 0 \\
 0 & 0 & 0 & 0 & 0 & 0 & 0 & -\tfrac{1}{\sqrt{15}}\lambda_{45\bar{5}} \\
\end{pmatrix} \nonumber\\
&+ W_{210}
\left(
\begin{smallmatrix}
0 & -\tfrac{3 \eta_{1} \lambda_{134}}{4} & \tfrac{5}{4 \sqrt{3}}\eta_{2} \lambda_{134} & \tfrac{1}{2} \sqrt{\tfrac{5}{3}} \eta_{2} \lambda_{14\bar{5}} & \tfrac{1}{2}
   \sqrt{\tfrac{3}{5}} \eta_{1} \lambda_{145} & 0 & 0 & 0 \\
 -\tfrac{3}{4} \eta_{1} \lambda_{134}& \tfrac{1}{2} \eta_{1} \lambda_{334}& \tfrac{5}{6 \sqrt{3}} \eta_{2} \lambda_{334}& -\tfrac{1}{8} \sqrt{\tfrac{5}{3}}
   \eta_{2} \lambda_{34\bar{5}} & \tfrac{3}{8} \sqrt{\tfrac{3}{5}} \eta_{1} \lambda_{345} & 0 & 0 & 0 \\
 \tfrac{5}{4 \sqrt{3}} \eta_{2} \lambda_{134} & \tfrac{5}{6 \sqrt{3}} \eta_{2} \lambda_{334}& \tfrac{7}{18} \eta_{4} \lambda_{334}& -\tfrac{1}{24 \sqrt{5}} \eta_{5}
   \lambda_{34\bar{5}}& \tfrac{1}{24} \sqrt{5} \eta_{2} \lambda_{345} & 0 & 0 & \tfrac{1}{3} \sqrt{\tfrac{5}{6}} \lambda_{345} \\
 \tfrac{1}{2} \sqrt{\tfrac{5}{3}} \eta_{2} \lambda_{145} & -\tfrac{1}{8} \sqrt{\tfrac{5}{3}} \eta_{2} \lambda_{345} & -\tfrac{1}{24 \sqrt{5}} \eta_{5} \lambda_{345}&
   \tfrac{1}{3} \sqrt{\tfrac{5}{3}} \eta_{6} \lambda_{45\bar{5}} & 0 & 0 & 0 & 0 \\
 \tfrac{1}{2} \sqrt{\tfrac{3}{5}} \eta_{1} \lambda_{14\bar{5}} & \tfrac{3}{8} \sqrt{\tfrac{3}{5}} \eta_{1} \lambda_{34\bar{5}} & \tfrac{1}{24} \sqrt{5} \eta_{2} \lambda_{34\bar{5}}
   & 0 & \tfrac{\eta_{1} \lambda_{45\bar{5}}}{\sqrt{15}} & 0 & 0 & 0 \\
 0 & 0 & 0 & 0 & 0 & \tfrac{1}{2} \eta_{1} \lambda_{444}& 0 & 0 \\
 0 & 0 & 0 & 0 & 0 & 0 & \tfrac{3}{2 \sqrt{10}}\eta_{1} \lambda_{46\bar{6}} & 0 \\
 0 & 0 & \tfrac{1}{3} \sqrt{\tfrac{5}{6}} \lambda_{34\bar{5}} & 0 & 0 & 0 & 0 & \tfrac{1}{3 \sqrt{15}} \lambda_{45\bar{5}}\\
\end{smallmatrix}\right) \nonumber\\
&+ Z_{210}
\left(
\begin{smallmatrix}
 0 & 0 & -\sqrt{\tfrac{5}{3}} \eta_{7} \lambda_{134} & \tfrac{1}{\sqrt{3}}\eta_{7} \lambda_{14\bar{5}} & 0 & 0 & 0 & \sqrt{\tfrac{2}{3}} \lambda_{145} \\
 0 & 0 & \tfrac{1}{3} \sqrt{\tfrac{5}{3}} \eta_{7} \lambda_{334} & \tfrac{1}{2 \sqrt{3}}\eta_{7} \lambda_{34\bar{5}} & 0 & 0 & 0 & 0 \\
 -\sqrt{\tfrac{5}{3}} \eta_{7} \lambda_{134} & \tfrac{1}{3} \sqrt{\tfrac{5}{3}} \eta_{7} \lambda_{334} & \tfrac{2}{9} \sqrt{5} \eta_{8} \lambda_{334} &
   \tfrac{1}{3}\eta_{6} \lambda_{34\bar{5}} & -\tfrac{1}{6}\eta_{7} \lambda_{345} & 0 & 0 & \tfrac{1}{3} \sqrt{\tfrac{2}{3}} \lambda_{345} \\
 \tfrac{1}{\sqrt{3}}\eta_{7} \lambda_{145} & \tfrac{1}{2 \sqrt{3}} \eta_{7} \lambda_{345}& \tfrac{1}{3}\eta_{6} \lambda_{345} & \tfrac{1}{3 \sqrt{3}} 2 \eta_{6} 
 \lambda_{45\bar{5}} & 0 & 0 & 0 & 0 \\
 0 & 0 & -\tfrac{1}{6}\eta_{7} \lambda_{34\bar{5}} & 0 & 0 & 0 & 0 & -\tfrac{1}{3} \sqrt{\tfrac{2}{3}} \lambda_{45\bar{5}} \\
 0 & 0 & 0 & 0 & 0 & 0 & 0 & 0 \\
 0 & 0 & 0 & 0 & 0 & 0 & 0 & 0 \\
 \sqrt{\tfrac{2}{3}} \lambda_{14\bar{5}} & 0 & \tfrac{1}{3} \sqrt{\tfrac{2}{3}} \lambda_{34\bar{5}} & 0 & -\tfrac{1}{3} \sqrt{\tfrac{2}{3}} \lambda_{45\bar{5}} & 0 & 0 & \tfrac{2 \lambda_{45\bar{5}}}{3
   \sqrt{3}} \\
\end{smallmatrix}\right).
\label{eq:MDT}
\end{align}
}
\endgroup

The way to interpret the compactly written doublet/triplet mass matrix $M_{D|T}$ in Eq.~\eqref{eq:MDT} is the following:
\begin{itemize}
\item \underline{Doublets:} obtain $M_D$ by crossing out the last row and column of $M_{D|T}$ and take $\eta_i=1$. 
\item \underline{Triplets:} obtain $M_T$ from $M_{D|T}$ by taking 
\begin{align}
\eta_{1}&=-\tfrac{2}{3},&
\eta_{2}&=-\tfrac{2}{\sqrt{3}},&
\eta_{3}&=-\tfrac{2}{13},&
\eta_{4}&=\tfrac{2}{7},&
\eta_{5}&=5,\nonumber\\
\eta_{6}&=0,&
\eta_{7}&=\tfrac{1}{\sqrt{3}},&
\eta_{8}&=2,&
\eta_{9}&=-2.&&\label{eq:triplet-Clebsch}
\end{align}
\end{itemize}
We have labeled the SM singlet VEVs in Eq.~\eqref{eq:MDT} by their $\mathrm{SU}(5)$ origin with letters $V$, $W$ and $Z$
for the representations $\mathbf{1}$, $\mathbf{24}$ and $\mathbf{75}$, respectively. Their
$\mathrm{SO}(10)$ origin is indicated in the index. The Clebsch coefficients $\eta$ are present only in contributions from $\mathrm{SU}(5)$ non-singlets, i.e.~the $W$ and $Z$ VEVs.
Note: this notation for the VEVs in the $\mathbf{45}$ and $\mathbf{210}$ is different than the one used in Section~\ref{sec:operators}; such representations in the Higgs sector relevant for DT splitting can be unrelated to those
in the Yukawa sector operators. 

In addition to the operators in Eq.~\eqref{eq:DT-superpotential-onecopy}, there are $7$ more renormalizable operators (completely) anti-symmetric in the representations of the same dimensions, such that multiple independent copies of a representation are required for the invariants to be non-zero:

\begin{align}
\begin{split}
\mathcal{O}^{\alpha\beta}_{112}&=(\mathbf{10}^\alpha)_{i}\,(\mathbf{10}^\beta)_j\,\mathbf{45}^{ij},\\
\mathcal{O}^{\alpha\beta}_{233}&=(\mathbf{120}^\alpha)^{ijk}\,(\mathbf{120}^\beta)_{ijl}\,\mathbf{45}_{k}{}^{l},\\
\mathcal{O}^{\alpha\beta}_{334}&=(\mathbf{120}^\alpha)^{ijk}\,(\mathbf{120}^\beta)^{lmn}\,\mathbf{210}^{abcd}\,\varepsilon_{ijklmnabcd},\\
\mathcal{O}^{\alpha\beta}_{244}&=(\mathbf{210}^\alpha)_{ijkl}\,(\mathbf{210}^\beta)^{ijkm}\,\mathbf{45}^{l}{}_{m},\\
\mathcal{O}^{\alpha\beta\gamma}_{444}&=(\mathbf{210}^\alpha)^{abcd}\,(\mathbf{210}^\beta)^{ijko}\,(\mathbf{210}^{\gamma})^{lmn}{}_{o}\,\varepsilon_{abcdijklmn},\\
\mathcal{O}^{\alpha\beta}_{366}&=(\mathbf{16}^\alpha)_A\,\mathbf{120}^{A}{}_{B}\,(\mathbf{16}^\beta)^{B},\\
\mathcal{O}^{\alpha\beta}_{3\bar{6}\bar{6}}&=(\mathbf{\overline{16}}^\alpha)_A\,\mathbf{120}^{A}{}_{B}\,(\mathbf{\overline{16}}^\beta)^{B}.\\
\end{split}\label{eq:antisymmetric-operators-list}
\end{align}

\noindent
The indices $\alpha$, $\beta$ and $\gamma$ label the copy of a representation. The above anti-symmetric operators yield by explicit computation the following doublet and triplet terms:
\begin{align}
\mathcal{O}^{\alpha\beta}_{112}&=\quad 
\left(
	D^\beta_{\mathbf{10}\supset\mathbf{5}}\overline{D}^\alpha_{\mathbf{10}\supset\mathbf{\overline{5}}}
	+T^\beta_{\mathbf{10}\supset\mathbf{5}}\overline{T}^\alpha_{\mathbf{10}\supset\mathbf{\overline{5}}}
\right)\,\left(
	\tfrac{1}{\sqrt{10}}\,V_{45}-\tfrac{\sqrt{3}}{2\sqrt{5}}\,\eta_1\,W_{45}
\right)+\cdots,\label{eq:antisymmetric-invariants-begin}\\
\mathcal{O}^{\alpha\beta}_{233}&=\quad
\left(
	D^\beta_{\mathbf{120}\supset\mathbf{5}\phantom{4}}\,\overline{D}^\alpha_{\mathbf{120}\supset\mathbf{\overline{5}}\phantom{4}}+
	T^\beta_{\mathbf{120}\supset\mathbf{5}\phantom{4}}\,\overline{T}^\alpha_{\mathbf{120}\supset\mathbf{\overline{5}}\phantom{4}}
\right)\,\left(
	\tfrac{1}{3\sqrt{10}}V_{45}-\tfrac{1}{2\sqrt{15}}\eta_1\,W_{45}
\right)\nonumber\\
&\quad +
\left(
	D^\beta_{\mathbf{120}\supset\mathbf{45}}\,\overline{D}^\alpha_{\mathbf{120}\supset\mathbf{\overline{45}}}+
	T^\beta_{\mathbf{120}\supset\mathbf{45}}\,\overline{T}^\alpha_{\mathbf{120}\supset\mathbf{\overline{45}}}
\right)\,\left(
	\tfrac{1}{3\sqrt{10}}V_{45}-\tfrac{1}{2\sqrt{15}}\eta_1\,W_{45}
\right)+\cdots,\\
\mathcal{O}^{\alpha\beta}_{334}&=\quad
\left(
	D^\beta_{\mathbf{120}\supset\mathbf{5}\phantom{4}}\,\overline{D}^\alpha_{\mathbf{120}\supset\mathbf{\overline{5}}\phantom{4}}+
	T^\beta_{\mathbf{120}\supset\mathbf{5}\phantom{4}}\,\overline{T}^\alpha_{\mathbf{120}\supset\mathbf{\overline{5}}\phantom{4}}
\right)\,\big(2\,V_{210}+\eta_1\,W_{210}\big)\,\Big(-18\sqrt{\tfrac{3}{5}}\Big)\nonumber\\
&\quad +
\left(
	D^\beta_{\mathbf{120}\supset\mathbf{45}}\,\overline{D}^\alpha_{\mathbf{120}\supset\mathbf{\overline{45}}}+
	T^\beta_{\mathbf{120}\supset\mathbf{45}}\,\overline{T}^\alpha_{\mathbf{120}\supset\mathbf{\overline{45}}}
\right)\left(
	12\,V_{210}+26\eta_3\, W_{210}+8\sqrt{5}\eta_9\,Z_{210}
\right)\sqrt{\tfrac{3}{5}}\nonumber\\
&\quad +
\left(
	D^\beta_{\mathbf{120}\supset\mathbf{5\phantom{4}}}\,\overline{D}^\alpha_{\mathbf{120}\supset\mathbf{\overline{45}}}+
	T^\beta_{\mathbf{120}\supset\mathbf{5\phantom{4}}}\,\overline{T}^\alpha_{\mathbf{120}\supset\mathbf{\overline{45}}}
\right)\,\left(
	\sqrt{5} \eta_2\,W_{210} + 2\eta_7 \,Z_{210}
\right)\,6\nonumber\\
&\quad +
\left(
	D^\beta_{\mathbf{120}\supset\mathbf{45}}\,\overline{D}^\alpha_{\mathbf{120}\supset\mathbf{\overline{5}\phantom{4}}}+
	T^\beta_{\mathbf{120}\supset\mathbf{45}}\,\overline{T}^\alpha_{\mathbf{120}\supset\mathbf{\overline{5}\phantom{4}}}
\right)\,\left(
	\sqrt{5} \eta_2\,W_{210} + 2\eta_7 \,Z_{210}
\right)\,6 +\cdots,\\
\mathcal{O}^{\alpha\beta}_{244}&=\quad 
\left(
	D^\alpha_{\mathbf{210}\supset\mathbf{5}}\,\overline{D}^\beta_{\mathbf{210}\supset\mathbf{\overline{5}}}+
	T^\alpha_{\mathbf{210}\supset\mathbf{5}}\,\overline{T}^\beta_{\mathbf{210}\supset\mathbf{\overline{5}}}
\right)\,\left(\tfrac{1}{\sqrt{10}}\,V_{45}+\tfrac{\sqrt{3}}{8\sqrt{5}}\eta_1\,W_{45}\right)+\cdots,\\
\mathcal{O}^{\alpha\beta\gamma}_{444}&=\quad
\left(D^\alpha_{\mathbf{210}\supset\mathbf{5}}\,\overline{D}^\beta_{\mathbf{210}\supset\mathbf{\overline{5}}}+T^\alpha_{\mathbf{210}\supset\mathbf{5}}\,\overline{T}^\beta_{\mathbf{210}\supset\mathbf{\overline{5}}}\right)
\,\left(4\,V^\gamma_{210} - 3\eta_1\,W^\gamma_{210}\right)\,3\sqrt{\tfrac{3}{5}}+\cdots,\\
\mathcal{O}^{\alpha\beta}_{366}&=\quad
\left( D_{\mathbf{120}\supset\mathbf{5}}\,\overline{D}^\alpha_{\mathbf{16}\supset\mathbf{\overline{5}}} 
+ T_{\mathbf{120}\supset\mathbf{5}}\,\overline{T}^\alpha_{\mathbf{16}\supset\mathbf{\overline{5}}}
\right)\;\tfrac{1}{2}V^\beta_{16}
+\cdots,\\
\mathcal{O}^{\alpha\beta}_{3\bar{6}\bar{6}}&=\quad
\left( \overline{D}^\beta_{\mathbf{\overline{16}}\supset\mathbf{5}}  \,  \overline{D}_{\mathbf{120}\supset\mathbf{\overline{5}}}
+ T^\beta_{\mathbf{\overline{16}}\supset\mathbf{5}}  \,  \overline{T}_{\mathbf{120}\supset\mathbf{\overline{5}}}
\right)\;\tfrac{1}{2}V^\alpha_{\overline{16}}+\cdots.
\label{eq:antisymmetric-invariants-end}
\end{align}

In the expressions above, the ellipsis symbol signifies the addition of terms with permuted indices $\alpha\beta$ or $\alpha\beta\gamma$, such that the expressions become completely anti-symmetric in these indices. Furthermore, the factors with Clebsch coefficients $\eta_i$ have different numerical values depending on whether they are multiplying a $D\overline{D}$ or $T\overline{T}$ term: $\eta_i=1$ for  doublets, while the values for triplets are given in Eq.~\eqref{eq:triplet-Clebsch}. The notation for some doublets, triplets and VEVs now also carries a multiplicity index in a straightforward extension of the notation from Eq.~\eqref{eq:labels-doublets} and \eqref{eq:labels-triplets}.

This completes all the data needed to compute $M_D$ and $M_T$ with any renormalizable potential. If multiple copies of a representation are used, one can distinguish between two cases: when the invariant is completely symmetric or anti-symmetric in the copies.\footnote{In the case of cubic invariants of a single type of representation, with $\mathbf{210}^3$ the only such example in our case, there is no independent mixed symmetry invariant.} The data for the anti-symmetric cases is obviously given in Eq.~\eqref{eq:antisymmetric-invariants-begin}-\eqref{eq:antisymmetric-invariants-end}. The symmetric case, on the other hand, can be reconstructed from the data in Eq.~\eqref{eq:MDT} noting that the invariants in \eqref{eq:DT-superpotential-onecopy} turn out to be the invariants completely symmetric in the representation factors of the same type.

\subsection{Discussion and the most predictive cases\label{section:DT-discussion}}

Given the model building tools for the Higgs sector and DT splitting developed in Section~\ref{sec:DT-tools}, we now reiterate how to make use of these tools step by step. The procedure to build a concrete Higgs sector and determine its impact on the Yukawa sector in the MSSM effective theory below the GUT scale consists of the following steps:
\begin{enumerate}
\item Choose the $\mathrm{SO}(10)$ Higgs representations involved in the model; this impacts which $\mathbf{H}$ are available for the Yukawa operators in Section~\ref{sec:operators}. 
\item Given the choice of representations, determine the available operators in the superpotential $W$, perhaps forbidding some with discrete symmetries. The
list of available operators at the renormalizable level is given in Eq.~\eqref{eq:DT-superpotential-onecopy} and \eqref{eq:antisymmetric-operators-list}. 
\item Given the representations and operators from the prior two steps, determine $M_D$ and $M_T$. The data for renormalizable operators is found in Eq.~\eqref{eq:MDT} and \eqref{eq:antisymmetric-invariants-begin}-\eqref{eq:antisymmetric-invariants-end}.
\item Make sure that DT splitting is achieved by confirming that $\det\,M_D=0$ and $\det\,M_{T}\neq 0$. We will return to the issue how to achieve this later.
\item Compute the normalized left and right null modes for $M_D$; when the basis for rows and columns consists of $D_k$ and $\overline{D}_l$, respectively, the obtained left and right null vectors correspond to the  coefficients $a_{k}$ and $b_{l}$, respectively. This determines the Yukawa sector operators in the MSSM effective theory via Eq.~\eqref{eq:Hu-Hd-content}.
\end{enumerate}

We turn now to the issue of achieving DT splitting in step 4.
One possibility is to make use of one of the mechanisms for DT splitting in the literature, e.g. the missing partner mechanism \cite{Masiero:1982fe,Babu:2006nf,Babu:2011tw} or Dimopoulos-Wilczek mechanism in $\mathrm{SO}(10)$ \cite{Dimopoulos:1981xm,Babu:1993we}. The mechanisms work by having a very particular form of the matrices $M_D$ and $M_T$ or a particular alignment of VEVs in these matrices. This always requires a very particular set-up in the Higgs sector, i.e.~very special choices in steps 1-3. We shall not consider this possibilities further in this paper.

Another possibility is \textit{DT splitting by fine-tuning}. While this option 
can typically be employed in any model, it is considered a less elegant solution to the DT splitting problem. One simply computes $\det M_D$ 
and imposes that the resulting expression vanishes, implying a
strict relation between the independent parameters of the Higgs sector, i.e.~it requires fine-tuning one of them. If $\det M_T\neq 0$ after imposing the fine-tuning relation, DT splitting was successfully achieved.

\begin{table}[htb]
\begin{center}
	\caption{Examples of models of Higgs sectors with DT splitting giving predictive values with no free parameters for the (anti)doublet VEVs $H^{u,d}_{i}$ defined in Eq.~\eqref{eq:hu-list}. For the predictions in the Yukawa sector, one can replace $\propto$ with $=$, see main body text for details. The scenarios are listed by their $\#$ number; we specify for each the available representations $\mathbf{H}$, the other representations in the Higgs sector,
the operators in the superpotential, the fine-tuning condition for DT splitting, and finally the location of MSSM Higgses $H_{u,d}$ in the flavor states $H_{i}^{u,d}$.\label{table:DT-models}}
	\begin{tabular}{llllll}
	\toprule
	\multicolumn{1}{p{0.5cm}}{\#}&
	\multicolumn{1}{p{1.6cm}}{$\mathbf{H}$}&
	\multicolumn{1}{p{2.2cm}}{other reps}&
	\multicolumn{1}{p{3cm}}{operators}&
	\multicolumn{1}{p{4.1cm}}{fine-tuning}&
	\multicolumn{1}{p{2.5cm}}{Higgs location}\\
	\midrule
	$1$&$\mathbf{10}$&$\mathbf{54}$&$m_{10}$,$\lambda'_{112}$&$\lambda'_{112}=-\tfrac{m_{10}}{W_{54}}$&$H^{u,d}_{1}=H_{u,d}$\\
	\midrule
	$2$&$\mathbf{120}$&$\mathbf{54}$&$m_{120}$,$\lambda'_{233}$&$\lambda'_{233}=-\tfrac{m_{120}}{6W_{54}}$&$H^{u,d}_{3}=-\tfrac{1}{2}H_{u,d}$\\
	&&&&&$H^{u,d}_{4}=\tfrac{\sqrt{3}}{2}H_{u,d}$\\
	\midrule
	$3$&$\mathbf{120}$&$\mathbf{54}$&$m_{120}$,$\lambda'_{233}$&$\lambda'_{233}=\tfrac{3m_{120}}{2W_{54}}$&$H^{u,d}_{3}=\tfrac{\sqrt{3}}{2}H_{u,d}$\\
	&&&&&$H^{u,d}_{4}=\tfrac{1}{2}\,H_{u,d}$\\
	\midrule
	$4$&$\mathbf{\overline{126}}$&$\mathbf{126},\mathbf{54}$&$m_{126}$,$\lambda'_{255}$,$\lambda'_{2\bar{5}\bar{5}}$&$\lambda'_{255}=\tfrac{m_{126}^2}{W_{54}^2\lambda'_{2\bar{5}\bar{5}}}$&$H^{u,d}_{2}\propto H_{u,d}$\\
	\midrule
	$5$&$\mathbf{10}$,$\mathbf{120}$&$\langle\mathbf{45}\rangle=\widetilde{X}_{1}$&$m_{10}$,$m_{120}$,$\lambda_{123}$&$\lambda_{123}=\tfrac{2\sqrt{2}i}{\widetilde{X}_{1}\sqrt{5}}\sqrt{m_{10}m_{120}}$&$H^{u}_{1}\propto H_u$\\
	&&&&&$H^{d}_{1}\propto -H_d$\\
	&&&&&$H^{u,d}_{3}\propto -H_{u,d}$\\
	&&&&&$H^{u,d}_{4}\propto \sqrt{3} H_{u,d}$\\
	\midrule
	$6$&$\mathbf{10}$,$\mathbf{120}$&$\langle\mathbf{45}\rangle=\widetilde{X}_{2}$&$m_{10}$,$m_{120}$,$\lambda_{123}$&$\lambda_{123}=\tfrac{2i}{\widetilde{X}_{2}\sqrt{5}}\sqrt{m_{10}m_{120}}$&$H^{u}_{1}\propto H_u$\\
	&&&&&$H^{d}_{1}\propto -H_d$\\
	&&&&&$H^{u,d}_{3}\propto\sqrt{3} H_{u,d}$\\
	&&&&&$H^{u,d}_{4}\propto H_{u,d}$\\
	\bottomrule
	\end{tabular}
\end{center}
\end{table}

We discuss now how DT splitting in general impacts the predictivity of the Yukawa operators in Eq.~\eqref{eq:invariant-explicit} for the different choices of $\mathbf{H}$. Using the tools of Section~\ref{sec:DT-tools}, the discussion culminates in a list of some simple predictive scenarios in Table~\ref{table:DT-models}. We make the following observations: 
\begin{itemize}
\item 

The coefficients $a_k$ and $b_l$ from step 5 are functions of the parameters present in $M_D$. An important goal is to find predictive scenarios, where the parameter values in the Higgs sector do not modify the Yukawa predictions.
We approach this goal by searching for scenarios, where the mass matrix $M_D$ is small and contains as few parameters as possible. This implies choosing the smallest possible number of operators and representations for the Higgs sector of the model. One parameter is eliminated by fine-tuning.
\item To achieve DT splitting, we need at least one $\mathrm{SO}(10)$ representation with a SM singlet VEV which is not an $\mathrm{SU}(5)$ VEV, so that $M_D\neq M_T$ due to the $\eta_i$ coefficients. This conclusion holds true even when using $\mathbf{126}$ and $\mathbf{\overline{126}}$ (when the model contains more triplets than doublets), since the $\mathrm{SU}(5)$ singlet VEVs do not contribute to any mixing of the extra triplet with other triplets, i.e.~there are no off-diagonal terms in the last row and column involving the $V$ VEVs in Eq.~\eqref{eq:MDT}.
\item We find that when $M_D$ is symmetric, the left null mode will be equal to the right null mode, thus implying $a_k=b_k$ for all $k$. These relations already reduce the freedom in the coefficients by half, e.g.~only $a_k$ remain to be determined. It is easy to obtain a symmetric $M_D$ by using only real representations, i.e.~in the scenarios $\mathbf{H}=\mathbf{10}$ or $\mathbf{H}=\mathbf{120}$.
\item A straightforward possibility to obtain a fully predictive Yukawa sector is for $a_k$ and $b_l$ to all be fixed numbers independent of the parameter values in the Higgs sector. Such scenarios are possible, e.g.~see cases $\# 1$-$3$ in Table~\ref{table:DT-models}. From the point of view of yielding a predictive MSSM Yukawa sector, however, it is sufficient that the ratios of $a_k$ and $b_l$ 
within each irreducible $\mathrm{SO}(10)$ representation $\mathbf{H}$ are fixed numbers. In such a case, the parameter-dependent common factor in all $a_k$ and $b_l$ in a given $\mathbf{H}$ can simply be absorbed into the Yukawa coupling coefficient in front of the operator. Examples include cases $\# 4$-$6$ in Table~\ref{table:DT-models}, further commented on below.
\item The Higgs representation $\mathbf{H}=\mathbf{10}$ has only one $D\overline{D}$ pair. A symmetric $M_D$ is sufficient to determine their VEV ratio and lead to a predictive Yukawa sector. This is implemented in case $\# 1$ of Table~\ref{table:DT-models}.
\item The Higgs representation $\mathbf{H}=\mathbf{120}$ has two $D\overline{D}$ pairs. A symmetric $M_D$ and a fixed ratio of the two relevant $a_i$s is thus sufficient to determine all their VEV ratios and give a predictive Yukawa operators involving $\mathbf{H}=\mathbf{120}$. A model containing only the doublets from $\mathbf{120}$ is implemented in cases $\# 2$ and $3$ of Table~\ref{table:DT-models}.
\item The implementation of a predictive case when $\mathbf{H}=\mathbf{\overline{126}}$ is a  bit more tricky, since it is a complex representation. It contains one $D\overline{D}$ pair, but $D$ and $\overline{D}$ are part of different $\mathrm{SU}(5)$ representations $\mathbf{5}$ and $\mathbf{\overline{45}}$, respectively. Furthermore, to balance the $D$-term vacuum equations in SUSY, the conjugate representation $\mathbf{126}$ typically needs to be introduced to the model. Nevertheless, a simple predictive scenario where $H_{2}^{u,d}=a H_{u,d}$ with the same number $a$ for both the $u$ and $d$ case was found, cf.~case $\# 4$ in Table~\ref{table:DT-models}, with the Higgs location written with a proportionality symbol $\propto$.
\item In cases $\# 5$ and $6$ in Table~\ref{table:DT-models}, we construct predictive models with $2$ available representations $\mathbf{H}$ for the operators. The proportionality symbols in the Higgs relations imply for case $\#2$ that $H^{u}_1=a H_u$, $H^{d}_1=-a H_d$, $H^{u,d}_3=-b H_{u,d}$ and $H_{4}^{u,d}=\sqrt{3}bH_{u,d}$ for some numbers $a$ and $b$. Notice that in each $\mathbf{H}$ the only ambiguity in the Higgses is the overall coefficient, i.e.~the $H^{u,d}_1$ in the $\mathbf{10}$ have the overall factor $a$ and the $H^{u,d}_{3,4}$ in the $\mathbf{120}$ have the overall factor $b$. This leads to a predictive scenario, since $a$ and $b$ can be absorbed into the Yukawa couplings operator by operator. Analogous relations hold for case $\# 6$ based on the Higgs relations in the table.
\end{itemize}

The list in Table~\ref{table:DT-models} contains merely some of the simplest options for predictive models. The reader can use the tools of Section~\ref{sec:DT-tools} to attempt their own constructions. The provided list, however, covers cases where $\mathbf{H}$ can be each of the representations $\mathbf{10}$, $\mathbf{120}$ and $\mathbf{\overline{126}}$; predictive scenarios thus include all possible $\mathbf{H}$ choices in the Yukawa operators of Eq.~\eqref{eq:invariant-general}. The cases when the flavor states $H^{u,d}_{i}$ are specified by the proportionality sign $\propto$  are those when an overall parameter-dependent factor for a given $\mathbf{H}$ representation can be absorbed into the Yukawa coefficient: for model building purposes, one can thus replace $\propto$ with the equal sign $=$ when interested in the predictions for the Yukawa sector at the level of the MSSM.

\section{Prescriptions for model building\label{sec:prescription-for-model-building}}

\subsection{General considerations\label{sec:prescription-for-model-building-general}}

We obtained explicit results for a single operator of the type as in Eq.~\eqref{eq:invariant-explicit} in Section~\ref{sec:operators}, and then analyzed the location of the MSSM Higgs fields in Section~\ref{sec:Higgs-location}.
We now provide a summary how to construct an entire model with these operators, based on single operator dominance in each Yukawa entry.  We include some additional model building considerations, for which we provide justification and technical details in Appendix~\ref{appendix:operators-mediators}. 

We provide below a step by step guide for the construction of a complete model. We divide the discussion into two parts: the choices that determine the model, and the subsequent consistency check of their validity:
\vskip 1.0cm
\begin{enumerate}
\item \underline{Choices for model building in the Yukawa sector:}
	\begin{itemize}
	\item Build the entire Yukawa sector\footnote{The Majorana neutrino masses need to be constructed separately.} by specifying which operators from Eq.~\eqref{eq:invariant-general}  are present for each entry in the Yukawa matrices. For each operator, one makes a choice of the Higgs representation $\mathbf{H}$: $\mathbf{H}\in\{\mathbf{10},\mathbf{120},\mathbf{\overline{126}}\}$. Also, for each operator one makes a choice of the directions the VEVs of each copy of the $\mathbf{45}$ and $\mathbf{210}$ take: 
	\begin{itemize}
	\item If all GUT VEVs have discrete directions aligned with the decompositions into maximal subgroups, the result in Eq.~\eqref{eq:result-discrete} applies. The required charges $q$ and normalizations $\mathcal{N}$ can be found in Tables~\ref{table:qnumbers-45} and \ref{table:qnumbers-210}, while the $\mathbf{H}$ dependent quantities are defined in Table~\ref{table:H-quantities}.
	\item If all $\mathbf{45}$ refer to the same copy, and the same holds for the $\mathbf{210}$, but both representations have an arbitrary GUT VEV direction, then the result in Eq.~\eqref{eq:result-continuous} applies. The definitions of the polynomials $P$ and $R$ can be found in Eq.~\eqref{eq:polynomial-p} and \eqref{eq:polynomial-r}.
	\item If one mixes multiple discrete and continuous directions, Eq.~\eqref{eq:result-continuous} can be extended by replacing the powers of polynomials $P$ and $R$ with a product, where each polynomial has its own direction, i.e.~its own value of the $\kappa$ variable(s). Discrete directions in any given polynomial can be obtained by applying the values of the $\kappa$ variables from Table~\ref{table:ratios-alignments}. 
	\end{itemize}	
	In a $3$ family model, the Yukawa operators are $3\times 3$ matrices. Operators connecting $\mathbf{16}_{I}$ and $\mathbf{16}_{J}$ for $I\neq J$ generate both the $I$-$J$ and $J$-$I$ off-diagonal Yukawa entries. Single operator dominance in every Yukawa entry then puts an upper limit of including at most $6$ such operators. In $\mathrm{SO}(10)$, that specifies the Yukawa couplings in all fermion sectors, including the Dirac type Yukawa for neutrinos.  More operators can of course be added if one relaxes the single operator dominance assumption.
	\item The Yukawa sector is fully determined once the locations of the MSSM Higgs doublets are specified, i.e.~the values $H^{u,d}_i$ in Table~\ref{table:H-quantities} are specified in terms of the MSSM Higgses $H_{u,d}$. These are in turn determined by the $a_k$ and $b_l$ coefficients of the left and right null eigenvector of $M_D$, see Eq.~\eqref{eq:Hu-Hd-content}. One can either remain agnostic about the Higgs sector and choose $a_k$ and $b_l$ as free parameters, or build a model giving a predictive choice with the tools from Section~\ref{sec:DT-tools}. The simplest choices for predictive cases are provided in Table~\ref{table:DT-models}.
	\end{itemize}
\item \underline{Consistency checks and considerations:}
	\begin{itemize}
	\item Knowing the set of all operators in the model, one knows all the possible $\mathbf{H}$ and all possible GUT VEV representations $\mathbf{45}$ and $\mathbf{210}$ that should be present. The first consistency check is performed if the Higgs sector is specified explicitly: all $\mathbf{H}$ used in the Yukawa operators should indeed be available based on the choice of the Higgs sector. If one remains agnostic about the Higgs sector, this consistency check can be skipped, but $a_k$ and $b_l$ should then be taken as free parameters.
	\item Another consideration is 	how to naturally achieve in each Yukawa entry only the presence of the desired operator(s), and forbid all others. These considerations can be made at two levels:
	\begin{itemize} 
	\item \underline{External legs:}
	\par
	It is possible to forbid Yukawa operators constructed from undesired combinations of fields by imposing global symmetries (which can be e.g.~discrete). This involves suitable charge assignments for the representations $\mathbf{16}_I$, $\mathbf{45}_{\alpha_i}$, $\mathbf{210}_{\beta_i}$ and $\mathbf{H}$ (specify the charges $q_{I}$, $x_i$, $y_j$ and $h$ from Appendix~\ref{appendix:operators-mediators}), which 
	allow the construction only of those Yukawa operators, whose charges of the fields (i.e.~external legs)  amount to a net zero charge. 	
	\item \underline{Internal contractions:} 
	\par
	Global symmetries impose constraints only on the external legs. Given the same representations in an invariant, there may still be multiple possible ways to internally contract the indices. One possible approach for allowing only a subset of operators with the same external legs is by use of mediators. In such an approach, one assumes the non-renormalizable $\mathrm{SO}(10)$ operators arise from integrating out heavy mediator fields (from a possibly renormalizable theory). Imposing the global symmetry onto the UV theory requires charge assignments for the mediators, thus allowing or forbidding certain types of contractions.
	\par
	The invariants formed from Eq.~\eqref{eq:invariant-explicit} all have contractions along spinor indices, which require mediator pairs in the representations $\mathbf{16}\oplus\overline{\mathbf{16}}$. Without any other types of mediators, all other types of contractions, i.e.~along (anti)fundamental indices, are absent. This restricts the Yukawa sector exactly to the type of operators considered in this paper.
	\par 
	The only remaining freedom in a spinorially contracted invariant is the order
	in which the representations $\mathbf{45}$	and $\mathbf{210}$ enter on either side of the Higgs representation $\mathbf{H}$. As discussed in Section~\ref{sec:operators},  permuting the order has an effect on the MSSM Yukawa prediction only if transferring a GUT VEV over $\mathbf{H}$. By a suitable choice of charges for the mediators and external legs, it is usually possible to allow only one specific permutation of the GUT VEVs, thus leading to single operator dominance. The details can be found in Appendix~\ref{appendix:operators-mediators}, we will only state here a brief summary. The required charge assignments for mediators are uniquely determined by the order of the external legs, cf.~Eq.~\eqref{eq:charge-alpha}-\eqref{eq:charge-beta}. At the level of the individual operator, any operator can be singled out by mediator charge assignments, except in diagonal Yukawa cases when only a single copy of either $\mathbf{45}$ or $\mathbf{210}$ is present, and the external legs are not placed as symmetrically as possible, since the more symmetric operators are generated with the same mediators as well. To be specific, for only a single copy of $\mathbf{45}$, $I=J$ and $m=m'=0$, operators with $|n-n'|>1$ cannot be singled out. Analogously, with a single copy of $\mathbf{210}$, the operators with $I=J$ and $n=n'=0$ cannot be singled out if $|m-m'|>1$.
\par
We emphasize that the mediator approach is used to
further restrict the Yukawa sector, i.e.~forbid certain contractions, so that the result at the MSSM level is single operator dominance. Whether the mediators postulated for the purpose of one Yukawa entry do not interfere with operators in other entries (and allow contractions, which we would like to forbid), is something that needs to be checked in a given model. The problem is thus reduced to finding charge assignments for external particles, and specific orders of GUT VEVs in the operators, such that only a unique invariant contraction is possible in any given Yukawa entry; the interested reader should again consult Appendix~\ref{appendix:operators-mediators}. This is a complication that needs to be considered only when multiple contractions with the given external legs are possible.
	\end{itemize}
\end{itemize}
\end{enumerate}

We briefly now turn to a discussion which model building choices yield the most predictive Yukawa sector, i.e.~as few free parameters as possible. It is clearly preferable to choose a predictive DT splitting scenario, e.g.~one of the choices from Table~\ref{table:DT-models}. Furthermore, the number of continuous parameters is further minimized by either using only discrete alignments of GUT VEVs, or as few continuous alignments as possible, essentially reducing ourselves to the two cases considered in Sections~\ref{sec:discrete-directions} and \ref{sec:arbitrary-directions}.

\subsection{Toy models \label{sec:toy-models}}

For illustrative purposes, we consider $3$ example models which exhibit the various aspects of model building that were just discussed. For all examples we consider a simplified setup with only two operators contributing to the Yukawa sector: $\mathcal{O}_{IJ}$ for $I=J=2$ and $I=J=3$, where $I$ and $J$ are family indices. We thus consider only the 2nd and 3rd family, and neglect the mixing between them, a good starting point given the hierarchical structure of the quark and charged lepton masses and mixings at low energies. The Yukawa part of the superpotential of our toy models is thus
\begin{align}
W_{\text{Yuk}}&=\lambda_2\,\mathcal{O}_{22}+\lambda_3\,\mathcal{O}_{33},\label{eq:two-operators}
\end{align}
setting the following form for the Yukawa matrices at the GUT scale:
\begin{align}
	\mathbf{Y}_u&=\begin{pmatrix}
	0&0&0\\
	0&y_c&0\\
	0&0&y_t\\
	\end{pmatrix}&
	\mathbf{Y}_d&=\begin{pmatrix}
	0&0&0\\
	0&y_s&0\\
	0&0&y_b\\
	\end{pmatrix},&
	\mathbf{Y}_e&=\begin{pmatrix}
	0&0&0\\
	0&y_\mu&0\\
	0&0&y_\tau\\
	\end{pmatrix},&
	\mathbf{Y}_\nu&=\begin{pmatrix}
	0&0&0\\
	0&y_{\nu_\mu}&0\\
	0&0&y_{\nu_\tau}\\
	\end{pmatrix}.
	\label{eq:Yukawa-matrices-setup}
    \end{align}

For the 3rd family, the top Yukawa coupling $y_t$ should be $\mathcal{O}(1)$, which is simplest to achieve by taking $\mathcal{O}_{33}$ to be a renormalizable operator not suppressed by powers of $X/\Lambda$, where $\Lambda$ is the cutoff scale for the effective $\SO(10)$ theory we consider and $X$ a GUT-scale VEV. As a common feature for all toy models, we thus choose the $\mathbf{H}=\mathbf{10}$ and $m_1=m_2=n_1=n_2=0$ in Eq.~\eqref{eq:invariant-general}, i.e.
\begin{align}
\mathcal{O}_{33}=\mathbf{16}_{F3}\cdot \mathbf{16}_{F3}\cdot \mathbf{10}.\label{eq:models-33}
\end{align}
The models will thus differ only in the choice of the $\mathcal{O}_{22}$ operator (the representations and VEV directions), as well as the choice of the Higgs location.

Since the coefficients $\lambda_{2}$ and $\lambda_{3}$ in Eq.~\eqref{eq:two-operators} can be adjusted to set the overall scale of the 2nd and 3rd family Yukawa couplings, and any phases in the Yukawa couplings absorbed into the appropriate fermion fields, the physically relevant predictions of our models are only the absolute values of Yukawa ratios:
\begin{align}
\text{Directly observable:}\quad 
\bigg|\frac{y_t}{y_b}\bigg|,
\bigg|\frac{y_c}{y_s}\bigg|, 
\bigg|\frac{y_\tau}{y_b}\bigg|, 
\bigg|\frac{y_\mu}{y_s}\bigg|.\qquad
\text{Also predicted:}\quad
\bigg|\frac{y_{\nu_\tau}}{y_b}\bigg|, 
\bigg|\frac{y_{\nu_\mu}}{y_s}\bigg|.\label{eq:model-predictions}
\end{align}
Since the neutrino sector also involves the unspecified Majorana mass matrix, the 
physical observables we consider in any fit exclude the two ratios involving neutrinos. Since these ratios are still predicted by the model, we will nevertheless specify them as predictions. 

We connect the model predictions of the Yukawa entries at the GUT scale to their experimental values at low scales by making use of data tables provided in \cite{Antusch:2013jca}. These tables provide GUT-scale values obtained by running the RG equations for the Yukawa couplings and mixings from their measured values at the scale $M_{Z}$ all the way to the GUT scale fixed at $M_{GUT}=2\cdot 10^{16}\,\mathrm{GeV}$. In between, a transition from the SM running to the MSSM running (and the $\overline{\text{MS}}$ renormalization scheme to the $\overline{\text{DS}}$ scheme) takes place at $M_{SUSY}=\mathrm{3}\,\mathrm{TeV}$, where the effects of the unknown SUSY spectrum are parametrized using threshold effect parameters $\eta_q$ and $\eta_b$. They are defined by performing the SM to MSSM matching via

\begin{align}
\begin{split}
\mathbf{Y}_u^{\text{MSSM}}&=\mathbf{Y}_u^{\text{SM}}\,\frac{1}{\sin\beta},\\
\mathbf{Y}_d^{\text{MSSM}}&=\mathrm{diag}\left(\tfrac{1}{1+\eta_q},\tfrac{1}{1+\eta_q},\tfrac{1}{1+\eta_b}\right)\,\mathbf{Y}_d^{\text{SM}}\,\frac{1}{\cos\beta},\\
\mathbf{Y}_e^{\text{MSSM}}&=\mathbf{Y}_e^{\text{SM}}\,\frac{1}{\cos\beta}.\\
\end{split}
\end{align}

These matching conditions include only $\tan\beta$ enhanced contributions and are a good approximation if $\tan\beta\gtrsim 5$. Furthermore, they hold for the Yukawa matrices written in the left-right (LR) convention (i.e.~the first family index of the Yukawa matrix refers to a left-handed field $Q$ or $L$) and for a basis where $\mathbf{Y}_u^{\text{SM}}$ and $\mathbf{Y}_e^{\text{SM}}$ are diagonal at the matching scale, while $\mathbf{Y}_d^{\text{SM}}$ contains only left mixing angle rotations, i.e. at the matching scale the SM Yukawa matrices should take the form
\begin{align}
\begin{split}
\mathbf{Y}_u^{\text{SM}}&=\mathrm{diag}(y_u,y_c,y_t),\\
\mathbf{Y}_d^{\text{SM}}&=\mathbf{V}_{\text{CKM}}^\dagger\,\mathrm{diag}(y_d,y_s,y_b),\\
\mathbf{Y}_{e}^{\text{SM}}&=\mathrm{diag}(y_e,y_\mu,y_\tau).\\
\end{split}
\end{align}
In our models the Yukawa matrices already take this canonical form due to the $2$ operator setup of Eq.~\eqref{eq:Yukawa-matrices-setup}.
Furthermore, we make a simplification by neglecting the threshold effects in the charged lepton sector\footnote{The threshold effect for the third family in charged leptons can be absorbed into $\beta$, then simply rewriting the parameters $\beta\mapsto \bar{\beta}$, $\eta_q\mapsto \bar{\eta}_q$ and $\eta_b\mapsto \bar{\eta}_b$, see \cite{Antusch:2013jca} for details. What we thus really neglect is the threshold effects in the 1st and 2nd family of charged leptons.}. 

The free parameters determining the Yukawa ratios of Eq.~\eqref{eq:model-predictions} of a model thus consists of the following:
\begin{align}
\text{Model free parameters:}\qquad \tan\beta,\ \eta_b,\ \eta_q,\ \text{others (model specific)},\label{eq:model-parameters}
\end{align}
where the model specific parameters may include parameters determining the direction of GUT-scale VEVs, or coefficients determining the presence of the MSSM Higgs doublets in the doublet flavor-eigenstates.

We emphasize that simple parameter and observable counting based on Eq.~\eqref{eq:model-predictions} and \eqref{eq:model-parameters} is not really meaningful, since a determination of the SUSY threshold correction parameters $\eta_b$ and $\eta_q$ puts constraints on the spectrum of the SUSY sparticles in a non-trivial way. A model thus predicts more observables than one might naively assume, 
but we shall not pursue the determination of the SUSY spectrum further in this paper, see e.g.~\cite{Antusch:2015nwi,Antusch:2017ano,Antusch:2019gmc} for examples.

We now present our example models.

\subsubsection{Example 1: discrete VEV directions and predictive DT\label{sec:model1}}
The first example consists of a predictive DT scenario and discrete VEV directions.

In particular, since $\mathcal{O}_{33}$ involves $\mathbf{H}=\mathbf{10}$, the simplest choice for a predictive DT scenario is case $\# 1$ from Table~\eqref{table:DT-models}, implying $H^{u,d}_1=H_{u,d}$.
 This yields a $t$-$b$-$\tau$ unification prediction
for the Yukawa couplings based on Eq.~\eqref{eq:Yukawa-ratios-discrete} and Table~\ref{table:DT-models-factors}:
\begin{align}
\bigg|\frac{y_t}{y_b}\bigg|&=\bigg|\frac{y_\tau}{y_b}\bigg|=\bigg|\frac{y_{\nu_{\tau}}}{y_b}\bigg|=1.\label{eq:model1-prediction}
\end{align}
It is known that such a scenario requires a $\tan\beta\approx 50$.

\begin{figure}[htb]            
            \begin{center}
            \includegraphics[width=0.8\textwidth]{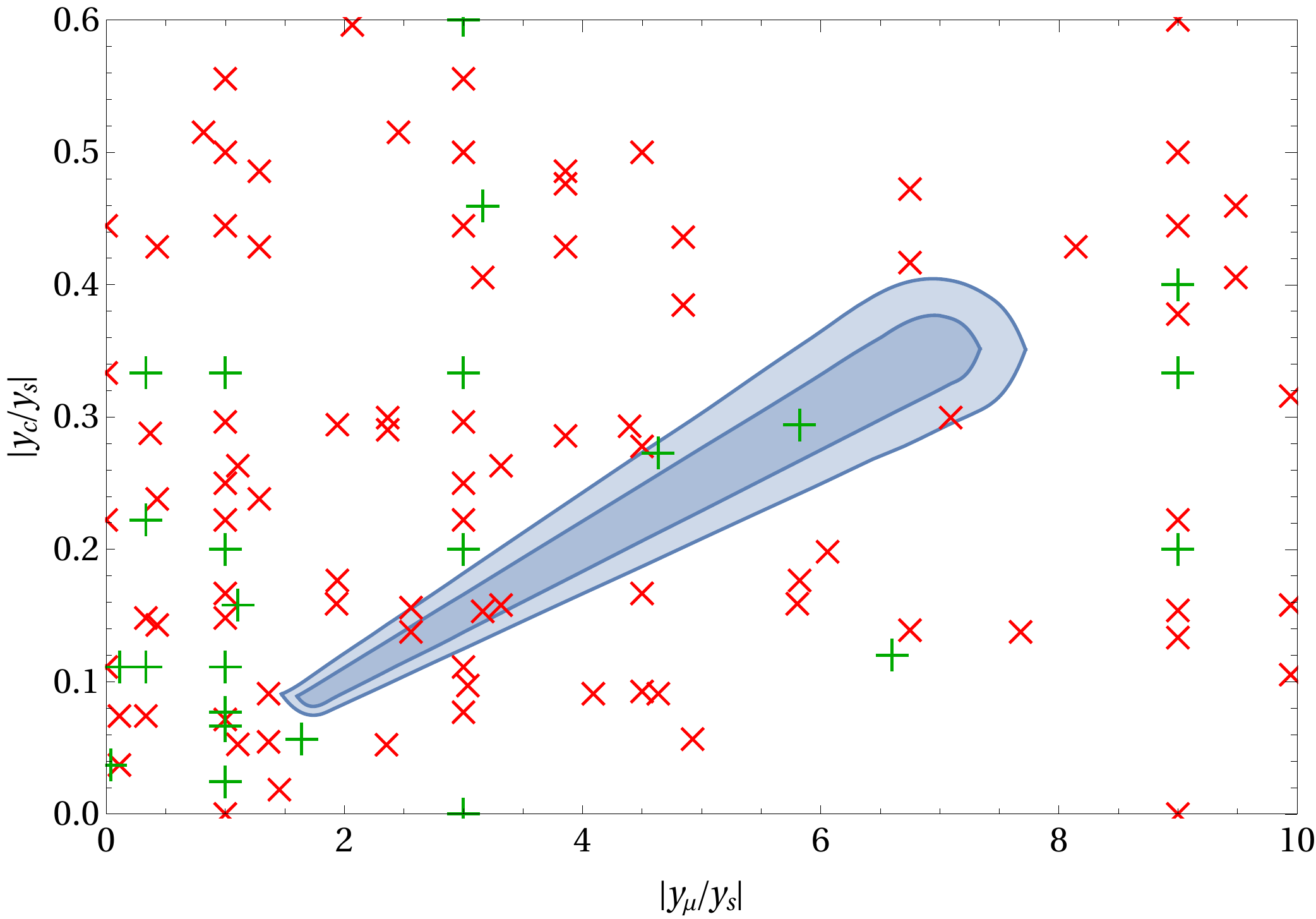}
            \par
            \caption{The plane of Yukawa ratios $|y_\mu/y_s|$ and $|y_c/y_s|$ at the GUT scale assuming Yukawa unification for the $3$rd family. The darker blue area indicates the region where the fit of the $4$ directly observable ratios has $\chi^2<1$, and the lighter blue area where $\chi^2<4$. Furthermore, the crosses indicate predicted ratios by the model in Example 1 (see main body text): the green ones correspond to a $\mathcal{O}_{22}$ operator with only the representations $\mathbf{45}$ and no $\mathbf{210}$, while the red crosses correspond to operators with at least one $\mathbf{210}$ factor. \label{fig:model1-ratios}}
            \end{center}
\end{figure}

For the $\mathcal{O}_{22}$ operator, we consider any non-renormalizable operator in Eq.~\eqref{eq:invariant-general} with $\mathbf{H}=\mathbf{10}$ of dimension at most $7$ in the superpotential and with discrete GUT VEV direction, i.e.~we demand $n+n'+m+m'\leq 4$ and the VEV direction for each of the $\mathbf{45}$ or $\mathbf{210}$ factors is independently chosen to be one of those from Eq.~\eqref{eq:VEVs-SU5} or \eqref{eq:VEVs-ps}. In other words, $\alpha_i, \beta_j, \alpha'_k, \beta'_l$ are chosen independently: 
\begin{align}
\mathcal{O}_{22}:\quad \mathbf{H}=\mathbf{10};\ n+n'+m+m'\leq 4;\ \alpha_i,\beta_j,\alpha'_k,\beta'_l\  \text{discrete directions}.
\end{align}
The results for the predicted 2nd family Yukawa ratios can be computed using Eq.~\eqref{eq:Yukawa-ratios-discrete}.
The large set of operators under consideration thus includes also all the cases of Table~\ref{table:yukawa_ratios} (or equivalently Figure~\ref{fig:model1-ratios}).

\begin{table}[htb]
\begin{center}
\caption{Operators with up to $4$ GUT-scale VEVs in discrete directions, which can be used for the $\mathcal{O}_{22}$ operator to provide a good fit $\chi^2<1$
in Figure~\ref{fig:model1-ratios}.\label{tab:ratios-for-22}}
{\begin{small}
\renewcommand{\arraystretch}{1.25}
\begin{tabular}{rrllll}
\toprule
Nr & $\big(\frac{(\mathbf{Y}_e)_{II}}{(\mathbf{Y}_d)_{II}},\frac{(\mathbf{Y}_u)_{II}}{(\mathbf{Y}_d)_{II}},\frac{(\mathbf{Y}_
u)_{II}}{(\mathbf{Y}_d)_{II}}\big)$ & \multicolumn{4}{l}{$\big(\{\mathbf{45}_{\alpha_i},\mathbf{210}_{\beta_j}\}|\{\mathbf{45}_{\alpha^\prime_k},\mathbf{210}_{\beta^\prime_l}\}\big)$} \\
\midrule
$1$ & $(-\frac{279}{109},-\frac{15}{109},-\frac{243}{109})$ & $(X_1,Z_2,Z_2,\tilde{Z}_1|\cdot)$\,,\hspace{0.3cm}$(X_1,Z_2,Z_2|\tilde{Z}_1)$ \\
$2$ & $(\frac{117}{37},-\frac{17}{111},\frac{405}{37})$ & $(X_1,Z_2,Z_2|X_1)$ \\
$3$ & $(\frac{63}{19},-\frac{3}{19},\frac{81}{19})$ & $(X_1,Z_2,\tilde{Z}_2|\cdot)$\,,\hspace{0.3cm}$(X_1,Z_2|\tilde{Z}_2)$\,,\hspace{0.3cm}$(X_1,X_1,X_2,\tilde{Z}_2|\cdot)$\,,\hspace{0.3cm}$(X_1,\tilde{X}_2,Z_2,\tilde{Z}_1|\cdot)$\,,\vspace{-0.1cm} \\
& & $(X_1,X_1,X_2|\tilde{Z}_2)$\,,\hspace{0.3cm}$(X_1,\tilde{X}_2,Z_2|\tilde{Z}_1)$\,,\hspace{0.3cm}$(X_1,Z_2,\tilde{Z}_1|\tilde{X}_2)$\,,\hspace{0.3cm}$(X_1,Z_2|\tilde{X}_2,\tilde{Z}_1)$ \\
$4$ & $(-\frac{99}{17},-\frac{5}{17},-\frac{81}{17})$ & $(X_1,\tilde{X}_2,Z_2|\cdot)$\,,\hspace{0.3cm}$(X_1,Z_2|\tilde{X}_2)$\,,\hspace{0.3cm}$(X_1,X_1,X_2,\tilde{X}_2|\cdot)$\,,\hspace{0.3cm}$(X_1,Z_2,\tilde{Z}_1,\tilde{Z}_2|\cdot)$\,,\vspace{-0.1cm} \\
& & $(X_1,Z_2,\tilde{Z}_1|\tilde{Z}_2)$\,,\hspace{0.3cm}$(X_1,Z_2,\tilde{Z}_2|\tilde{Z}_1)$\,,\hspace{0.3cm}$(X_1,Z_2|\tilde{Z}_1,\tilde{Z}_2)$\,,\hspace{0.3cm}$(\tilde{X}_2|X_1,X_1,X_2)$ \\
\bottomrule
\end{tabular}
\renewcommand{\arraystretch}{1.0}
\end{small}}
\end{center}
\end{table}

Since the location of the Higgses is predictive  and the VEV directions are discrete, there are no new model specific parameters in Eq.~\eqref{eq:model-parameters}, while the direct observables are the Yukawa ratios from Eq.~\eqref{eq:model-predictions}. We allow the free parameters to vary in the range
\begin{align}
\begin{split}
20\leq\tan\beta\leq70,\\
-0.6\leq\eta_b,\eta_q\leq0.6,\\
\end{split}
\end{align}
and compute for each parameter point a $\chi^2$ value based on the $4$ directly observable Yukawa ratios. The values for the ratios as well as the standard deviations\footnote{The tables in \cite{Antusch:2013jca} provide only the values of the Yukawa couplings and their errors. The relative errors for the ratio $x/y$ is computed by taking $\sqrt{\delta_{x}^2+\delta_{y}^2}$, where $\delta_{x}$ and $\delta_{y}$ are the relative errors of the quantities $x$ and $y$, respectively.} are computed from interpolating the data tables provided by \cite{Antusch:2013jca}. The parameters $\eta_b$ and $\tan\beta$ have to be chosen, such that the data gives the $|y_t/y_b|$ and $|y_\tau/y_b|$ close to $1$ according to the model prediction from Eq.~\eqref{eq:model1-prediction}. The prediction of the 2nd family ratios $|y_c/y_s|$ and
$|y_\mu/y_s|$ then still has the freedom of $\eta_q$.

The results of the fit of the data projected onto the two 2nd family ratios, as well as the possible predictions from the constructed operators, are shown in Figure~\ref{fig:model1-ratios}. The darker and lighter blue regions correspond to the values of ratios consistent with a fit giving $\chi^2<1$ and $\chi^2<4$, respectively. The green $+$ symbols correspond to model predictions from $\mathcal{O}_{22}$ operators containing only the representations $\mathbf{45}$ ($m=m'=0$), while the more numerous red crosses correspond to cases where at least one $\mathbf{210}$ in some VEV direction is also involved, i.e.~$m+m'>0$. The predicted Yukawa ratios from different operators may be equal, so a single symbol may represent identical predictions of more than one operator. Furthermore, preference was given to operators with no representations $\mathbf{210}$: if both a red cross and a green plus would need to be drawn in the same location in the figure, only the latter is shown. 

We can see that ratio-pairs of only a few operators fall into the low $\chi^2$ regions in Figure~\ref{fig:model1-ratios}. The operators falling into the best-fit region with $\chi^2<1$ are listed in Table~\ref{tab:ratios-for-22}. These operators are the most promising candidates for the 2nd family operator in further model building. They all contain either $3$ or $4$ GUT-scale VEVs, and include none of the candidates with $2$ GUT-scale VEVs or less from Table~\ref{table:yukawa_ratios}. We emphasize that all this holds only with the assumption of 3rd family Yukawa unification and case $\# 1$ for predictive DT splitting.

As an alternative model building approach, we can relax in the next examples either the discrete VEV assumption or the assumption of having a (most) predictive case of DT.

\subsubsection{Example 2: arbitrary VEV direction and predictive DT\label{sec:model2}}

We modify now Example 1 to an arbitrary VEV direction approach from Section~\ref{sec:arbitrary-directions}.

In particular, we again choose case $\# 1$ of Table~\ref{table:DT-models} for the predictive Higgs location scenario. The renormalizable operator $\mathcal{O}_{33}$ from Eq.~\eqref{eq:models-33} again yields the $t$-$b$-$\tau$ unification prediction of Eq.~\eqref{eq:model1-prediction}. For the $\mathcal{O}_{22}$
operator, we choose $\mathbf{H}=\mathbf{10}$ consistent with our DT splitting scenario, while choosing only a single copy of $\mathbf{45}$ to carry the VEV: $m=m'=0$ and $\alpha_i=\alpha'_k$ for all $i$ and $k$ in Eq.~\eqref{eq:invariant-general}. This choice is in accordance with Section~\ref{sec:arbitrary-directions} and minimizes the added number of model specific parameters: we only need to add the VEV ratio $\kappa$
in the single copy of the $\mathbf{45}$ as a free parameter. The choice of $\mathbf{210}$ would instead add $2$ new parameters, and we will not consider it here.

We wish to construct a model with single operator dominance, and since we are dealing with $I=J=2$ and only one type of GUT VEV, we are restricted to taking
operators with $|n-n'|\leq 1$, as was discussed in Section~\ref{sec:prescription-for-model-building-general} (and is derived in detail in Appendix~\ref{appendix:operators-mediators}). We shall consider only operators of dimension at most $8$ in the superpotenial, i.e.~$n+n'\leq 5$. The considered choices for the 2nd family operator are thus
\begin{align}
\mathcal{O}_{22}:\quad \mathbf{H}=\mathbf{10};\ m=m'=0;\ n+n'\leq 5;\ |n-n'|\leq 1;\ \alpha_i=\alpha'_k\ \forall i,k.
\end{align}

This leads to the possibilities of $n$ and $n'$ in Table~\ref{table:models-2} (referred to as ``models''), where $n\leq n'$ can be assumed without loss of generality. The 2nd family Yukawas ratios $|y_\mu/y_s|$, $|y_c/y_s|$ and $|y_{\nu_{\mu}}/y_s|$ are computed with the tools established in Section~\ref{sec:arbitrary-directions}, in particular by the use of Eq.~\eqref{eq:result-continuous}. Since only the representation $\mathbf{45}$ is involved, only the $P$-type polynomials are relevant in that equation. Case $\# 1$ for the DT scenario implies $H^{u,d}_{1}=H_{u,d}$. The Yukawa ratios are functions of the VEV ratio $\kappa=X_2/X_1$ in the $\mathbf{45}$. 

\begin{table}[htb]
\begin{center}
\caption{Models with different choices of $n$ and $n'$ in $\mathcal{O}_{22}$ and their predictions for Yukawa ratios.\label{table:models-2}}
\par
\begin{tabular}{lrrl@{\hbox to 1cm{}}rrr}
\toprule
model&$n_1$&$n_2$&&$|y_\mu/y_s|$&$|y_c/y_s|$&$|y_{\nu_\mu} / y_s|$\\
\midrule
model 0&$0$&$1$&&$1$&$1$&$1$\\\addlinespace
model 1a&$1$&$1$&\multirow{2}{*}{$\Big\}$ model 1}&
	\multirow{2}{*}{$\Big|\frac{9 \left(4 \sqrt{3} \kappa +\sqrt{2}\right)}{9 \sqrt{2}-4 \sqrt{3} \kappa }\Big|$}&
	\multirow{2}{*}{$\Big| \frac{3 \sqrt{2}-8 \sqrt{3} \kappa }{9 \sqrt{2}-4 \sqrt{3} \kappa } \Big|$}&
	\multirow{2}{*}{$\Big|\frac{45}{9-2\sqrt{6}\kappa}\Big|$}\\
model 1b&$1$&$2$&&&&\\\addlinespace
model 2a&$2$&$2$&\multirow{2}{*}{$\Big\}$ model 2}&
	\multirow{2}{*}{$\Big| \frac{81 \left(48 \kappa ^2+8 \sqrt{6} \kappa +2\right)}{\left(9 \sqrt{2}-4 \sqrt{3} \kappa \right)^2} \Big|$}&
	\multirow{2}{*}{$\Big| \frac{6 \left(32 \kappa ^2-8 \sqrt{6} \kappa +3\right)}{\left(9 \sqrt{2}-4 \sqrt{3} \kappa \right)^2} \Big|$}&
	\multirow{2}{*}{$\Big|\frac{1350}{(3\sqrt{6}-4\kappa)^2}\Big|$}\\
model 2b&$2$&$3$&&&&\\
\bottomrule
\end{tabular}
\end{center}
\end{table}

The obtained results in Table~\ref{table:models-2} indicate that the modulus of the Yukawa ratios for models 1a and 1b are the same, so we refer collectively to both models as ``model 1''. Similarly, the predictions from models 2a and 2b are the same, so we refer to them collectively as model 2. Model $0$ leads to $c$-$s$-$\mu$ unification at the GUT scale and is not a viable starting point for model building: considering just the $s$ to $\mu$ ratio at the GUT scale, it needs to be between $3$ and $6$ based on typical SUSY threshold corrections, see e.g.~\cite{Antusch:2013jca}. We are thus left with models $1$ and $2$ as viable candidates in this simplest setup.

A final model building consideration is whether one can allow only the two operators $\mathcal{O}_{33}$ and $\mathcal{O}_{22}$, and no others. We use the approach from Appendix~\ref{appendix:operators-mediators}, and assign $\mathrm{U}(1)$ charges to the representations. We label the charges of $\mathbf{16}_2$, $\mathbf{16}_3$, $\mathbf{45}$ and $\mathbf{10}$ by $q_2$, $q_3$, $x$ and $h$, respectively. The net zero charge of $\mathcal{O}_{33}$ and $\mathcal{O}_{22}$ demands
\begin{align}
2q_3+h&=0,&2q_2+h+(n+n')x&=0.\label{eq:model2-eq1}
\end{align}
 Forbidding a $\mathcal{O}_{32}$ operator for any power $k\in\mathbb{N}_0$ of the $\mathbf{45}$ then implies
\begin{align}
q_3+q_2+h+kx=(q_3-q_2)(2k-n-n')/(n+n')&\neq 0,\label{eq:model2-eq2}
\end{align}
where we inserted the expression for $h$ by solving Eq.~\eqref{eq:model2-eq1}.
The non-vanishing of the total charge in Eq.~\eqref{eq:model2-eq2} 
then holds for any $k\in\mathbb{N}_0$ provided $q_3\neq q_2$ and $n+n'$ is odd. From this perspective, models 1b and 2b in Table~\ref{table:models-2}
have an odd $n+n'$, so we can forbid in them off diagonal Yukawa couplings by imposing for example charges $q_3=h=0$, $q_2=-1$, $x=1$, leading to the addition of mediators $\mathbf{16}$ with charges $\alpha_1=\beta_1=0$ (and its conjugate). Strictly speaking, it is thus models 1b and 2b that we consider, at least when using techniques from Appendix~\ref{appendix:operators-mediators} for imposing single operator dominance.

The directly observable Yukawa ratios are those in Eq.~\eqref{eq:model-predictions}, while the model parameters consist of the ones in Eq.~\eqref{eq:model-parameters} with the additional complex parameter $\kappa$.
It turns out that in both models $1$ and $2$ it is possible to fit the observables. We now show this in a sequence of considerations:

\begin{enumerate}
	\item Our models predict $y_t/y_b=1$ and $y_\tau/y_b=1$. We search for these ratio values in the $\tan\beta$-$\eta_b$ plane. Figure~\ref{fig:tanbeta-etab} shows the regions where the data tables from \cite{Antusch:2013jca} give $y_t/y_b$ (red) and $y_\tau/y_b$ (blue) to be around $1$, with the error range for these quantities also provided by the data tables. We see that the two regions overlap; inside is a point where the ratios are exactly one, which determines our default values of $\tan\beta$ and $\eta_b$:
	\begin{align}
	\tan\beta&=50.159,&\eta_{b}&=-0.1573.\label{eq:anayltic-default-tanbeta-etab}
	\end{align} 
	The large $\tan\beta$ is expected due to $t$-$b$-$\tau$ unification.
\begin{figure}[htb]            
            \begin{center}
            \includegraphics[width=8cm]{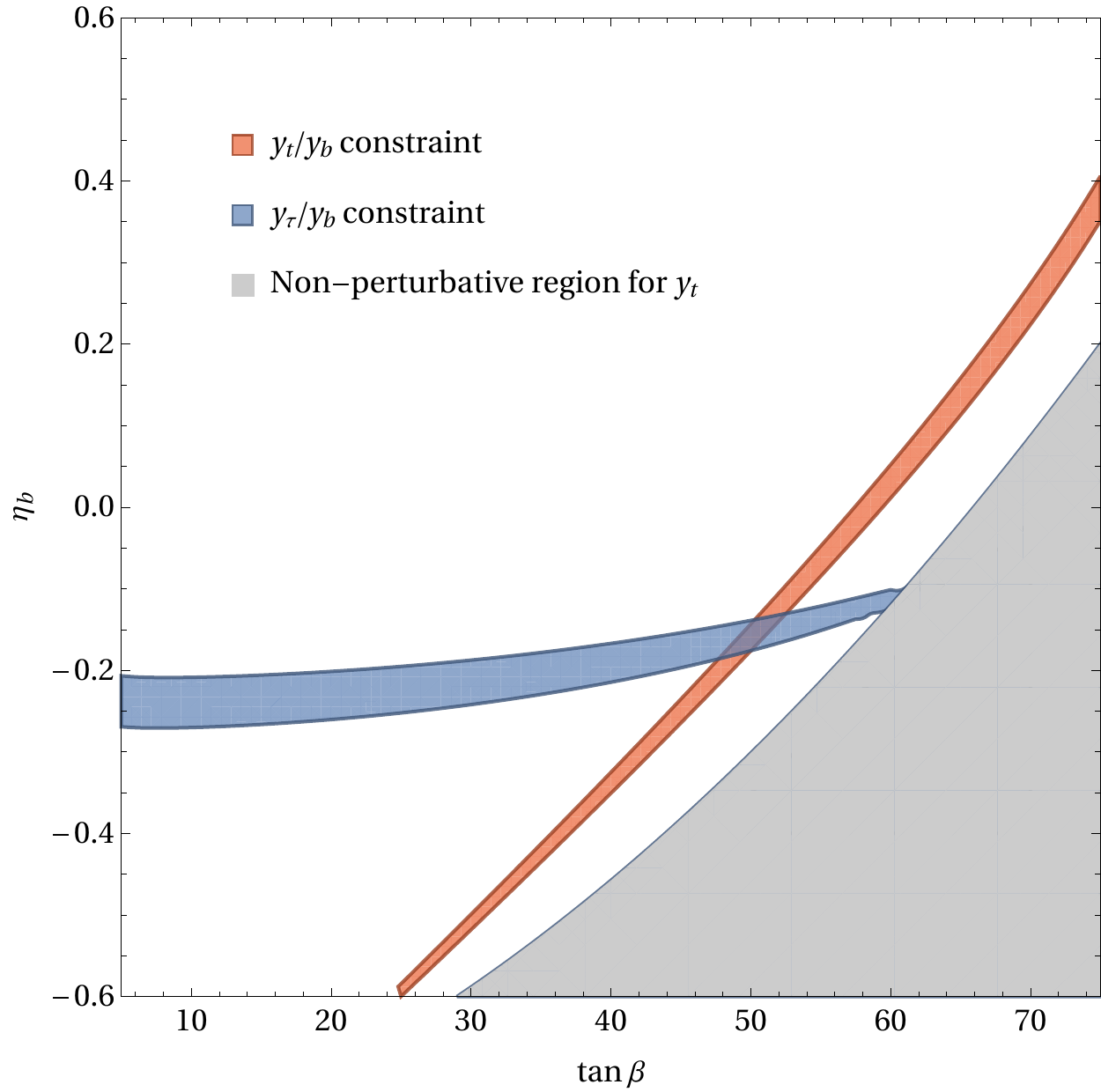}
            \par
            \caption{Allowed regions in the $\tan\beta$-$\eta_b$ plane consistent with Yukawa ratios of 3rd family being equal to $1$.\label{fig:tanbeta-etab}}
            \end{center}
\end{figure}
	\item Taking the values in Eq.~\eqref{eq:anayltic-default-tanbeta-etab}, the only remaining threshold parameter to be determined for the ratio $y_s/y_c$ is $\eta_q$. We plot this dependence in Figure~\ref{fig:etaq-values}. We read off the possible range for the ratio $y_s/y_c$ to be, for example, between $6.65$ and $3.58$, assuming the interval range $\eta_q\in(-0.3,0.3)$. 
\begin{figure}[htb]
            \begin{center}
            \includegraphics[width=8cm]{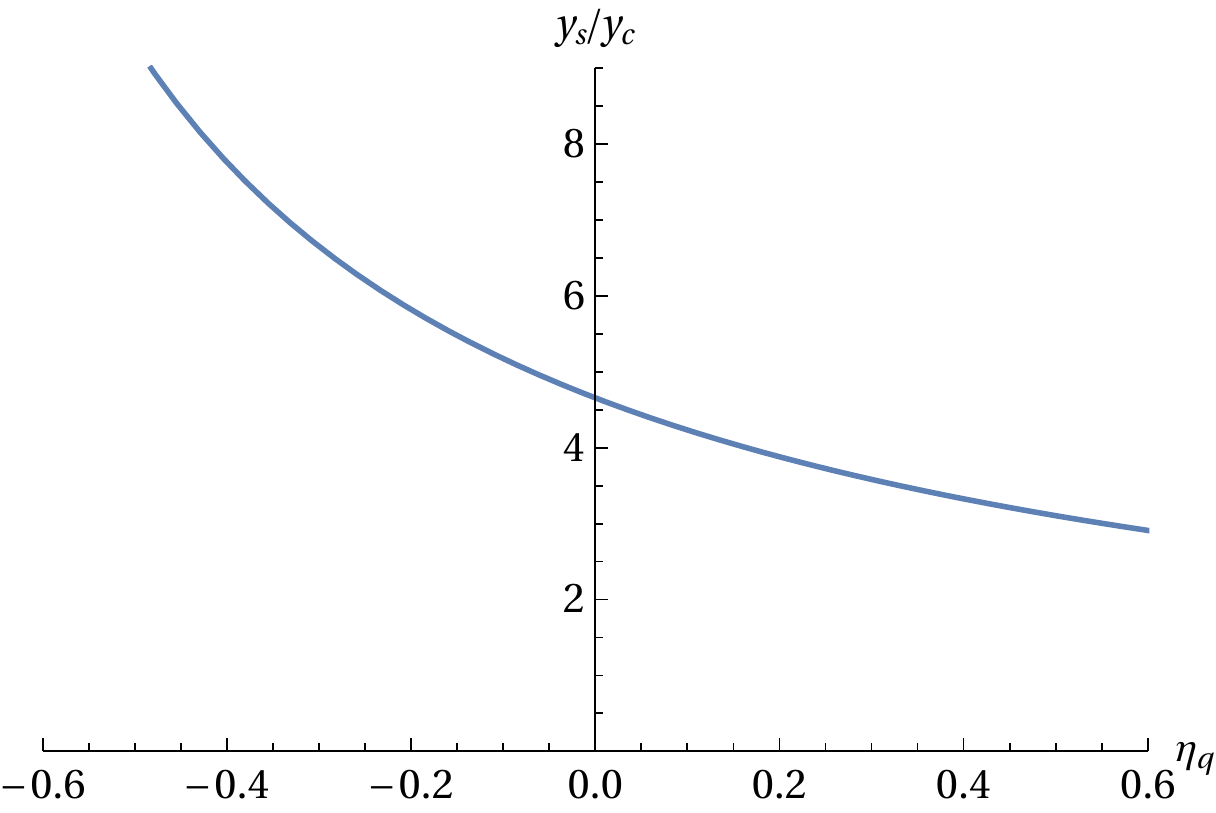}
            \par
            \caption{The values of the ratio $y_s/y_c$ for the second family as a function of $\eta_{q}$, assuming the default values $\tan\beta=50.159$ and $\eta_{b}=-0.1573$ consistent with Figure~\ref{fig:tanbeta-etab}.\label{fig:etaq-values}}
            \end{center}
\end{figure}      
	\item We plot regions in the complex $\kappa$-plane, where the ratios $y_s/y_c$ and $y_\mu/y_c$ give suitable values; the result is shown in Figure~\ref{fig:kappa-plane}. The models predicts the Yukawa ratios to be functions of $\kappa$, see Table~\ref{table:models-2}, which are then compared to the experiment-derived data tables of their GUT values in \cite{Antusch:2013jca}. Note that we are considering $y_\mu/y_c=(y_\mu/y_s)/(y_c/y_s)$ as one of the two ratios, since it does not depend on the threshold parameter $\eta_q$ (only $y_s$ depends on it). The allowable values for the $y_\mu/y_c$ ratio are computed from the data tables taking a combined relative error $\sqrt{(\delta_{y_\mu})^2+(\delta_{y_c})^2}$ at a fixed $\tan\beta$ and $\eta_b$ from Eq.~\eqref{eq:anayltic-default-tanbeta-etab}. The allowed values of $y_s/y_c$, on the other hand, are in the range specified by step 2. We can see that for both models 1 and 2 there exist overlap regions, where $\kappa$ values can fit well both 2nd family Yukawa ratios. Note that the pictures are symmetric with respect to the sign in the complex phase of $\kappa$, since replacing $\kappa$ with $\kappa^\ast$ does not change the absolute value of Yukawa ratios in Table~\ref{table:models-2}. 
\end{enumerate}
           
\begin{minipage}{\linewidth}
            \includegraphics[width=7.5cm]{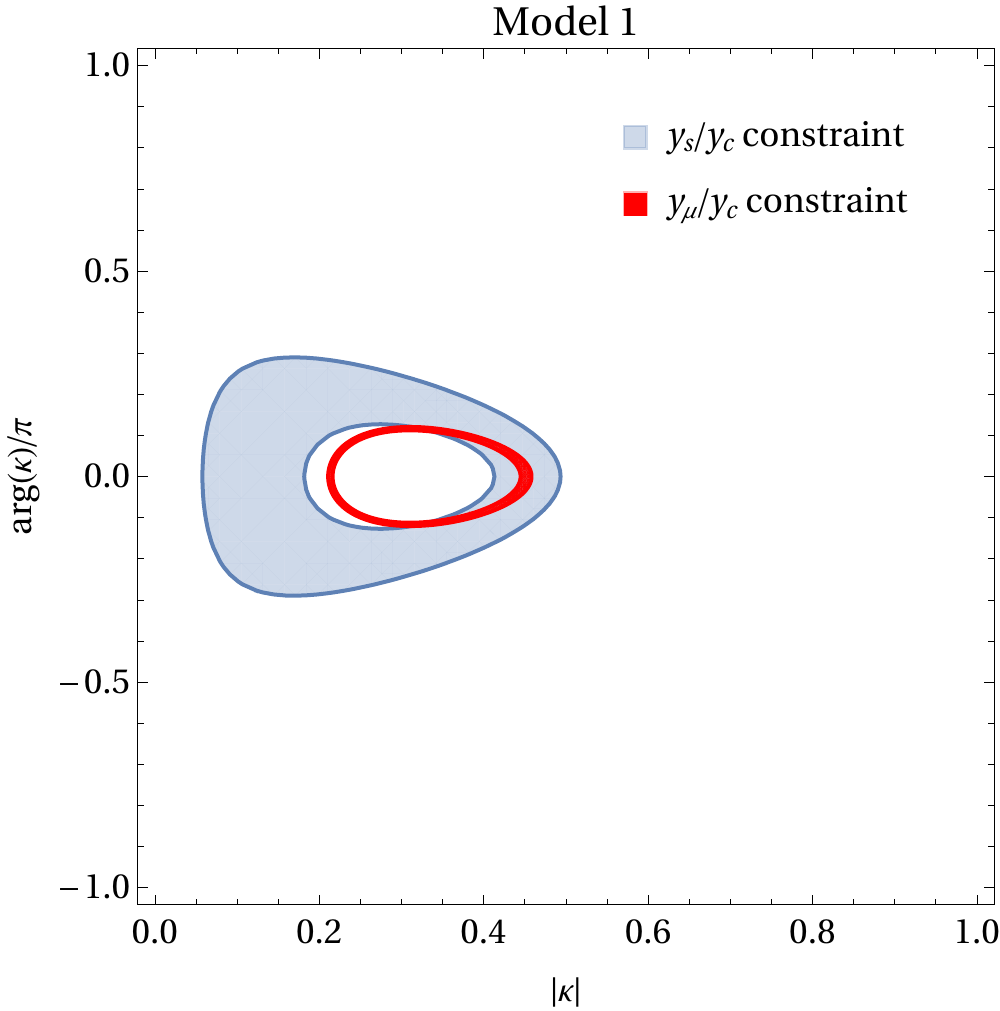}
            \hspace{0.5cm}
            \includegraphics[width=7.5cm]{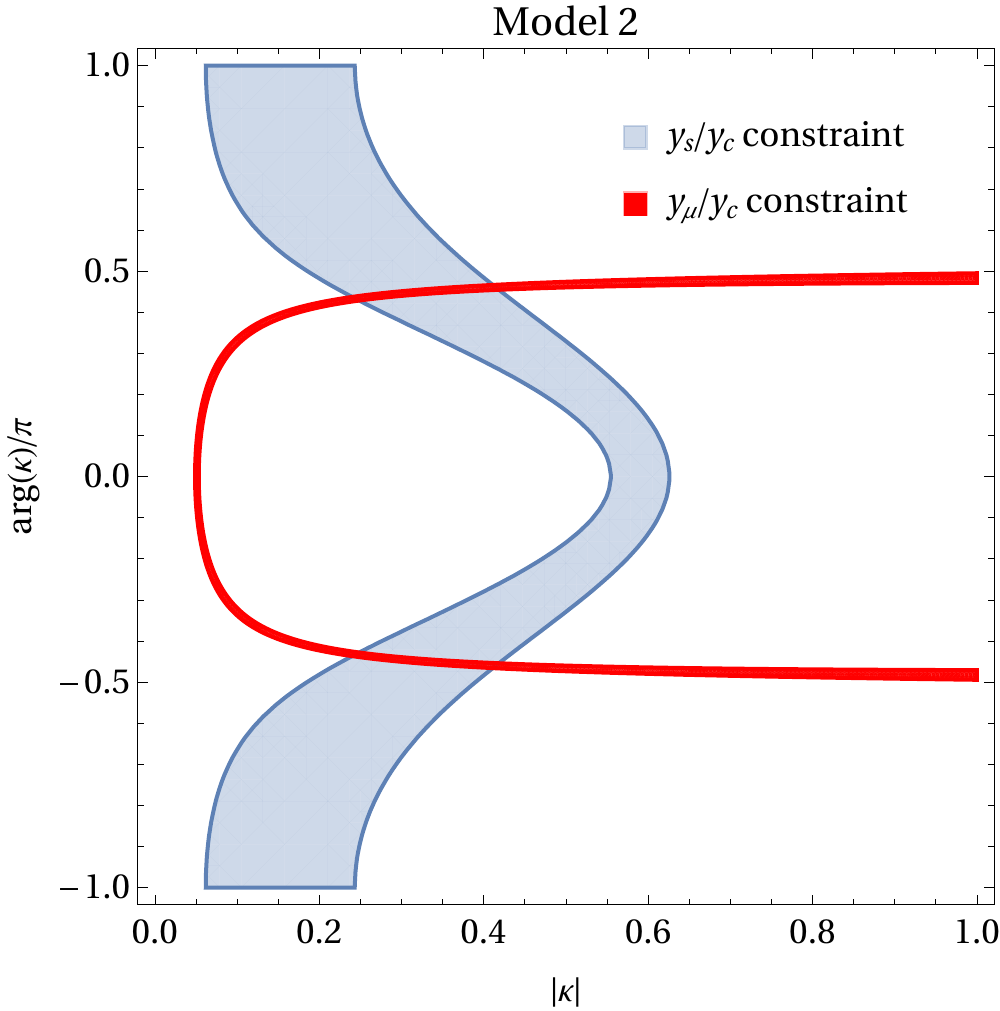}
            \captionof{figure}{Allowed regions of the 2nd family ratios $y_s/y_c$ (blue) and $y_\mu/y_c$ (red) in the modulus-phase plane for the complex $\kappa$ for model 1 (left) and model 2 (right) of Example 2. Notice that overlap regions exist in both models.\label{fig:kappa-plane}}
\end{minipage}

In this way we are able to fit all observables: points in the overlap region (in the $\tan\beta$-$\eta_b$ plane) in Figure~\ref{fig:tanbeta-etab} fit the 3rd family ratios, while points in the overlap region (in the $\kappa$ plane) in Figure \ref{fig:kappa-plane} fit the 2nd family ratios. 

\subsubsection{Example 3: discrete VEV directions and less-predictive DT\label{sec:model3}}
As a final example, we consider another possible modification of Example 1, in which the GUT VEV directions remain discrete, but we introduce additional freedom through the ambiguity in the location of the MSSM Higgses.

We could in principle remain agnostic about the doublet-triplet splitting  and introduce free parameters for the coefficients $a_i$ and $b_i$ as discussed in Section~\ref{section:DT-discussion}. It might be instructive, however, to explicitly construct an example of such a less-predictive DT splitting.

We consider the Higgs sector containing the doublets to consists for example of 
$\mathbf{10}\oplus\mathbf{126}\oplus\mathbf{\overline{126}}\oplus\mathbf{210}$,
where $\mathbf{H}=\mathbf{10}$ is used for $\mathcal{O}_{33}$, and $\mathbf{H}=\mathbf{\overline{126}}$ is used in $\mathcal{O}_{22}$. We keep the couplings $m_{10}$, $m_{126}$, $m_{210}$, $\lambda_{145}$, $\lambda_{14\bar{5}}$, $\lambda_{45\bar{5}}$, $\lambda_{444}$
in Eq.~\eqref{eq:DT-superpotential-onecopy}, such that we obtain the following doublet and triplet mass matrices: 

{\scriptsize
\begin{align}
M_D&=
\left(\begin{smallmatrix}
m_{10} & 0 & \sqrt{\frac{3}{5}} \lambda_{145} V_{210} & \lambda_{145} V_{126}
   \\
 0& m_{126}& 0 & 0 \\
 \sqrt{\frac{3}{5}} \lambda_{14\bar{5}} V_{210} & 0 & m_{126}+\frac{2}{\sqrt{15}} \lambda_{45\bar{5}} V_{210} & \lambda_{45\bar{5}} V_{126}
   \\
  \lambda_{14\bar{5}} V_{\overline{126}} & 0 & \lambda_{45\bar{5}} V_{\overline{126}}  & m_{210}+\lambda_{444}V_{210} \\
\end{smallmatrix}\right)\nonumber\\
&+\left(\begin{smallmatrix}
0 & \lambda_{14\bar{5}} \left(\frac{\sqrt{5}}{2\sqrt{3}} W_{210}+\frac{1}{\sqrt{3}} Z_{210}\right)  & \frac{1}{2} \sqrt{\frac{3}{5}} \lambda_{145} W_{210} & 0 \\
 \lambda_{145} \left(\frac{\sqrt{5}}{2\sqrt{3}} W_{210}+\frac{1}{\sqrt{3}} Z_{210}\right) & \lambda_{45\bar{5}} \left(\frac{\sqrt{5}}{3\sqrt{3}} W_{210}+\frac{2}{3\sqrt{3}} Z_{210}\right)  & 0 & 0 \\
 \frac{1}{2} \sqrt{\frac{3}{5}} \lambda_{14\bar{5}} W_{210}  & 0 & \frac{1}{\sqrt{15}} \lambda_{45\bar{5}} W_{210} & 0
   \\
 0 & 0 & 0 & \frac{1}{2} \lambda_{444}W_{210}\\
\end{smallmatrix}\right),\\
M_T&=\left(\begin{smallmatrix}
m_{10} & 0 & \sqrt{\frac{3}{5}} \lambda_{145} V_{210} & \lambda_{145}  V_{126} &
   0 \\
 0 & m_{126} & 0 & 0 & 0 \\
 \sqrt{\frac{3}{5}} \lambda_{14\bar{5}} V_{210} & 0 & m_{126}+\frac{2}{\sqrt{15}} \lambda_{45\bar{5}} V_{210} & V_{126} \lambda_{45\bar{5}} &
   0 \\
 \lambda_{14\bar{5}} V_{\overline{126}} & 0 & \lambda_{45\bar{5}} V_{\overline{126}} & m_{210}+ \lambda_{444} V_{210} & 0 \\
0 & 0 & 0 & 0 & m_{126}-\frac{1}{\sqrt{15}} \lambda_{45\bar{5}} V_{210}\\
\end{smallmatrix}\right)\nonumber\\
&+\left(\begin{smallmatrix}
0 & -\frac{1}{3} \lambda_{14\bar{5}} \left(\sqrt{5} W_{210}- Z_{210}\right) & 
-\frac{1}{\sqrt{15}} \lambda_{145} W_{210} & 0 &
   \sqrt{\frac{2}{3}} \lambda_{145} Z_{210} \\
 -\frac{1}{3} \lambda_{145} \left(\sqrt{5} W_{210}-Z_{210}\right) & 0 & 0 & 0 & 0 \\
 -\frac{1}{\sqrt{15}} \lambda_{14\bar{5}} W_{210} & 0 & -\frac{2}{3\sqrt{15}} \lambda_{45\bar{5}} W_{210} & 0 &
   -\frac{1}{3} \sqrt{\frac{2}{3}} \lambda_{45\bar{5}} Z_{210}  \\
 0 & 0 & 0 & -\frac{1}{3} \lambda_{444} W_{210} & 0 \\
 \sqrt{\frac{2}{3}} \lambda_{14\bar{5}} Z_{210} & 0 & -\frac{1}{3} \sqrt{\frac{2}{3}}  \lambda_{45\bar{5}} Z_{210} & 0 & \lambda_{45\bar{5}} 
 \left(\frac{1}{3 \sqrt{15}} W_{210}+\frac{2}{3\sqrt{3}}
   Z_{210}\right)  \\
\end{smallmatrix}\right).
\end{align}
}
They are written in the (sub)bases
\begin{align}
\begin{split}
D_{i}&=\begin{pmatrix}
D_{\mathbf{10}\supset\mathbf{5}}&
D_{\mathbf{126}\supset\mathbf{45}}&
D_{\mathbf{\overline{126}}\supset\mathbf{5}}&
D_{\mathbf{210}\supset\mathbf{5}}\\
\end{pmatrix},\\
\overline{D}_{i}&=\begin{pmatrix}
\overline{D}_{\mathbf{10}\supset\mathbf{\overline{5}}}&
\overline{D}_{\mathbf{\overline{126}}\supset\mathbf{\overline{45}}}&
\overline{D}_{\mathbf{126}\supset\mathbf{\overline{5}}}&
\overline{D}_{\mathbf{210}\supset\mathbf{\overline{5}}}\\
\end{pmatrix},\\
T_{i}&=\begin{pmatrix}
T_{\mathbf{10}\supset\mathbf{5}}&
T_{\mathbf{126}\supset\mathbf{45}}&
T_{\mathbf{\overline{126}}\supset\mathbf{5}}&
T_{\mathbf{210}\supset\mathbf{5}}&
T_{\mathbf{\overline{126}}\supset\mathbf{\mathbf{50}}}\\
\end{pmatrix},\\
\overline{T}_{i}&=\begin{pmatrix}
\overline{T}_{\mathbf{10}\supset\mathbf{\overline{5}}}&
\overline{T}_{\mathbf{\overline{126}}\supset\mathbf{\overline{45}}}&
\overline{T}_{\mathbf{126}\supset\mathbf{\overline{5}}}&
\overline{T}_{\mathbf{210}\supset\mathbf{\overline{5}}}&
\overline{T}_{\mathbf{126}\supset\mathbf{\overline{50}}}\\
\end{pmatrix}.\\
\end{split}\label{eq:model3-basis}
\end{align}
Based on these explicit matrices, we can perform fine-tuning in the parameter $\lambda_{444}$, such that 
\begin{align}
\det M_D&\approx 0,&
\det M_T&\neq 0.
\end{align}
Note that the $M_T$ matrix has an additional row and column compared to $M_D$, but additional triplet states get mixed with the others only in the second term.

The left (right) null eigenmode of $M_D|_{\lambda_{444}}$, where the vertical bar denotes the insertion of the fine-tuned expression $\lambda_{444}$, can then be solved for analytically. It corresponds to the MSSM Higgs $H_u$ ($H_d$), and has the components $a_i$ ($b_i$) in the basis $D_i$ ($\overline{D}_i$), where $i$ goes from $1$ to $4$. The coefficients $a_i$ and $b_i$ are functions of the parameters and VEVs, and are properly normalized by $\sum_i |a_i|^2=\sum_i |b_i|^2=1$. We omit here the resulting very complicated analytic expressions for $a_i$ and $b_i$, but we checked explicitly that their expressions are independent, so that they can indeed be treated as free parameters.

The (anti)doublets involved in the Yukawa sector are located in the $\mathbf{10}$ and the $\mathbf{\overline{126}}$. According to Table~\ref{table:H-quantities} we have
\begin{align}
H^{u}_{1}&:=H_{u}^{\mathbf{10}}=H_{\nu}^{\mathbf{10}},\\
H^{d}_{1}&:=H_{d}^{\mathbf{10}}=H_{e}^{\mathbf{10}},\\
H^{u}_{2}&:=H_{u}^{\mathbf{\overline{126}}}=H_{\nu}^{\mathbf{\overline{126}}},\\
H^{d}_{2}&:=H_{d}^{\mathbf{\overline{126}}}=H_{e}^{\mathbf{\overline{126}}},
\end{align}
and according to our basis labels $D_i$ and $\overline{D}_i$ in Eq.~\eqref{eq:model3-basis} and the definitions of $H^{u,d}_i$ in Eq.~\eqref{eq:hu-list} and \eqref{eq:hd-list} we have
\begin{align}
H^{u}_{1} & =a_1^\ast \,H_u+\ldots,\\
H^{d}_{1} & =b_1^\ast \,H_d+\ldots,\\
H^{u}_{2} & =a_3^\ast \,H_u+\ldots,\\
H^{d}_{2} & =b_2^\ast \,H_d+\ldots.
\end{align}
With these equations, the DT splitting and MSSM Higgs locations are completely determined, up to the values of $a_i$ and $b_i$, which we take as free parameters.

We now choose the 2nd family Yukawa operator $\mathcal{O}_{22}$; 
for simplicity we take simply $\mathbf{H}=\mathbf{\overline{126}}$, $m=m'=0$, $n=1$, $n'=0$, with the discrete direction $\alpha_1=X_1$. Using Eq.~\eqref{eq:result-discrete}, this setup gives the superpotential
\begin{align}
W&=\frac{\lambda_2}{\Lambda} \;\mathbf{16}_2\cdot \langle \mathbf{45}\rangle_{X_1}\cdot\mathbf{\overline{126}}\cdot\mathbf{16}_2 + \lambda_3 \mathbf{16}_3\cdot\mathbf{10}\cdot\mathbf{16}_3\\[6pt]
&=\phantom{+}\lambda_3 \;\Big(\;Q_3 \; u^c_3 \; H^{\mathbf{10}}_{u} \; (C_{ud}^{\mathbf{10}}+s^\mathbf{10} C_{ud}^{\mathbf{10}}) + Q_3 \; d^c_3\; H^{\mathbf{10}}_{d}\; (C_{ud}^{\mathbf{10}}+s^\mathbf{10} C_{ud}^{\mathbf{10}})\nonumber\\
&\qquad\qquad + L_{3}\;e^c_3\;H_d^{\mathbf{10}}\;(C_{e\nu}^{\mathbf{10}}+s^\mathbf{10} C_{e\nu}^{\mathbf{10}}) + L_3\; \nu^c_3\; H_u^{\mathbf{10}}\;(C_{e\nu}^{\mathbf{10}}+s^\mathbf{10} C_{e\nu}^{\mathbf{10}})\Big)\nonumber\\
&\quad\! + \frac{\lambda_2}{\Lambda}\; \mathcal{N}_{X_1}\;\Big(
Q_2 \; u^c_2 \; H^{\mathbf{\overline{126}}}_{u} \; 
(C_{ud}^{\mathbf{\overline{126}}}\;q_{X_1}(Q)+s^\mathbf{\overline{126}} C_{ud}^{\mathbf{\overline{126}}}\;q_{X_1}(u^c))\nonumber\\ 
&\qquad\qquad 
+ Q_2 \; d^c_2\; H^{\mathbf{\overline{126}}}_{d}\; 
(C_{ud}^{\mathbf{\overline{126}}}\;q_{X_1}(Q)+s^\mathbf{\overline{126}} C_{ud}^{\mathbf{\overline{126}}}\;q_{X_1}(d^c))\nonumber\\
&\qquad\qquad 
+ L_{2}\;e^c_2\;H_d^{\mathbf{\overline{126}}}\;
(C_{e\nu}^{\mathbf{\overline{126}}}\;q_{X_1}(Q)+s^\mathbf{\overline{126}} C_{e\nu}^{\mathbf{\overline{126}}}\;q_{X_1}(e^c))\nonumber\\
&\qquad\qquad
+ L_2\; \nu^c_2\; H_u^{\mathbf{\overline{126}}}\;
(C_{e\nu}^{\mathbf{\overline{126}}}\;q_{X_1}(L)+s^\mathbf{\overline{126}} C_{e\nu}^{\mathbf{\overline{126}}}\;q_{X_1}(\nu^c))\Big)\\[6pt]
&=2\sqrt{2}\,\lambda_3\,
(a_1^\ast\, Q_3\, u^c_3\, H_u + b_1^\ast \, Q_3\,d^c_3\,H_d + b_1^\ast L_3\,e^c_3\,H_d + a_1^{\ast}\,L_3\,\nu^c_3\,H_u)\nonumber\\
&\quad - 40\sqrt{2}\, \frac{\lambda_2}{\Lambda}
(a_3^\ast\, Q_3\, u^c_3\, H_u - b_2^\ast \, Q_3\,d^c_2\,H_d + 3b_2^\ast L_3\,e^c_3\,H_d -3 a_3^{\ast}\,L_2\,\nu^c_2\,H_u).
\end{align}
We ignored the $\overline{\Delta}$ term of right-handed neutrinos. The last line of the superpotential expression, written only with MSSM fields,
clearly gives the following Yukawa ratios:
\begin{align}
|y_\tau/y_b|&=1,&|y_t/y_b|&=|a_1/b_1|,&|y_{\nu_\tau}/y_b|&=|a_1/b_1|,\nonumber\\
|y_\mu/y_s|&=3,&|y_c/y_s|&=|a_3/b_2|,&|y_{\nu_\mu}/y_s|&=3|a_3/b_2|.
\end{align}
The directly observable predictions are the $4$ Yukawa ratios not involving the neutrinos. The ratios of the up and down sectors $y_t/y_b$ and $y_c/y_s$ involve the ratios $|a_1/b_1|$ and $|a_3/b_2|$, respectively, which can 
have arbitrary values (depending on the values of the parameters $m_{10}$, $m_{126}$, $m_{210}$, $\lambda_{145}$, $\lambda_{14\bar{5}}$ and $\lambda_{45\bar{5}}$ of the doublet sector, while $\lambda_{444}$ is fine-tuned).
The free parameters in this example are thus those in Eq.~\eqref{eq:model-parameters} and the additional two $|a_1/b_1|$ and $|a_3/b_2|$.

\begin{figure}[htb]            
            \begin{center}
            \includegraphics[width=7cm]{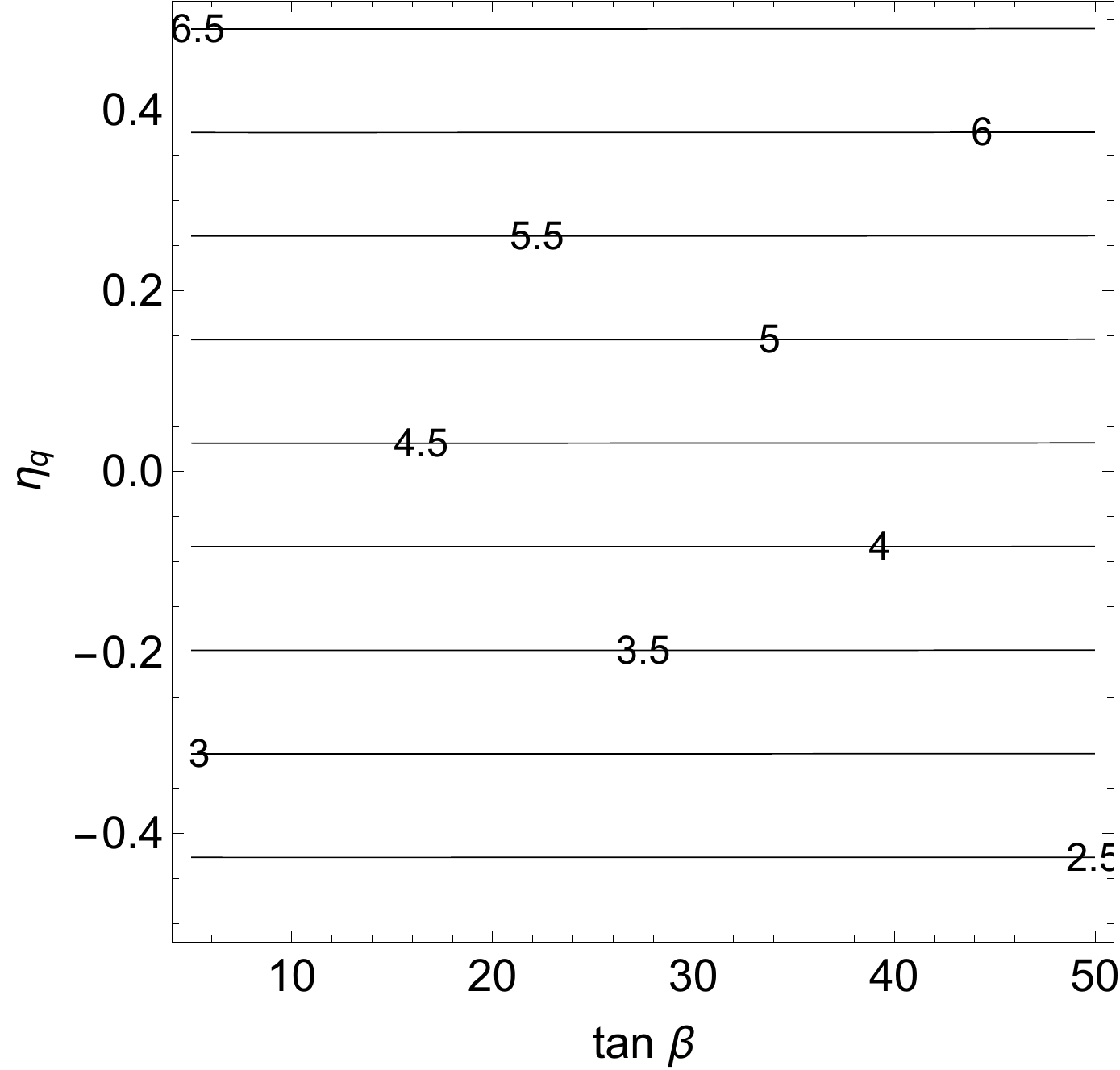}
            \par
            \caption{Contour plot for the value of $y_\mu/y_s$ in the $\tan\beta$-$\eta_q$ plane obtained from the data tables of \cite{Antusch:2013jca} by taking $\eta_b=-0.2$.\label{fig:tanbeta-etaq}}
            \end{center}
\end{figure}

The concrete predictions of this example model are thus only the charged lepton to down sector ratios $y_\tau/y_b$ and $y_\mu/y_s$. We already know that the $y_\tau/y_b$ ratio can be fit to $1$ from Figure~\ref{fig:tanbeta-etab}, from which we can read off $\eta_{b}\approx -0.2$. Using this value, we can construct a contour plot for the $y_\mu/y_s$ ratio values in the $\tan\beta$-$\eta_q$ plane given the $\eta_b$ value we specified, see Figure~\ref{fig:tanbeta-etaq}. It is clear from the figure that $y_\mu/y_s=3$ can be reached by $\eta_{q}\approx -0.31$, independent of $\tan\beta$. This shows the 2nd and 3rd Yukawa family can be successfully fit in Example 3.

\section{Conclusions}
We investigated in this paper a class of non-renormalizable $\mathrm{SO}(10)$ Yukawa operators of the form $(\mathbf{45}^{n}\cdot\mathbf{210}^{m}\cdot \mathbf{16}_I)\cdot\mathbf{H}\cdot(\mathbf{45}^{n'}\cdot\mathbf{210}^{m'}\cdot\mathbf{16}_J)$, where $\mathbf{H}\in\{\mathbf{10},\mathbf{120},\mathbf{\overline{126}}\}$. Their predictions in the up, down, charged-lepton and neutrino sectors were computed. It was determined which ingredients are needed for (predictive) model building, and we constructed three toy models as examples.

The model building approach taken was that of single operator dominance, so that in each entry of the Yukawa matrices a contribution from only one operator dominates.
In contrast to $\mathrm{SU}(5)$, where only the down and charged-lepton sectors are connected via operators, the $\mathrm{SO}(10)$ case connected all fermion sectors amongst themselves, i.e.~the operator predicts $3$ ratios of Yukawa entries, making such models potentially more predictive than $\mathrm{SU}(5)$ models. A complication arises due to $\mathrm{SO}(10)$ representations acquiring GUT-scale VEVs containing more than one SM singlet state, i.e.~the $\mathbf{45}$ contains $2$ and $\mathbf{210}$ contains $3$. We have considered $2$ cases in Section~\ref{sec:operators}: when these fields acquire a VEV in a discrete direction aligned with a state with well-defined transformation properties under one of the maximal subgroups ($G_{51}$ or $G_{422}$), or to allow for an arbitrary direction in singlet space, thus introducing new free parameters. Even in the latter case the model could still be predictive, since each operator predicts $3$ ratios of Yukawa entries, as well as the possibility of using the same GUT VEV fields in multiple operators, thus minimizing the number of introduced parameters. The main general results for the operators of Eq.~\eqref{eq:invariant-explicit} are collected in Eq.~\eqref{eq:result-discrete} of Section~\ref{sec:discrete-directions} for the discrete case and  Eq.~\eqref{eq:result-continuous} of Section~\ref{sec:arbitrary-directions} for the arbitrary direction case, with all the supporting definitions found in Section~\ref{sec:operators}. Some concrete examples for discrete alignments are shown in Figure~\ref{fig:yukawa_ratios} and Table~\ref{table:yukawa_ratios}.

Furthermore, we found that computing the resulting terms of the $\mathrm{SO}(10)$ operators is not sufficient, since there is an additional ambiguity in the location of the MSSM Higgses. The location of $H_u$ and $H_d$ is specified only when they are solved for as the left and right light eigenstate of the doublet-antidoublet mass matrix $M_D$, which is intimately connected to the issues of doublet-triplet splitting. The Higgs location then crucially influences the Yukawa predictions, cf.~Section~\ref{sec:Higgs-location}. This is another feature of the models not found in the $\mathrm{SU}(5)$ case. It arises in $\mathrm{SO}(10)$ for two reasons: because the predictions involve all Yukawa sectors, in particular sectors involving both $H_u$ and $H_d$ in the MSSM, as well as possibly having Higgs representations (the $\mathbf{120}$ in particular), which contain more than one doublet-antidoublet pair, which then couple differently to different Yukawa sectors. We provide the necessary tools for the reader to perform DT splitting and determine the Higgs location for the model of their choosing in Section~\ref{sec:DT-tools}, and suggest a list of predictive scenarios (not involving any additional free parameters) in Table~\ref{table:DT-models} of Section~\ref{section:DT-discussion}.

We considered both model building elements --- the operator choice and the Higgs location --- then in Section~\ref{sec:prescription-for-model-building} and discussed how to approach $\mathrm{SO}(10)$ flavor model building.
Achieving single operator dominance for example requires to forbid all operators except the desired one in any Yukawa matrix entry. These restrictions may be imposed by assigning charges under an extra (e.g.~global) symmetry: assigning charges to the external legs of the operators introduces constraints on the representations used in each Yukawa entry, while introducing $\mathbf{16}\oplus\mathbf{\overline{16}}$ mediators with charges restricts the types of internal contractions allowed in the operators. The mediator restrictions have been explored in detail in Appendix~\ref{appendix:operators-mediators}. The contractions leading to different 
Yukawa predictions correspond to choosing for each $\mathbf{45}$ and $\mathbf{210}$ factor with which of the two fermionic $\mathbf{16}_F$ representations it contracts, i.e.~on which side of $\mathbf{H}$ it is located. This allows for single operator dominance to be retained in almost all cases, except for some particular cases with only one type of $\mathbf{45}$ or $\mathbf{210}$ with an arbitrary direction VEV. 

We also presented $3$ examples of toy models involving only the 2nd and 3rd family Yukawa entries in Section~\ref{sec:toy-models}. Example 1 involves the most predictive possibility, when only discrete VEV directions and a predictive DT scenario are considered, and we identify the most promising candidate operators for the 2nd family. We then modify this example and further investigate an arbitrary VEV direction approach in Example 2, and a less-predictive DT scenario in Example 3. Both Examples 2 and 3 can also be successfully fit to the data. Overall, the $3$ examples show all the model building aspects and approaches discussed in this paper. 

Beyond this work, a complete $\mathrm{SO}(10)$ model would require a few more ingredients. The details of the entire Higgs sector were not studied, i.e.~it was not considered how $\mathrm{SO}(10)$ spontaneously breaks to the SM gauge group. Also, it was not considered how the VEV directions of the $\mathbf{45}$ and $\mathbf{210}$ in the Yukawa operators are achieved. These issues are from a model building perspective orthogonal to the predictions in the Yukawa sector, which this paper is concerned with, so we do not address them here.

A missing ingredient in the Yukawa sector, however, are the operators providing Majorana masses for right-handed neutrinos. An example of such an operator could for example be $(\mathbf{16}_F\,\mathbf{\overline{16}})^{2}$, with $\mathbf{\overline{16}}$ acquiring a GUT-scale VEV. Since the choice of such operators is again model dependent, we have not considered them in the analysis.

A potential limitation of this type of $\mathrm{SO}(10)$ model building is that Landau poles of the unified gauge coupling can occur well within one order of magnitude above the GUT scale, especially if one introduces many different representations $\mathbf{210}$ to the model. From this point of view there is a preference for representations of lower dimensionality, and a preference for a smaller number of them, i.e.~a preference for simpler models of this type. This is a consideration in building any complete model that should not be neglected.

In summary, this paper provides all the necessary model building tools and results for constructing flavor $\mathrm{SO}(10)$ GUTs, at least those based on the wide class of non-renormalizable Yukawa operators we considered. Applying these tools opens up new routes towards $\mathrm{SO}(10)$ flavor GUT models, in a similar spirit to single operator dominance models in $\mathrm{SU}(5)$. While there are more aspects to consider in $\mathrm{SO}(10)$, the model builder can be rewarded with an even more predictive Yukawa sector.

\section*{Acknowledgements}
The work of S.A., C.H.~and V.S.~has been supported by the Swiss National Science Foundation.
\appendix

\section{Conventions for $\SO(10)$ invariants\label{appendix:conventions}}

We review in this section the relevant conventions that we use for constructing and writing down $\mathrm{SO}(10)$ invariants. External references for some of these constructions, but not necessarily with the same conventions, can be found e.g.\ in \cite{Slansky:1981yr,Wilczek:1981iz,Bajc:2004xe,Aulakh:2000sn}.

The $45$ generators of $\mathrm{SO}(10)$ are labelled by $\Sigma_{pq}$, with antisymmetric indices $p$ and $q$ ($p,q=1,...,10$), and they fulfil the commutation relation
\begin{align}
[\Sigma_{pq},\Sigma_{rs}] = -i(\delta_{pr}\Sigma_{qs} + \delta_{qs}\Sigma_{pr} - \delta_{ps}\Sigma_{qr} - \delta_{qr}\Sigma_{ps}). \label{eq:commutator-generators-so10}
\end{align}
These commutation relations can be realized by $10 \times 10$ matrices defined by
\begin{align}
(\Sigma_{pq})_{rs} = i(\delta_{pr}\delta_{qs}-\delta_{ps}\delta_{qr}), \label{generators-fundamental}
\end{align}
normalized\footnote{This is different compared to the GUT normalization where the Dynkin index of the fundamental representation $\mathbf{10}$ of $\mathrm{SO}(10)$ is $1$.} to Dynkin index $2$. We refer to this basis of the $10$-dimensional irreducible representation as the \textit{real basis}; in the context of tensor methods in group theory, it has lower indices only and its components are written as $\mathbf{10}_p$. To form invariants it is convenient to introduce a \textit{complex basis}, where all states have well defined SM quantum numbers. This basis can be conveniently written by considering the $\mathbf{10}=\mathbf{5}\oplus\mathbf{\overline{5}}$ decomposition under the $\mathrm{SU(5)}$ subgroup, and it is written with an upper or a lower index $i$:
\begin{align}
\mathbf{10}^i :\; \begin{pmatrix} \mathbf{5} \\ \overline{\mathbf{5}} \end{pmatrix},\quad \mathbf{10}_i :\; \begin{pmatrix} \overline{\mathbf{5}} \\ \mathbf{5}\end{pmatrix}.
\label{eq:fundamental-complex-components}
\end{align}

We choose an $\mathrm{SU}(5)$ embedding where the entries in the $\mathbf{5}$ are of the form $\frac{1}{\sqrt{2}}(x_p + i x_{p+1})$, where the components are $x_p = \mathbf{10}_p$ and $p$ is odd; the $\overline{\mathbf{5}}$ contains the same entries, but with $-i$ instead of $i$. According to this embedding, the transition from the complex basis to the real basis is performed by the unitary $10\times10$ matrix $P_{pi}\equiv\mathbf{P}$, where the first five columns corresponding to the images of the basis vectors of $\mathbf{5}$ are of the form $\frac{1}{\sqrt{2}}\begin{matrix} (0 & ... & 0 & 1 & +i & 0 & ... & 0)^\textsf{T} \end{matrix}$, where $1$ and $i$ are in the locations $p$ and $p+1$ for odd $p$, and the last five columns, corresponding to the $\overline{\mathbf{5}}$, are of the same form but with $-i$ instead of $i$. This defines all other transition matrices between the real basis and one of the two complex bases (written with an upper or lower complex index, respectively):
\begin{align}
&\text{\phantom{anti}complex to real:}&P_{pi}&\equiv \mathbf{P},\nonumber\\
&\text{anticomplex to real:}&P_{p}{}^{i}&\equiv \mathbf{P}^*,\nonumber\\
&\text{real to \phantom{anti}complex:}&P^{i}{}_{p} & \equiv \mathbf{P}^{-1} = \mathbf{P}^\dagger,\nonumber\\
&\text{real to anticomplex:}&P_{ip} &\equiv (\mathbf{P}^\ast)^{-1} = \mathbf{P}^\textsf{T}, 
\label{eq:bases-change-fundamenal}
\end{align}
where we used the index notation for components on the left-hand side and matrix notation on the right-hand side. The indices $p,q,r,...$ are referred to as real indices, and $i,j,k,...$ as complex fundamental or antifundamental indices, depending on whether they are upper or lower, respectively. Observe that in the component notation all matrices are denoted by $P$, but have different index types and placement. For example, $P_{pi}$ transforms the complex basis to the real basis, i.e.~$\mathbf{10}_p=P_{pi}\,\mathbf{10}^{i}$, while $P^{i}{}_{p}$ is its inverse and transforms the real basis into the complex basis, i.e.~$\mathbf{10}^{i}=P^{i}{}_{p}\,\mathbf{10}_p$. The real indices are always lower, so the Einstein summation convention applies to them if two lower ones are repeated.

Eq.~\eqref{eq:bases-change-fundamenal} gives the following unitarity relations:
\begin{align}
P^{i}{}_{p}\,P_{pj}&=\delta^{i}{}_{j},&P_{pi}\,P^{i}{}_{q}&=\delta_{pq},\nonumber\\
P_{ip}\,P_{p}{}^{j}&=\delta_{i}{}^{j},&P_{p}{}^{i}\,P_{iq}&=\delta_{pq}.
\end{align}
Furthermore, the complex index $i$ can be raised or lowered by the matrices
\begin{align}
\begin{split}
P^{ij} &= {P^i}_p {P_p}^j \equiv \begin{pmatrix} 0 & \mathbb{1}_{5\times5} \\ \mathbb{1}_{5\times5} & 0 \end{pmatrix},\\
P_{ij} &= P_{ip} P_{pj} = (P^{ij})^* \equiv \begin{pmatrix} 0 & \mathbb{1}_{5\times5} \\ \mathbb{1}_{5\times5} & 0 \end{pmatrix}. \\
\end{split}
\label{eq:raise-lower-fundamental}
\end{align}
The representation $\mathbf{10}$ is real, so it contains $10$ real degrees of freedom $x_p$; in the complex basis, the $\mathbf{5}$ contains $5$ complex degrees of freedom, while the $\mathbf{\overline{5}}$ has those same $5$ complex degrees of freedom, but complex conjugated. If the $\mathbf{10}$ is complexified, i.e.~the $x_p$ are complex, then $\mathbf{5}$ and $\mathbf{\overline{5}}$ carry independent complex degrees of freedom.

The spinor representation of $\mathrm{SO}(10)$ is a reducible $32$-dimensional representations, which consists of the direct sum $\mathbf{16} \oplus \overline{\mathbf{16}}$. The components of the $\mathbf{32}$ are labelled by an upper spinor index $A$, where the first and second 16 components belong to the $\mathbf{16}$ and $\overline{\mathbf{16}}$, respectively. An alternative form is written with a lower spinor index, having the $\mathbf{16}$ and $\overline{\mathbf{16}}$ switched:
\begin{align}
\mathbf{32}^A :\; \begin{pmatrix} \mathbf{16} \\ \overline{\mathbf{16}} \end{pmatrix},\quad \mathbf{32}_A :\; \begin{pmatrix} -\overline{\mathbf{16}} \\ \phantom{-}\mathbf{16} \end{pmatrix}. \label{eq:spinor-components}
\end{align}
The minus sign in the definition of $\mathbf{32}_A$ is consistent with lowering the index of $\mathbf{32}^A$ by use of the charge conjugation matrix $C$, which shall be defined later.
The main building blocks in the construction of the spinor representation are the ten $32 \times 32$-dimensional gamma matrices $\Gamma_p$, which form an orthogonal basis of the Clifford algebra defined by the following anti-commutation relations:
\begin{align}
\{\Gamma_p,\Gamma_q\} = 2\delta_{pq}\mathbb{1}. \label{eq:commutator-gamma-so10}
\end{align}
An explicit form of the gamma matrices is given by the Kronecker product of five $2 \times 2$-dimensional matrices
\begin{align}
\begin{split}
\Gamma_{2k-1} &:= \mathbb{1} \otimes ... \otimes \mathbb{1} \otimes \tau_1 \otimes \tau_3 \otimes ... \otimes \tau_3, \\
\Gamma_{2k} &:= \underbrace{\mathbb{1} \otimes ... \otimes \mathbb{1}}_{k-1} \otimes \tau_2 \otimes \underbrace{\tau_3 \otimes ... \otimes \tau_3}_{5-k},
\end{split} \label{eq:gamma-construction}
\end{align}
where $k=1,...,5$ and $\tau_i$ are the Pauli matrices
\begin{align}
\tau_1 := \begin{pmatrix} 0 & \phantom{-}1 \\ 1 & \phantom{-}0 \end{pmatrix},\quad \tau_2 := \begin{pmatrix} 0 & -i \\ i & \phantom{-}0 \end{pmatrix},\quad \tau_3 := \begin{pmatrix} 1 & \phantom{-}0 \\ 0 & -1 \end{pmatrix}.
\end{align}
With this construction the gamma matrices are hermitian, i.e. $\Gamma_p = (\Gamma_p)^\dagger$. In the basis corresponding to $\mathbf{32}^A$ (see Eq.~\eqref{eq:spinor-components}) the gamma matrices have the index structure ${(\Gamma_p)^A}_B$, with the real fundamental index $p=1,...,10$ and the spinor indices $A,B=1,...,32$. In the same basis the block structure of the gamma matrices is 
\begin{align}
\Gamma_p = \begin{pmatrix}
\square & \blacksquare \\
\blacksquare & \square
\end{pmatrix},
\label{eq:gamma-block}
\end{align}
where each square represents a block of size $16\times 16$. The entries belonging to the white squares vanish, which implies a block off-diagonal structure. Thus, a product of an even or odd number of gamma matrices have the following block forms:
\begin{equation}
\text{odd number of }\Gamma_p\text{:}\; \begin{pmatrix}
\square & \blacksquare \\
\blacksquare & \square
\end{pmatrix},\quad
\text{even number of }\Gamma_p\text{:}\; \begin{pmatrix}
\blacksquare & \square \\
\square & \blacksquare
\end{pmatrix}.
\label{eq:gamma-product-block}
\end{equation}
The Hermitian generators in the spinor representation are defined as
\begin{align}
\Sigma_{pq}^\text{spinor} := \frac{i}{4}[\Gamma_p,\Gamma_q], \label{generators-spinor}
\end{align}
inheriting the spinor indices ${}^A{}_B$ and the block diagonal structure from the product of two gamma matrices, thus confirming the split $\mathbf{16}\oplus\mathbf{\overline{16}}$. Assuming the Clifford algebra relation of Eq.~\eqref{eq:commutator-gamma-so10}, the generators satisfy the commutation relations of Eq.~\eqref{eq:commutator-generators-so10}.

By definition, the charge conjugation matrix $C$ of $\mathrm{SO}(10)$ fulfils the identity
\begin{align}
C \Gamma_p C^{-1} = -(\Gamma_p)^\textsf{T} \label{eq:gamma-c-relation}
\end{align}
for all gamma matrices. Taking the hermiticity of the generators into account, this identity leads to the equation
\begin{align}
C \Sigma_{pq}C^{-1} = -(\Sigma_{pq})^*, \label{eq:generator-c-relation}
\end{align}
meaning that the spinor representation is rotated into its conjugate representation. In the basis $\mathbf{32}^A$ of Eq.~\eqref{eq:spinor-components} the matrix $C \equiv C_{AB}$ and its inverse $C^{-1} \equiv (C_{AB})^{-1} = C^{AB}$ have the following form
\begin{align}
C_{AB} \equiv \begin{pmatrix} 0 & -\mathbb{1}_{16\times16} \\ +\mathbb{1}_{16\times16} & \phantom{-}0 \end{pmatrix},\quad C^{AB} \equiv \begin{pmatrix} 0 & +\mathbb{1}_{16\times16} \\ -\mathbb{1}_{16\times16} & \phantom{-}0 \end{pmatrix}.
\label{eq:charge-conjugation-matrix}
\end{align}
Therefore $C_{AB}$ and $C^{AB}$ are used to lower and raise spinor indices, i.e.
\begin{align}
\mathbf{32}_A &= C_{AB}\,\mathbf{32}^B\,,\quad \mathbf{32}^A = C^{AB} \,\mathbf{32}_B,\label{eq:lower-raise-spinor}
\end{align}
which is consistent with the definitions in Eq.~\eqref{eq:spinor-components}.

When forming invariants, it is more practical to use the gamma matrices $\Gamma_i$ with a complex antifundamental index instead of a real one, determined via 
\begin{align}
(\Gamma_i)^{A}{}_{B}&=P_{ip}\,(\Gamma_p)^{A}{}_{B}.
\end{align}
Since only the basis of the fundamental index is changed, the block structures shown in Eq.~\eqref{eq:gamma-block} and \eqref{eq:gamma-product-block} are not affected.

The irreducible representations of $\mathrm{SO}(10)$ are located in tensor products of the $\mathbf{10}$, $\mathbf{16}$ and $\overline{\mathbf{16}}$. It is convenient to embed the $\mathbf{16}$ and the $\overline{\mathbf{16}}$ into the reducible representation $\mathbf{32}$ by setting the components of one of the parts to zero in Eq.~\eqref{eq:spinor-components}. Thus, the spinor index $A$ is used to label the components of both $\mathbf{16}$ and $\overline{\mathbf{16}}$, i.e.
\begin{align}
\mathbf{16}^A :\; \begin{pmatrix} \mathbf{16} \\ 0 \end{pmatrix},\quad
\overline{\mathbf{16}}^A :\; \begin{pmatrix} 0 \\ \overline{\mathbf{16}} \end{pmatrix}.
\end{align}
The lowest dimensional irreducible representations of $\mathrm{SO}(10)$ then have the following index structure:
\begin{align}
\mathbf{10}^i,\,\mathbf{16}^A,\,\overline{\mathbf{16}}^A,\,\mathbf{45}^{[ij]},\,\mathbf{54}^{(ij)},\,\mathbf{120}^{[ijk]},\,\mathbf{126}^{[ijklm]},\,\overline{\mathbf{126}}^{[ijklm]},\,\mathbf{144}^{Ai},\,\overline{\mathbf{144}}^{Ai},\,\mathbf{210}^{[ijkl]},\,\mathbf{210}'^{(ijk)},
\end{align}
where parentheses indicate symmetrization and square brackets indicate anti-symmetrization of the indices. The irreducible representations carrying only fundamental indices are real representations, whereas the ones with a spinor index are complex. We use a notation for the representations where all indices are by default upper.

In order to specify the components of the $\mathbf{54}$, $\mathbf{210}'$, $\mathbf{144}$ and $\overline{\mathbf{144}}$, lower dimensional representations have to be projected out of the tensor products
\begin{align}
\begin{split}
(\mathbf{10}\otimes\mathbf{10})^{(ij)} &= (\mathbf{54}\oplus\mathbf{1})^{(ij)},\\
(\mathbf{10}\otimes\mathbf{10}\otimes\mathbf{10})^{(ijk)} &= (\mathbf{210}'\oplus\mathbf{10})^{(ijk)},\\
(\mathbf{16}\otimes\mathbf{10})^{Ai} &= (\mathbf{144}\oplus\mathbf{16})^{Ai},\\
(\overline{\mathbf{16}}\otimes\mathbf{10})^{Ai} &= (\mathbf{144}\oplus\overline{\mathbf{16}})^{Ai},\\
\end{split}
\end{align}
by imposing the relations
\begin{align}
\mathbf{54}^{ij}\,P_{ij}&=0,&\mathbf{210}'^{ijk}\,P_{jk}&=0,&\Gamma_{i}{}^{A}{}_{B}\,\mathbf{144}^{Bi}&=0.
\end{align}
The $\mathbf{126}$ and $\overline{\mathbf{126}}$ are determined by a restriction to entries satisfying the (anti)self-duality identities
\begin{align}
\mathbf{126}_{[pqrst]} &= -\frac{i}{5!} \epsilon_{pqrstvwxyz} \mathbf{126}_{[vwxyz]},&
\overline{\mathbf{126}}_{[pqrst]} &= +\frac{i}{5!} \epsilon_{pqrstvwxyz} \overline{\mathbf{126}}_{[vwxyz]},
\end{align}
where $\epsilon$ denoting the rank $10$ completely anti-symmetric Levi-Civita symbol, and the real basis must be taken.

Invariant terms are formed by contracting indices of the same type. While in the case of complex fundamental and spinor indices an upper index has to be combined with a lower one, there is no such restriction in the real fundamental basis, where all indices are of the same height. In order to raise and lower indices, the matrices $P^{ij}$, $P_{ij}$, defined in Eq.~\eqref{eq:raise-lower-fundamental}, and the charge conjugation matrices $C^{AB}$, $C_{AB}$, defined in Eq.~\eqref{eq:charge-conjugation-matrix}, are used. Fundamental indices are transformed into spinor indices by the contraction with gamma matrices. The representations with anti-symmetric fundamental indices have a simple form in spinor space, since they can be written as $32 \times 32$ matrices with an upper-lower spinor index pair:
\begin{align}
{\mathbf{10}^A}_B &:= \mathbf{10}^i\,{{(\Gamma_i)}^A}_B, \label{eq:spinor-10}\\
{\mathbf{45}^A}_B &:= \mathbf{45}^{[ij]}\,{{(\Gamma_i \Gamma_j)}^A}_B, \label{eq:spinor-45}\\
{\mathbf{120}^A}_B &:= \mathbf{120}^{[ijk]}\,{{(\Gamma_i \Gamma_j \Gamma_k)}^A}_B, \label{eq:spinor-120}\\
{\mathbf{210}^A}_B &:= \mathbf{210}^{[ijkl]}\,{{(\Gamma_i \Gamma_j \Gamma_k \Gamma_l)}^A}_B, \label{eq:spinor-210}\\
{\mathbf{126}^A}_B &:= \mathbf{126}^{[ijklm]}\,{{(\Gamma_i \Gamma_j \Gamma_k \Gamma_l \Gamma_m)}^A}_B, \label{eq:spinor-126}\\
{\overline{\mathbf{126}}^A}_B &:= \overline{\mathbf{126}}^{[ijklm]}\,{{(\Gamma_i \Gamma_j \Gamma_k \Gamma_l \Gamma_m)}^A}_B. \label{eq:spinor-126bar}
\end{align}
They adhere to the block structure in Eq.~\eqref{eq:gamma-product-block}. In spinor space, contractions in products of such representations are thus simply done by matrix  multiplication. This notation is especially useful when forming invariants involving the $\mathbf{16}$ or $\overline{\mathbf{16}}$. For example, the invariant with two $\mathbf{16}$s, one $\mathbf{10}$ and one $\mathbf{45}$, can then be written as
\begin{align}
\mathbf{10}\otimes\mathbf{16}\otimes\mathbf{16}\otimes\mathbf{45} &\supset \mathbf{16}^A\, C_{AB}\, {\mathbf{45}^B}_C\, {\mathbf{10}^C}_D\, \mathbf{16}^D.
\end{align}

As a final aid for group theoretic considerations, we provide decompositions of the lowest dimensional irreducible representations of $\mathrm{SO}(10)$ under the maximal subgroups $G_{51}$ and $G_{422}$, listed in Tables~\ref{tab:decompositions-su5} and \ref{tab:decompositions-patisalam}. In addition, Table~\ref{tab:number-singlets-doublets} shows the number of SM singlets and weak doublets in each of these representations, which is relevant information for DT splitting considered in Section~\ref{sec:Higgs-location}.

\begin{table}[htb]
\caption{Decomposition of the lowest dimensional irreducible representations of $\mathrm{SO}(10)$ under the maximal subgroup $G_{51}$. The $\mathrm{U}(1)$ charges need to be multiplied with an additional factor $(2\sqrt{5})^{-1}$ for proper normalization corresponding to a Dynkin index $2$. Terms with a straight underline contain one SM singlet, while terms with a wavy underline contain one weak doublet, with an additional label $+$ or $-$ identifying the representation as $(\mathbf{1},\mathbf{2},\pm\frac{1}{2})$ of the SM gauge group $G_{321}$.\label{tab:decompositions-su5}}
\vskip -0.5 cm
\begin{align*}
\begin{array}{rcl}
\toprule
\mathrm{SO}(10) & \supset & \mathrm{SU}(5)\times\mathrm{U}(1) \\
\midrule
\mathbf{10} & = & \uwave{\mathbf{5}(+2)}\uset{+} \oplus \uwave{\overline{\mathbf{5}}(-2)}\uset{-} \\
\mathbf{16} & = & \uline{\mathbf{1}(-5)} \oplus \uwave{\overline{\mathbf{5}}(+3)}\uset{-} \oplus \mathbf{10}(-1) \\
\mathbf{45} & = & \uline{\mathbf{1}(0)} \oplus \mathbf{10}(+4) \oplus \overline{\mathbf{10}}(-4) \oplus \uline{\mathbf{24}(0)} \phantom{\uwave{(}}\\
\mathbf{54} & = & \mathbf{15}(+4) \oplus \overline{\mathbf{15}}(-4) \oplus \uline{\mathbf{24}(0)} \phantom{\uwave{(}}\\
\mathbf{120} & = & \uwave{\mathbf{5}(+2)}\uset{+} \oplus \uwave{\overline{\mathbf{5}}(-2)}\uset{-} \oplus \mathbf{10}(-6) \oplus \overline{\mathbf{10}}(+6) \oplus \uwave{\mathbf{45}(+2)}\uset{+} \oplus \uwave{\overline{\mathbf{45}}(-2)}\uset{-} \\
\mathbf{126} & = & \uline{\mathbf{1}(-10)} \oplus \uwave{\overline{\mathbf{5}}(-2)}\uset{-} \oplus \mathbf{10}(-6) \oplus \overline{\mathbf{15}}(+6) \oplus \uwave{\mathbf{45}(+2)}\uset{+} \oplus \overline{\mathbf{50}}(-2) \\
\mathbf{144} & = & \uwave{\mathbf{5}(-3)}\uset{+} \oplus \uwave{\overline{\mathbf{5}}(-7)}\uset{-} \oplus \overline{\mathbf{10}}(+1) \oplus \overline{\mathbf{15}}(+1) \oplus \uline{\mathbf{24}(+5)} \oplus \overline{\mathbf{40}}(+1) \oplus \uwave{\mathbf{45}(-3)}\uset{+} \\
\mathbf{210} & = & \uline{\mathbf{1}(0)} \oplus \uwave{\mathbf{5}(-8)}\uset{+} \oplus \uwave{\overline{\mathbf{5}}(+8)}\uset{-} \oplus \mathbf{10}(+4) \oplus \overline{\mathbf{10}}(-4) \oplus \uline{\mathbf{24}(0)} \oplus \mathbf{40}(-4) \oplus \overline{\mathbf{40}}(+4) \oplus \uline{\mathbf{75}(0)} \\
\mathbf{210}' & = & \mathbf{35}(-6) \oplus \overline{\mathbf{35}}(+6) \oplus \uwave{\mathbf{70}(+2)}\uset{+} \oplus \uwave{\overline{\mathbf{70}}(-2)}\uset{-} \\
\bottomrule
\end{array}
\end{align*}
\end{table}

\begin{table}
\caption{Decomposition of the lowest dimensional irreducible representations of 
$\mathrm{SO}(10)$ under the maximal subgroup $G_{422}$ (Pati-Salam group). Terms with a straight underline contain one SM singlet, whereas terms with a wavy underline contain one or two weak doublets, with a label of $+$ or $-$ discriminating between $(\mathbf{1},\mathbf{2},\pm\frac{1}{2})$ of the SM gauge group $G_{321}$ (with the symbol $\pm$ if both are present).\label{tab:decompositions-patisalam}}
\vskip -0.5cm
\begin{align*}
\begin{array}{rcl}
\toprule
\mathrm{SO}(10) & \supset & \mathrm{SU}(4)\times\mathrm{SU}(2)_L\times\mathrm{SU}(2)_R \\
\midrule
\mathbf{10} & = & \uwave{(\mathbf{1},\mathbf{2},\mathbf{2})}\uset{\pm} \oplus (\mathbf{6},\mathbf{1},\mathbf{1}) \\
\mathbf{16} & = & \uwave{(\mathbf{4},\mathbf{2},\mathbf{1})}\uset{-} \oplus \uline{(\overline{\mathbf{4}},\mathbf{1},\mathbf{2})} \\
\mathbf{45} & = & \uline{(\mathbf{1},\mathbf{1},\mathbf{3})} \oplus (\mathbf{1},\mathbf{3},\mathbf{1}) \oplus (\mathbf{6},\mathbf{2},\mathbf{2}) \oplus \uline{(\mathbf{15},\mathbf{1},\mathbf{1})} \phantom{\uwave{(}}\\
\mathbf{54} & = & \uline{(\mathbf{1},\mathbf{1},\mathbf{1})} \oplus (\mathbf{1},\mathbf{3},\mathbf{3}) \oplus (\mathbf{6},\mathbf{2},\mathbf{2}) \oplus (\mathbf{20}',\mathbf{1},\mathbf{1}) \phantom{\uwave{(}}\\
\mathbf{120} & = & \uwave{(\mathbf{1},\mathbf{2},\mathbf{2})}\uset{\pm} \oplus (\mathbf{6},\mathbf{1},\mathbf{3}) \oplus (\mathbf{6},\mathbf{3},\mathbf{1}) \oplus (\mathbf{10},\mathbf{1},\mathbf{1}) \oplus (\overline{\mathbf{10}},\mathbf{1},\mathbf{1}) \oplus \uwave{(\mathbf{15},\mathbf{2},\mathbf{2})}\uset{\pm} \\
\mathbf{126} & = & (\mathbf{6},\mathbf{1},\mathbf{1}) \oplus \uline{(\mathbf{10},\mathbf{1},\mathbf{3})} \oplus (\overline{\mathbf{10}},\mathbf{3},\mathbf{1}) \oplus \uwave{(\mathbf{15},\mathbf{2},\mathbf{2})}\uset{\pm} \\
\mathbf{144} & = & \uline{(\mathbf{4},\mathbf{1},\mathbf{2})} \oplus \uwave{(\overline{\mathbf{4}},\mathbf{2},\mathbf{1})}\uset{+} \oplus (\mathbf{4},\mathbf{3},\mathbf{2}) \oplus \uwave{(\overline{\mathbf{4}},\mathbf{2},\mathbf{3})}\uset{\pm} \oplus (\mathbf{20},\mathbf{1},\mathbf{2}) \oplus (\overline{\mathbf{20}},\mathbf{2},\mathbf{1}) \\
\mathbf{210} & = & \uline{(\mathbf{1},\mathbf{1},\mathbf{1})} \oplus (\mathbf{6},\mathbf{2},\mathbf{2}) \oplus \uwave{(\mathbf{10},\mathbf{2},\mathbf{2})}\uset{-} \oplus \uwave{(\overline{\mathbf{10}},\mathbf{2},\mathbf{2})}\uset{+} \oplus \uline{(\mathbf{15},\mathbf{1},\mathbf{1})} \oplus \uline{(\mathbf{15},\mathbf{1},\mathbf{3})} \oplus (\mathbf{15},\mathbf{3},\mathbf{1}) \\
\mathbf{210}' & = & \uwave{(\mathbf{1},\mathbf{2},\mathbf{2})}\uset{\pm} \oplus (\mathbf{1},\mathbf{4},\mathbf{4}) \oplus (\mathbf{6},\mathbf{1},\mathbf{1}) \oplus (\mathbf{6},\mathbf{3},\mathbf{3}) \oplus (\mathbf{20}',\mathbf{2},\mathbf{2}) \oplus (\mathbf{50},\mathbf{1},\mathbf{1}) \\
\bottomrule
\end{array}
\end{align*}
\end{table}

\begin{table}
\caption{Number of SM singlets and weak doublets contained in the lowest dimensional irreducible representations of $\mathrm{SO}(10)$, according to the decompositions in Tables~\ref{tab:decompositions-su5} and \ref{tab:decompositions-patisalam}. The labels $S$, $D$ and $\overline{D}$ indicate the representations $(\mathbf{1},\mathbf{1},0)$, $(\mathbf{1},\mathbf{2},+\frac{1}{2})$ and $(\mathbf{1},\mathbf{2},-\frac{1}{2})$ of the SM gauge group $G_{321}$.\label{tab:number-singlets-doublets}}
\setlength{\arraycolsep}{0.5cm}
\renewcommand{\arraystretch}{1.3}
\vskip -0.5cm
\begin{align*}
\begin{array}{rccc}
\toprule
\mathrm{SO}(10) & S & D & \overline{D} \\
\midrule
\mathbf{10} & 0 & 1 & 1 \\
\mathbf{16} & 1 & 0 & 1 \\
\mathbf{45} & 2 & 0 & 0 \\
\mathbf{54} & 1 & 0 & 0 \\
\mathbf{120} & 0 & 2 & 2 \\
\mathbf{126} & 1 & 1 & 1 \\
\mathbf{144} & 1 & 2 & 1 \\
\mathbf{210} & 3 & 1 & 1 \\
\mathbf{210}' & 0 & 1 & 1 \\
\bottomrule
\end{array}
\end{align*}
\renewcommand{\arraystretch}{1}
\end{table}

\section{Construction of operators via mediators\label{appendix:operators-mediators}}

In this appendix we investigate the conditions under which \textit{single operator dominance} in model building in the Yukawa sector can be justified without simply setting the coefficients in front of unwanted operators to zero.

We focus on the class of non-renormalizable superpotential operators in Eq.~\eqref{eq:invariant-explicit} that are of relevance to this paper. These operators generate Yukawa terms once the SM singlets acquire GUT-scale VEVs. The generic form of the operators is 
	\begin{align}
	W \supset \mathbf{16}_I\cdot\mathbf{16}_J\cdot\mathbf{H}\cdot\mathbf{45}^{n+n'}\cdot\mathbf{210}^{m+m'},
	\label{eq:generic-superpotential-operator}
	\end{align}
where the SM fermions of each family are embedded into a representation $\mathbf{16}$ with family indices $I,J\in\{1,2,3\}$, and where $\mathbf{H}$ is a Higgs representation. The options for $\mathbf{H}$ are $\mathbf{10}$, $\mathbf{120}$ or $\overline{\mathbf{126}}$ and it contains (some part of) the MSSM Higgs doublet/antidoublet $(\mathbf{1},\mathbf{2},\pm1/2)$ with an EW scale VEV, cf.~Section~\ref{sec:Higgs-location}. Furthermore, the $n+n'$ representations $\mathbf{45}$ and $m+m'$ representations $\mathbf{210}$ acquire GUT-scale VEVs in their SM singlet components, yielding the Yukawa terms. 

Crucially, specifying the powers $n+n'$ and $m+m'$ in Eq.~\eqref{eq:generic-superpotential-operator}  does not determine the operator uniquely. The order of the GUT-scale VEVs and $\mathbf{H}$ can for example be reshuffled in Eq.~\eqref{eq:invariant-explicit}. Furthermore, there are alternative ways to contract the indices of representations to those of Eq.~\eqref{eq:invariant-explicit}. For example, in the case $\mathbf{H} = \mathbf{10}$, $n=1$ and $m=m'=n'=0$, the two possible independent ways of contracting the indices are the following:
\begin{align}
{\mathbf{45}^A}_D\, (\mathbf{16}_I)^D\, C_{AB}\, {\mathbf{10}^B}_E\, (\mathbf{16}_J)^E, \label{eq:index-contraction-ex1}\\
(\mathbf{16}_I)^A\, C_{AB}\, {(\mathbf{45}\cdot\mathbf{10})^B}_E\, (\mathbf{16}_J)^E, \label{eq:index-contraction-ex2}
\end{align}
where
\begin{align}
{(\mathbf{45}\cdot\mathbf{10})^B}_E = \mathbf{45}^{ij}\, P_{jk}\, \mathbf{10}^k\, {(\Gamma_i)^B}_E, \label{eq:index-contraction-ex2-1}
\end{align}
with all the necessary definitions found in Appendix~\ref{appendix:conventions}. The difference in the contractions is whether the $\mathbf{10}$ contracts to spinor form via $\Gamma_i$ in Eq.~\eqref{eq:index-contraction-ex1}, or contracts though the $\mathbf{45}$ via a fundamental index in Eq.~\eqref{eq:index-contraction-ex2}. The two operators in Eq.~\eqref{eq:index-contraction-ex1} and \eqref{eq:index-contraction-ex2} turn out to form independent invariants while containing the same representations. Imposing extra symmetries (e.g.~global) on these fields cannot discriminate between such contraction ambiguities, thus implying the presence of all such operators in the superpotential, which is at odds with single operator dominance.

\begin{figure}
\begin{center}
\includegraphics[scale=0.56]{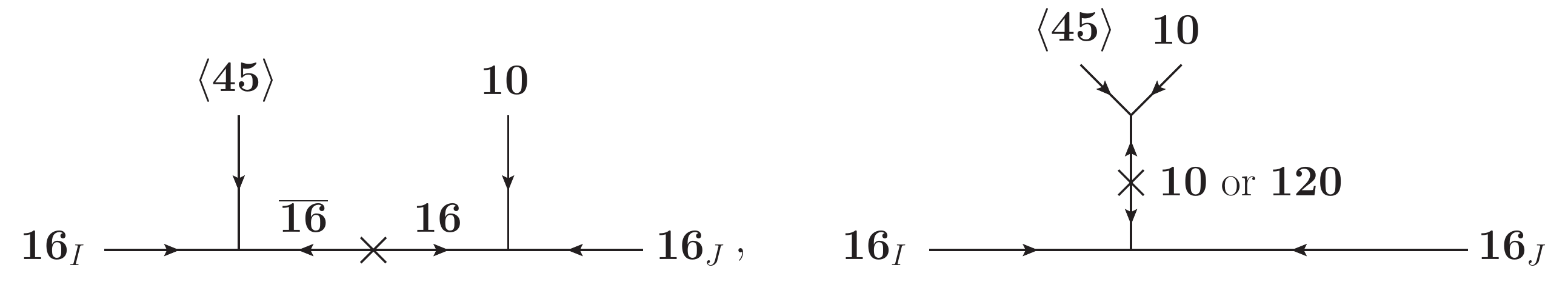}
\end{center}
\caption{Non-renormalizable superpotential operators formed via mass insertions of heavy mediator fields from renormalizable operators. The two diagrams with the same external fields, but different mediators, lead to different non-renormalizable operators, i.e.~those in Eq.~\eqref{eq:index-contraction-ex1} and \eqref{eq:index-contraction-ex2}. The mass insertions are labelled by a $\times$. The angle brackets denote the SM singlet GUT-scale VEV of the scalar component of the $\mathbf{45}$, such that each diagram forms a Yukawa term at the MSSM level.}
\label{fig:mediator-ex}
\end{figure}

A way to circumvent this limitation is to impose symmetries in an extended theory containing additional mediator fields, which effectively control the allowed contractions (analogous to what can be done in $\mathrm{SU}(5)$, see~\cite{Antusch:2009gu,Antusch:2013rxa}). We could imagine that the extended theory is renormalizable, while the non-renormalizable operators under study arise as effective operators once the heavy mediator fields are integrated out above the GUT scale. For example, in the extended theory the operator in Eq.~\eqref{eq:index-contraction-ex1} is formed via mediators in the representations $\mathbf{16}$ and $\mathbf{\overline{16}}$, and a mass insertion from $\mathbf{16}\cdot\overline{\mathbf{16}}$. The operator in Eq.~\eqref{eq:index-contraction-ex2}, on the other hand, requires a mediator $\mathbf{10}$, and a mass insertion from $\mathbf{10}\cdot\mathbf{10}$. A mediator $\mathbf{120}$ is also possible, but it does not yield an expression linearly independent from the other two. The construction of these operators is presented graphically in Figure~\ref{fig:mediator-ex}. 

\begin{figure}
\begin{center}
\includegraphics[scale=0.56]{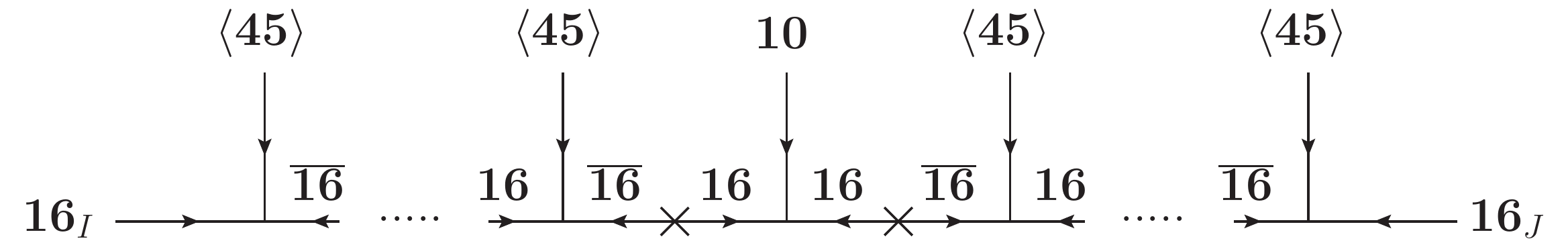}
\end{center}
\caption{A generic non-renormalizable superpotential operator of the form $\mathbf{16}_I\cdot\mathbf{16}_J\cdot\mathbf{10}\cdot\mathbf{45}^{n+n'}$ written using renormalizable interaction terms with mediators of the type $\mathbf{16}$ and $\overline{\mathbf{16}}$. The mass insertions are labelled by a $\times$. The angle brackets denote the SM singlet GUT-scale VEVs of the scalar component of the $\mathbf{45}$s, such that the diagram forms a Yukawa term. Neighboring VEV insertions commute among each other, but not with the insertion of the EW scale VEV of the Higgs field representation. Thus, it is crucial whether a $\langle\mathbf{45}\rangle$ is located on the left- or right-hand side of the $\mathbf{10}$.}
\label{fig:mediator-16-16bar}
\end{figure}

In Figure~\ref{fig:mediator-16-16bar} a generic operator of the form $\mathbf{16}_I\cdot\mathbf{16}_J\cdot\mathbf{10}\cdot\mathbf{45}^{n+n'}$, i.e. $\mathbf{H} = \mathbf{10}$ and $m=m'=0$, is written by using only renormalizable interactions with the $\mathbf{16}$ and $\overline{\mathbf{16}}$ mediators. In general each $\mathbf{45}$ factor is a different field with a different VEV alignment. If $m,m'\neq 0$, the diagram stays the same, but with some of the $\langle\mathbf{45}\rangle$ replaced by $\langle\mathbf{210}\rangle$. Additionally, the $\mathbf{10}$ can be replaced by a $\mathbf{120}$ or $\overline{\mathbf{126}}$ if we make a choice of a different Higgs field representation. Since in spinor notation the matrices ${\langle\mathbf{45}\rangle^A}_B$ and ${\langle\mathbf{210}\rangle^A}_B$ are diagonal, these VEVs commute among each other, but they do not commute with the EW scale VEV of the Higgs representation ${\mathbf{H}^A}_B$. The relative order of the GUT-scale VEVs on each side of $\mathbf{H}$ is thus irrelevant, in the sense that they yield the same low energy Yukawa operators. The only relevant aspect is whether a VEV is located on the left- or on the right-hand side of the Higgs representation $\mathbf{H}$.

\begin{figure}
\begin{center}
\includegraphics[scale=0.56]{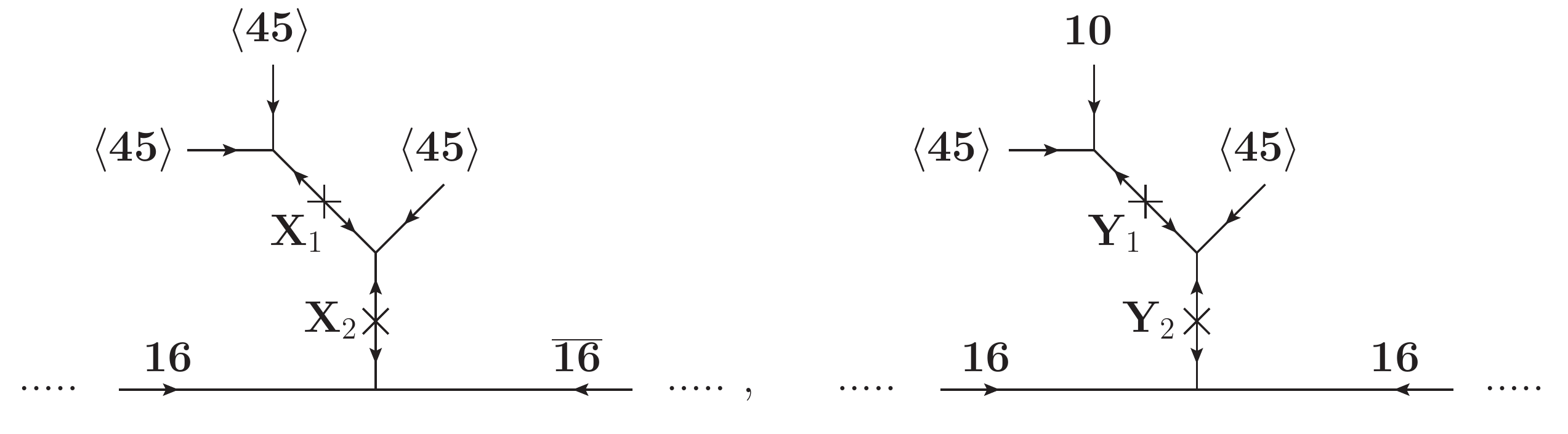}
\end{center}
\caption{Examples of how external legs from Figure~\ref{fig:mediator-16-16bar} can have a complicated tree substructure if representations other than $\mathbf{16}$ and $\mathbf{\overline{16}}$ are considered for mediators. A VEV external leg is shown on the left (only $\langle\mathbf{45}\rangle$), whereas a Higgs external leg with a $\mathbf{10}$ is shown on the right.
An incomplete list of examples: $(\mathbf{X}_{1},\mathbf{X}_2)\in\{(\mathbf{45},\mathbf{45}),(\mathbf{45},\mathbf{210}),(\mathbf{54},\mathbf{45})\}$, and $(\mathbf{Y}_{1},\mathbf{Y}_2)\in\{(\mathbf{10},\mathbf{10}),(\mathbf{120},\mathbf{\overline{126}}),(\mathbf{10},\mathbf{120}))\}$. The mediator labels are written schematically; the mass insertion of $\mathbf{X}_i$ is actually a vertex $\mathbf{X}_{i}\cdot\mathbf{\overline{X}}_i$, so both $\mathbf{X}_i$ and its conjugate representation need to be used, and analogous for $\mathbf{Y}_i$. All mediator representations of the tree substructure except for $\mathbf{\overline{126}}$ are real though, in the sense that they form quadratic invariants. }
\label{fig:general-external-legs}
\end{figure}

If in addition non-spinor representations are used for mediators alongside $\mathbf{16}\oplus\mathbf{\overline{16}}$, the external legs with a $\langle\mathbf{45}\rangle$ or $\mathbf{10}$ can form complicated tree graph structures, as shown in Figure~\ref{fig:general-external-legs}. Considering only the mediators $\mathbf{16}\oplus\overline{\mathbf{16}}$ thus limits us to diagrams with simple external legs as in Figure~\ref{fig:mediator-16-16bar} and  prevents complications from tree substructures of external legs connecting to the ``fermion line'' of $\mathbf{16}$s as in Figure~\ref{fig:general-external-legs}. This simple case of a fermion line of $\mathbf{16}$ and $\mathbf{\overline{16}}$ representations, to which all other external legs attach, exactly corresponds to the explicit contractions through spinor indices in Eq.~\eqref{eq:invariant-explicit} of the operators we consider in this paper.

\begin{figure}
\begin{center}
\includegraphics[scale=0.56]{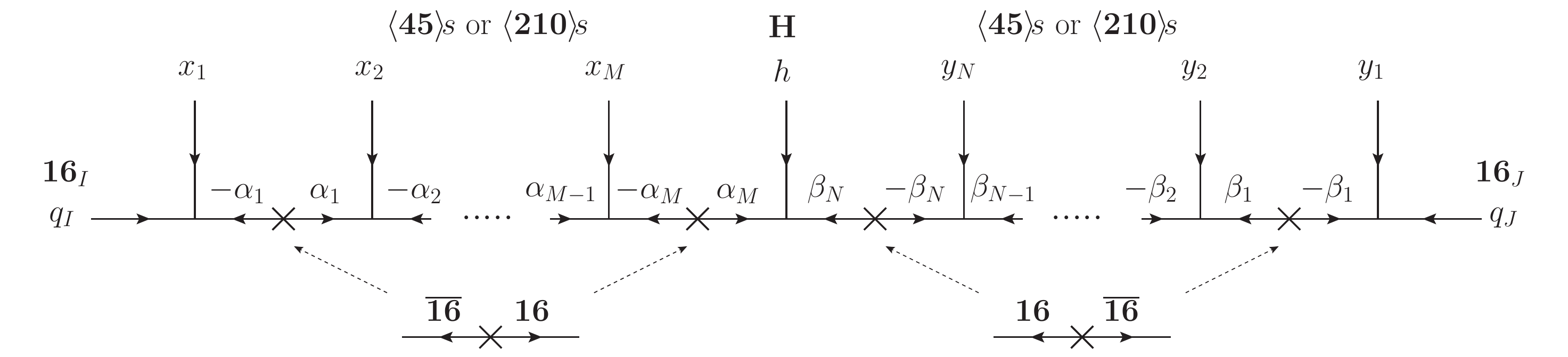}
\end{center}
\caption{The global charge assignments to the fields in Figure~\ref{fig:mediator-16-16bar}: $h$ and $q_{I(J)}$ label the charges of the $\mathbf{H}$ and $\mathbf{16}_{I(J)}$, respectively. On the left-hand side of the $\mathbf{H}$, the $\mathbf{45}$s or $\mathbf{210}$s have charges $x_i$ and the mediators $\mathbf{16}$, $\overline{\mathbf{16}}$ carry charges $\alpha_i$, $-\alpha_i$. Analogous labels are used on the right-hand side of $\mathbf{H}$ with replacements $x_i \mapsto y_i$ and $\alpha_i \mapsto \beta_i$.}
\label{fig:mediator-16-16bar-charges}
\end{figure}

We are interested in constructing operators which lead to unique predictions for the Yukawa couplings via single operator dominance. Even using just the $\mathbf{16}\oplus\mathbf{\overline{16}}$ mediators, we still need to control the order in which the $\mathbf{45}$s and $\mathbf{210}$s attach to the fermion line 
in Figure~\ref{fig:mediator-16-16bar}. One can introduce a global $\mathrm{U}(1)$ symmetry, or a suitable discrete subgroup, to distinguish between these cases. The assignment of the charges to the fields is shown in Figure~\ref{fig:mediator-16-16bar-charges}: we assume there are $M$ VEV legs to the left of $\mathbf{H}$, and $N$ VEV legs to the right, so that $M=m+n$ and $N=m'+n'$. If the mediators are integrated out, the diagram in Figure~\ref{fig:mediator-16-16bar-charges} corresponds to the non-renormalizable operator
\begin{align}
\Big(({\mathbf{X}_M)^{A_M}}_{A_{M-1}}\,...\,{(\mathbf{X}_1)^{A_1}}_D\, (\mathbf{16}_I)^D\Big)\, C_{A_{M}B}\; {\mathbf{H}^B}_{E_N}\,\Big(({\mathbf{Y}_N)^{E_N}}_{E_{N-1}}\,...\,{(\mathbf{Y}_1)^{E_1}}_F\, (\mathbf{16}_J)^F\Big),
\label{eq:index-contraction-16-16bar-general}
\end{align}
where each $\mathbf{X}_i$ and $\mathbf{Y}_j$ is either a $\mathbf{45}$ or a $\mathbf{210}$. Note that a different position of the $C$ matrix in the product would give the same invariant up to a minus sign due to the commutation relation with gamma matrices in Eq.~\eqref{eq:gamma-c-relation}. The global charges of the mediators on the left- and right-hand side of the Higgs field are labelled by $\alpha_i$ and $\beta_j$, respectively.\footnote{Note that these charges are unrelated to the alignments, which are also labeled by $\alpha$ and $\beta$ in Eq.~\eqref{eq:invariant-general}.} Note that for each $\alpha_i$ and $\beta_j$ we have in principle a different pair of mediators $\mathbf{16}\oplus\mathbf{\overline{16}}$. The sum of the charges in each vertex needs to be zero, yielding a large system of equations. The mediator charges are computed to be
\begin{align}
\alpha_i &= q_I + \sum_{s=1}^i x_s\quad(i=1,...,M),\label{eq:charge-alpha}\\
\beta_j &= q_J + \sum_{t=1}^j y_t\quad(j=1,...,N),\label{eq:charge-beta}
\end{align}
where $q_{I(J)}$ are the charges of the $\mathbf{16}_{I(J)}$, and $x_s$ and $y_t$ are the charges of the $\mathbf{X}_s$ and $\mathbf{Y}_t$ representations in Eq.~\eqref{eq:index-contraction-16-16bar-general} on the left- and right-hand side of $\mathbf{H}$, respectively.
The charge $h$ of the Higgs field representation is chosen such that the total charge of the non-renormalizable operator vanishes:
\begin{align}
h = -\big(q_I + q_J + \sum_{s=1}^M x_s + \sum_{t=1}^N y_t\big).
\end{align}
Thus, specifying the charges of the $\mathbf{16}_{I(J)}$ and the $\mathbf{45}$s or $\mathbf{210}$s is sufficient for all the other charges (the charges of the Higgs representation and of the mediators) to be fixed as well. In general, the charge of the Higgs needs to be consistent in the wider context of multiple operators (since we populate multiple entries of the Yukawa matrix), which can be checked already at the level of external legs only. If we choose a discrete global $\mathbb{Z}_k$ symmetry instead of $\mathrm{U}(1)$, all the charge equations hold modulo $k$. 

In any given model, if the mediator mechanism is employed to impose single operator dominance, it needs to be checked with all mediators introduced into the model that they do not allow also undesired diagrams. In particular, mediators introduced for one operator may allow undesired contractions in another operator. This appendix has provided the reader with all the necessary considerations and tools to check the consistency explicitly in any given model. We also provide some general conclusions on forbidding undesired diagrams below, but only when considering an operator in a single Yukawa entry.

A diagram as in Figure~\ref{fig:mediator-16-16bar} is said to be \textit{protected} by a set of global charges of the fields if there exists no other diagram with the same external legs using the same set (or a subset) of mediators. This implies that permutations among the VEV legs or with the Higgs leg are forbidden for a protected diagram. As discussed earlier, only the relative position of the VEVs to the Higgs $\mathbf{H}$ impacts the MSSM predictions, so having a protected diagram is a sufficient (but not necessary) condition for single operator dominance. Whether a diagram can be protected or not is discussed for the following two cases:
\begin{itemize}
\item $I \neq J$: 
\vspace{-0.2cm}
	\begin{itemize}[leftmargin=0.5cm]
	\item Such diagrams can always be protected. For example, choose $x_s>0$, $y_t>0$ and $q_J=q_I+\sum_{s=1}^m x_s+\sum_{t=1}^n y_t$, where different $\mathbf{45}$s or $\mathbf{210}$s are assumed to have different charges $x_s$ and $y_t$, so that no one charge is a sum of some of the others. This then yields a unique allowed order of external legs, assuming no special relations between $x_s$ and $y_t$, i.e.~for a random choice of rational charges.
	\end{itemize}
\item $I=J$:
\vspace{-0.2cm}
	\begin{itemize}[leftmargin=0.5cm]
	\item If all $\mathbf{45}$s or $\mathbf{210}$s are the same field ($\mathbf{X}_i=\mathbf{Y}_j$ for all $i$ and $j$): Only the cases $|M-N|\leq 1$ can be protected, i.e. only the diagrams which are as symmetric as possible in the number of external legs left and right of $\mathbf{H}$. In all other cases the diagrams always come along with more symmetric diagrams, which potentially generate different Yukawa couplings, as illustrated in Figure~\ref{fig:mediator-16-16bar-ex2}.
	\item If not all $\mathbf{45}$s or $\mathbf{210}$s are the same field: for a generic order of external legs, there may be no protection possible. However, the values of the Yukawa couplings do not depend on the specific order of the $\mathbf{45}$s or $\mathbf{210}$s while permuting on the same side of $\mathbf{H}$. It was checked numerically that among the diagrams which are equivalent from the point of view of the Yukawa couplings, there always exists one which is protected. An example is given in Figure~\ref{fig:mediator-16-16bar-ex1}. Incidentally, this provides further motivation for our choice of operators under consideration in Eq.~\eqref{eq:invariant-explicit}, since all such operators (with multiple fields) can be protected for some internal reordering of $\{\alpha_i,\beta_j\}$ and some reordering of $\{\alpha'_k,\beta'_l\}$ while retaining the same MSSM Yukawa prediction.

	\begin{figure}
	\begin{center}
	\includegraphics[scale=0.56]{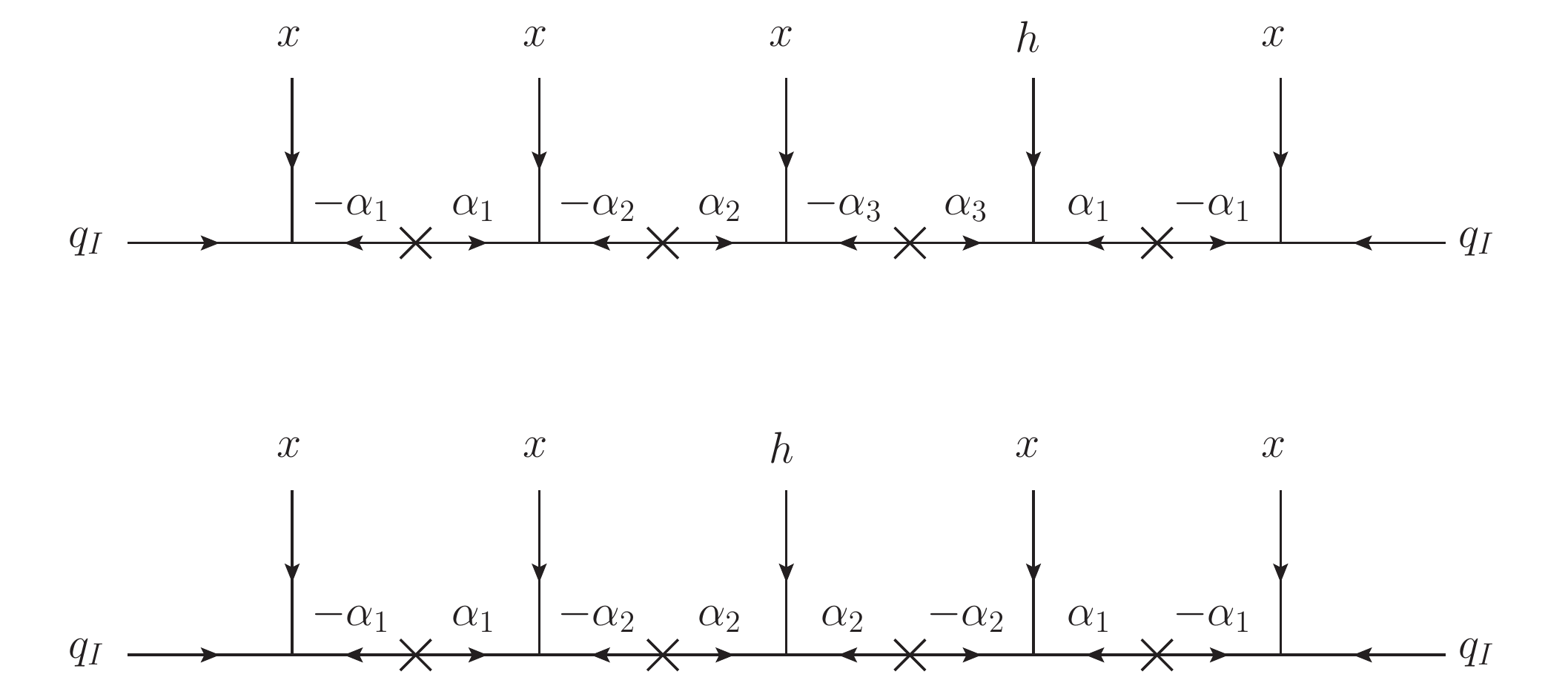}
	\end{center}
	\caption{An example of a diagram (upper) in the case of $I=J$ and only one different $\mathbf{45}$ or $\mathbf{210}$ field with global $\mathrm{U}(1)$ charge $x$, which cannot be protected. If the upper diagram is present, then the lower, more symmetric one can be constructed, using a subset of the mediators.}
	\label{fig:mediator-16-16bar-ex2}
	\end{figure}

	\begin{figure}
	\begin{center}
	\includegraphics[scale=0.56]{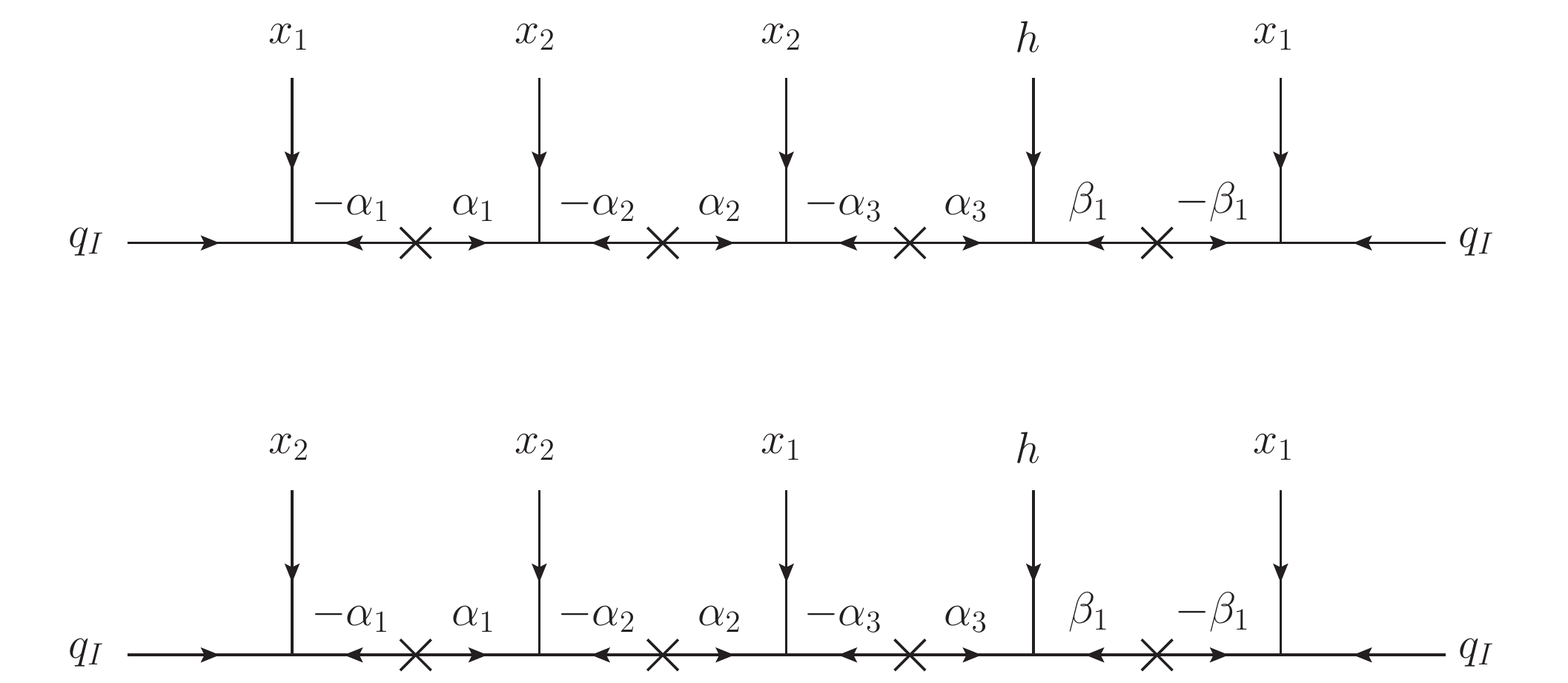}
	\end{center}
	\caption{Example of two diagrams in the case of $I=J$ providing the same Yukawa couplings, but where only one of them can be protected. They contain two different $\mathbf{45}$ or $\mathbf{210}$ fields with charges $x_1$ and $x_2$. For any choice of global charges in the first diagram, the symmetric diagram with a pair of legs $x_1$-$x_2$ on both sides of $\mathbf{H}$ can also be constructed. In contrast, the second diagram, where the fields on the left-hand side of the Higgs fields are permuted, can be protected by a suitable choice of charges, for example $q_I=0$, $x_1=3$ and $x_2=1$.}
	\label{fig:mediator-16-16bar-ex1}
	\end{figure}
	\end{itemize}
\end{itemize}
In summary, for any non-renormalizable operator as in Eq.~\eqref{eq:index-contraction-16-16bar-general}, there exists a diagram as in Figure~\ref{fig:mediator-16-16bar} which can be protected, and which leads to a unique Yukawa operator when integrating out the mediators. Exceptions occur only in the case of $I=J$ and only one different $\mathbf{45}$ or $\mathbf{210}$ field, where diagrams which are not as symmetric as possible ($|M-N|>1$) cannot be protected. Note that the protection we considered applies only when considering operators for a single Yukawa entry; adding more Yukawa entries (more operators) may introduce new mediators with charges, which interfere with the protection of the previous operators. The relevance of the concept of protected operators in concrete models is thus the following: operators that cannot be protected also cannot be used for single operator dominance; protected operators can be used freely in the preliminary model building stages, but subsequent consistency of allowing only a certain set of operators needs to be checked with all external fields and introduced mediators.

\vskip 1cm

\end{document}